\documentclass{aa}

\usepackage{graphicx}
\usepackage{color}

\usepackage{txfonts}
\usepackage[normalem]{ulem}
\usepackage{natbib,twoopt}
\usepackage[breaklinks=true]{hyperref}
\bibpunct{(}{)}{;}{a}{}{,} 
\makeatletter
\newcommandtwoopt{\citeads}[3][][]{\href{http://adsabs.harvard.edu/abs/#3}%
{\def\hyper@linkstart##1##2{}%
\let\hyper@linkend\@empty\citealp[#1][#2]{#3}}}
\newcommandtwoopt{\citepads}[3][][]{\href{http://adsabs.harvard.edu/abs/#3}%
{\def\hyper@linkstart##1##2{}%
\let\hyper@linkend\@empty\citep[#1][#2]{#3}}}
\newcommandtwoopt{\citetads}[3][][]{\href{http://adsabs.harvard.edu/abs/#3}%
{\def\hyper@linkstart##1##2{}%
\let\hyper@linkend\@empty\citet[#1][#2]{#3}}}
\newcommandtwoopt{\citeyearads}[3][][]%
{\href{http://adsabs.harvard.edu/abs/#3}
{\def\hyper@linkstart##1##2{}%
\let\hyper@linkend\@empty\citeyear[#1][#2]{#3}}}
\makeatother
\usepackage{amstext}

\newcommand{\Gaia}{\textit{Gaia}}
\newcommand{\Hipparcos}{\textsc{Hipparcos}}
\newcommand\TychoTwo{\textit{Tycho}-2}
\newcommand{\GDR}{\Gaia\ DR}
\newcommand{\G}{$G$}
\newcommand{\BP}{$G_{\rm BP}$}
\newcommand{\RP}{$G_{\rm RP}$}

\begin{document} 

   \title{Gaia Data Release 2}

   \subtitle{Photometric content and validation}

\author{
  D.~W.~Evans\inst{\ref{inst:ioa}}                      
  \and M.~Riello\inst{\ref{inst:ioa}}                   
  \and F.~De~Angeli\inst{\ref{inst:ioa}}                
  \and J.~M.~Carrasco\inst{\ref{inst:ub}}               
  \and P.~Montegriffo\inst{\ref{inst:oabo}}             
  \and C.~Fabricius\inst{\ref{inst:ub}}         
  \and C.~Jordi\inst{\ref{inst:ub}}                     
  \and L.~Palaversa\inst{\ref{inst:ioa}}                
  \and C.~Diener\inst{\ref{inst:ioa}}                   
  \and G.~Busso\inst{\ref{inst:ioa}}                    
  \and C.~Cacciari\inst{\ref{inst:oabo}}                
  \and F.~van~Leeuwen\inst{\ref{inst:ioa}}              
  \and P.~W.~Burgess\inst{\ref{inst:ioa}}               
  \and M.~Davidson\inst{\ref{inst:roe}}              
  \and D.~L.~Harrison\inst{\ref{inst:ioa},\ref{inst:kavli}}               
  \and S.~T.~Hodgkin\inst{\ref{inst:ioa}}               
  \and E.~Pancino\inst{\ref{inst:oaar},\ref{inst:ssdc}}               
  \and P.~J.~Richards\inst{\ref{inst:stfc}}               
  \and G.~Altavilla\inst{\ref{inst:oaro},\ref{inst:ssdc},\ref{inst:oabo}}
  \and L.~Balaguer-N\'u\~nez\inst{\ref{inst:ub}}
  \and M.~A.~Barstow\inst{\ref{inst:leicester}}
  \and M.~Bellazzini\inst{\ref{inst:oabo}}
  \and A.~G.~A.~Brown\inst{\ref{inst:leiden}}
  \and M.~Castellani\inst{\ref{inst:oaro}}
  \and G.~Cocozza\inst{\ref{inst:oabo}}
  \and F.~De~Luise\inst{\ref{inst:oaab}}
  \and A.~Delgado\inst{\ref{inst:ioa}}
  \and C.~Ducourant\inst{\ref{inst:bord}}
  \and S.~Galleti\inst{\ref{inst:oabo}}
  \and G.~Gilmore\inst{\ref{inst:ioa}}
  \and G.~Giuffrida\inst{\ref{inst:oaro}}
  \and B.~Holl\inst{\ref{inst:geneva},\ref{inst:ecogia}}
  \and A.~Kewley\inst{\ref{inst:ioa}} 
  \and S.~E.~Koposov\inst{\ref{inst:carnegieMellon},\ref{inst:ioa}}
  \and S.~Marinoni\inst{\ref{inst:oaro},\ref{inst:ssdc}}
  \and P.~M.~Marrese\inst{\ref{inst:oaro},\ref{inst:ssdc}}
  \and P.~J.~Osborne\inst{\ref{inst:ioa}}
  \and A.~Piersimoni\inst{\ref{inst:oaab}}
  \and J.~Portell\inst{\ref{inst:ub},\ref{inst:ieec}}
  \and L.~Pulone\inst{\ref{inst:oaro}}
  \and S.~Ragaini\inst{\ref{inst:oabo}}
  \and N.~Sanna\inst{\ref{inst:oaar}}
  \and D.~Terrett\inst{\ref{inst:stfc}}     
  \and N.~A.~Walton\inst{\ref{inst:ioa}}    
  \and T.~Wevers\inst{\ref{inst:nijmegen},\ref{inst:ioa}} 
  \and {\L}.~Wyrzykowski\inst{\ref{inst:warsaw},\ref{inst:ioa}} 
}

\institute{
Institute of Astronomy, University of Cambridge, Madingley Road, Cambridge CB3 0HA, UK\\
\email{dwe@ast.cam.ac.uk}
\label{inst:ioa}
\and
Institut del Ci\`encies del Cosmos (ICC), Universitat de Barcelona (IEEC-UB), c/ Mart\'{\i} i Franqu\`es, 1, 08028 Barcelona, Spain
\label{inst:ub}
\and
INAF -- Osservatorio  di Astrofisica e Scienza dello Spazio di Bologna, via Gobetti 93/3, 40129 Bologna, Italy
\label{inst:oabo}
\and 
Institute for Astronomy, University of Edinburgh, Royal Observatory, Blackford Hill, Edinburgh EH9 3HJ, UK
\label{inst:roe}
\and
Kavli Institute for Cosmology Cambridge, Madingley Road, Cambridge, CB3 0HA, UK
\label{inst:kavli}
\and
INAF - Osservatorio Astronofisico di Arcetri, Largo Enrico Fermi 5, 50125, Firenze, Italy
\label{inst:oaar}
\and
Space Science Data Center - ASI, Via del Politecnico SNC, 00133 Roma, Italy
\label{inst:ssdc}
\and
STFC, Rutherford Appleton Laboratory, Harwell, Didcot, OX11 0QX, UK
\label{inst:stfc}
\and
INAF-Osservatorio Astronomico di Roma, Via Frascati 33, 00078, Monte Porzio Catone (Roma), Italy
\label{inst:oaro}
\and
Leicester Institute of Space and Earth Observation and Department of Physics and Astronomy, University of Leicester, University Road, Leicester LE1 7RH, UK
\label{inst:leicester}
\and
Leiden Observatory, Leiden University, Niels Bohrweg 2, 2333 CA Leiden, The Netherlands
\label{inst:leiden}
\and
INAF-Osservatorio Astronomico d'Abruzzo, Via Mentore Maggini SNC, 64100 Teramo, Italy
\label{inst:oaab}
\and
Laboratoire d'Astrophysique de Bordeaux, Univ. Bordeaux, CNRS, B18N, All{\'e}e Geoffroy Saint-Hilaire, 33615 Pessac, France
\label{inst:bord}
\and
Department of Astronomy, University of Geneva, Ch. des Maillettes 51, CH-1290 Versoix, Switzerland
\label{inst:geneva}
\and 
Department of Astronomy, University of Geneva, Ch. d'Ecogia 16, CH-1290 Versoix, Switzerland
\label{inst:ecogia}
\and
McWilliams Center for Cosmology, Carnegie Mellon University, 5000 Forbes Ave, 15213, USA
\label{inst:carnegieMellon}
\and
Institut d'Estudis Espacials de Catalunya (IEEC), C/ Gran Capit\`a 2-4, E08034 Barcelona, Spain
\label{inst:ieec}
\and
Department of Astrophysics/IMAPP, Radboud University, P.O. Box 9010, 6500GL Nijmegen, The Netherlands
\label{inst:nijmegen}
\and
Warsaw University Astronomical Observatory, Al. Ujazdowskie 4, 00-478 Warszawa, Poland
\label{inst:warsaw}
}

\date{Received \textbf{Month Day, 201X}; accepted \textbf{Month Day, 201X}}

 
  \abstract
   {}
   {We describe the photometric content of the second data release of the \Gaia\ project (\GDR2) and its validation along with the quality of the data.}
    {The validation was mainly carried out using an internal analysis of the photometry. External comparisons were also made, but were
     limited by the
     precision and systematics that may be present in the external catalogues used.}
   {In addition to the photometric quality assessment, we present the best estimates of the three photometric passbands. Various colour-colour transformations
   are also derived to enable the users to convert between the \Gaia\ and commonly used passbands.}
   {The internal analysis of the data shows that the photometric calibrations can reach a
     precision as low as 2 mmag on individual CCD
measurements. Other tests show that systematic effects
   are present in the data at the 10 mmag level.}


   \maketitle
%

\section{Introduction}\label{Sect:Introduction}

The launch of the ESA \Gaia\ satellite mission \citepads{2016A&A...595A...1G} in December 2013 marked the start of an exciting period 
for more than 400 people that are part of the Data Processing and Analysis Consortium (DPAC) and for the 
astronomical scientific community in general. 
The first catalogue, released in September 2016 \citepads[\GDR1][]{2016A&A...595A...2G}, already showed the enormous potential of the \Gaia\ astrometric
and photometric data, leading to the publication of almost 300 refereed papers in the first year after
the release.
The \GDR1 photometric catalogue contained a measurement of the average flux in the 
\G\ band for over 1.1 billion sources \citepads{2017A&A...599A..32V}. 
No colour information was released at that time,
with the consequence that many photometric investigations required
cross-matching the \Gaia\ catalogue with other photometric catalogues to acquire photometry 
in some additional bands, with the effect of reducing the number of usable sources and possibly introducing 
some inconsistencies due to the different origins of the photometry.

The second data release (\GDR2) in April 2018 \citepads{DR2-DPACP-36} presents a significant advance in all 
photometric investigations by providing photometry in the \G\ band for approximately 1.7 billion sources
and in the integrated \BP\ and \RP\ bands for approximately 1.4 billion sources
calibrated to a consistent and homogeneous photometric system. The release also includes the results 
of two applications of the \Gaia\ photometry: a catalogue of astrophysical parameters \citepads[effective temperature and
line-of-sight extinction for stars brighter than $G=17$ and luminosity and radius whenever good
parallaxes were available, see][]{AstrophysicalParameters} and epoch data for a sample of variable stars
\citepads[more than 0.5 million sources classified as RR Lyrae, Cepheids, Delta Scuti, LPV, and SX Phe type stars,
see][]{VariabilityAnalysis}. Additional validation covering other aspects of the data release can be found in the general validation
paper \citepads{DR2-DPACP-39}.

This paper focusses on the photometric aspects of \GDR2.
In Sect. \ref{Sect:Content} we present the photometric content of the data release. Section \ref{Sect:InputData}
provides an overview of the data that contributed to the generation of the photometric catalogue with a focus on the
differences with respect to \GDR1.
The following sections show various aspects of the internal validation activities and provide a detailed view of the quality of
the photometric data.
Sections \ref{Sect:LsSsCoefficients} and \ref{Sect:LsConvergence} focus on the results of the calibration process.
Section \ref{Sect:Accumulations} presents a statistical analysis of the source photometry. Section \ref{Sect:ExternalComparisons}
shows the results of the validation of the \Gaia\ photometry with respect to other catalogues.
Some known issues are also described, and guidelines for users are given in the final sections. 
A metric that can help identifying sources that may be affected by blending due to image crowding, and faint-end
biases linked to background effects, is presented in Sect. \ref{Sect:BpRpExcess}.
The \GDR2 passbands and zeropoints and their validation is the topic of Sect. \ref{Sect:Passbands}.
The statistical properties of the three subsets of the photometric catalogue produced by slightly different calibration procedures are described in Sect.~\ref{Sect:GoldSilverBronze}.
Section \ref{Sect:EpochPhotometry} describes a few aspects of the epoch photometry released for variable sources.
Appendix \ref{Sect:Transformations} presents colour-colour transformations between the \Gaia\ passbands and
other photometric systems as implemented in various surveys. An approach to correct the \Gaia\ magnitudes
for effects of saturation at the bright end is recommended in Appendix \ref{Sect:Saturation}.
Finally, Appendix \ref{Sect:Acronyms} contains a list of acronyms and \Gaia-specific terminology used in this paper.

\section{Photometric content of \GDR2}\label{Sect:Content}

The second \Gaia\ photometric catalogue  contains the latest \G-band photometry for all sources and 
\BP\ and \RP\ photometry for 80\% of them. The mean photometric measurements have been obtained by 
processing epoch photometry collected over a period of about 670 days of mission operations.

The broad \G\ passband covers the range $[330, 1050]$ nm, and its definition is optimized to collect the maximum of light
for the astrometric measurements. The \BP\ and \RP\ photometry instead are derived from the integration of the blue and red
photometer (BP
and RP) low-resolution spectra covering the ranges $[330, 680]$ nm and $[630, 1050]$ nm. These
wavelength ranges are those defined in the pre-mission specification 
\citepads{2010A&A...523A..48J}.
The numerical values corresponding to the three passbands are part of \GDR2 together with the definition of the photometric zeropoints.
The passband definition is distributed in electronic tabular format as part of this paper and is available via the VizieR service.

One of the main challenges in the \Gaia\ photometric processing is due to the large number of different 
instrument configurations that are possible during the acquisition of an observation \citepads{PhotProcessing}. This results 
in effectively different instruments that need to be calibrated to a homogeneous system. As of today, no external catalogue is
available that would offer the accuracy and amount of data required for calibrating such 
a complex instrument. For this reason, the definition of a reference system that is homogeneous over the entire 
set of instrument configurations must rely on the \Gaia\ data itself. External data are only used to link 
the internal photometric reference system to the absolute one. For a detailed description of the principles
of the photometric calibration and the difference between internal and external calibration
see \citeads{2016A&A...595A...7C}.

The \G-band photometry included in \GDR2 is the result of a new reduction and
is based on a much extended period of mission operations (more than eight additional months) in comparison with
the \GDR1 catalogue. The internal photometric system was re-initialised using the available data and
is therefore expected to be different from the system defined for \GDR1. In particular, a colour term between
the two systems is expected.
The passbands published in \GDR2 are only applicable to the \GDR2 photometry.

It is important to keep in mind that the actual source catalogue is different and should be treated as
a new catalogue superseding the one published in \GDR1. Even though the two catalogues will have many
source identifiers in common, the corresponding sources may be defined by different lists of epoch
measurements, effectively leading to different sets of astrometric and photometric 
properties \citepads{DR2-DPACP-45}.

\begin{figure*}
\centering
\includegraphics[width=\hsize]{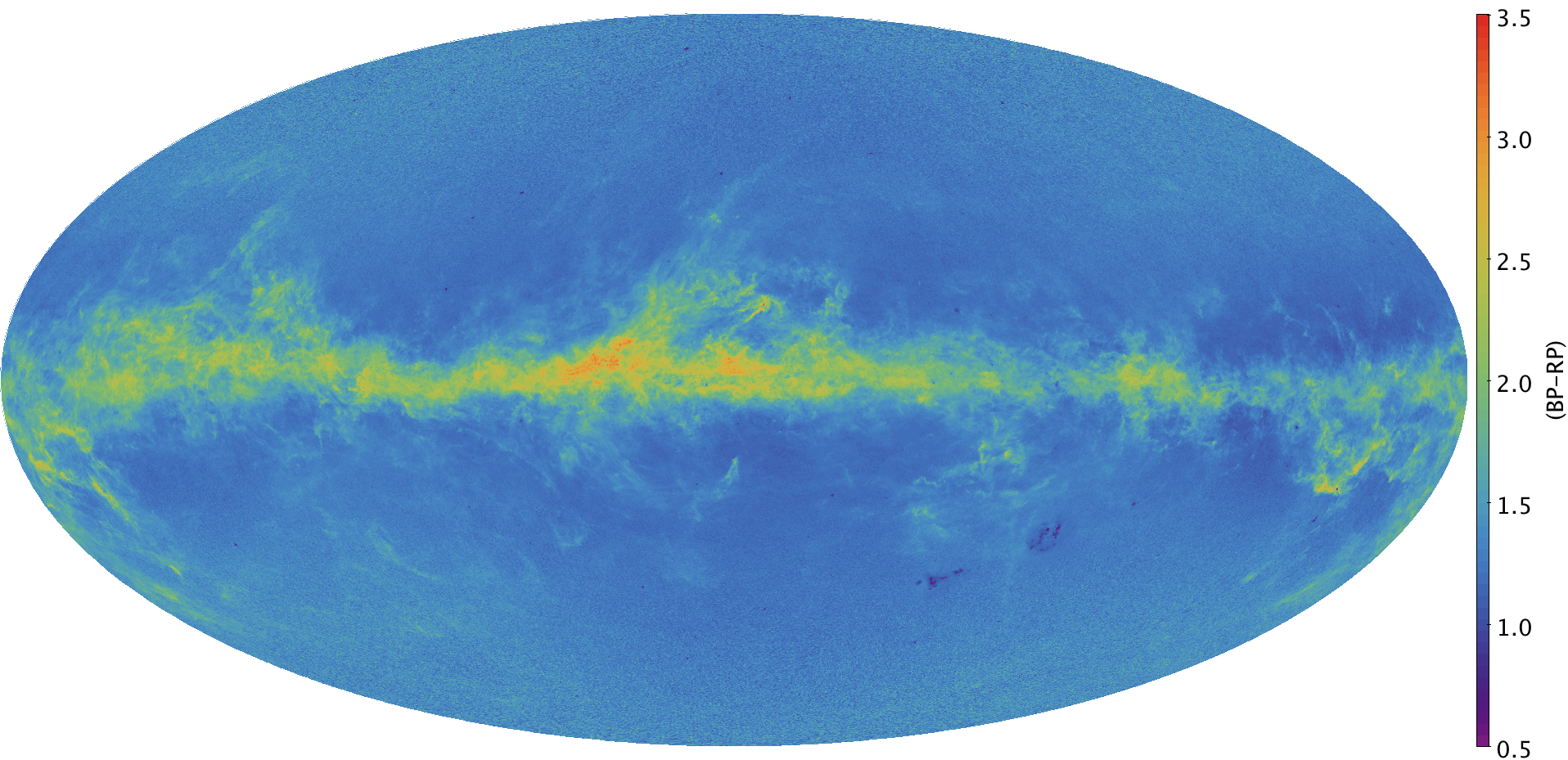}
\caption{Colour distribution as a function of sky position in Galactic coordinates. 
The pixelisation scheme adopted is HEALPix
level 8, implying a pixel size of approximately 190 square arcmin or 0.05 square degree.
Each pixel represents the median colour (\BP$-$\RP) of all sources 
with $G<19$ in that pixel.
}
\label{Fig:ColourSky}
\end{figure*}

Figure \ref{Fig:ColourSky} shows the all-sky view of the colour distribution based on the \GDR2
photometric catalogue. 
Each pixel in the colour sky map is colour-coded according to the median colour (\BP$-$\RP) of
all sources brighter than $G=19$ falling in the corresponding area in the sky.
The colour sky distribution offers a taste of the extraordinary potential and quality of the
photometric data released in \GDR2.
High-extinction star-forming regions close to the Galactic plane stand out in the colour map as a result of reddening effects. 

\section{Input data}\label{Sect:InputData}

To facilitate the understanding of the rest of this paper, we briefly introduce some characteristics of the \Gaia\ satellite and instruments. Many more details are provided in 
\citeads{2016A&A...595A...1G}. 
The \Gaia\ satellite scans the sky approximately along great circles. The scanning strategy has been optimised to maximise the scientific performance of the mission and does not cover the entire sky
homogeneously: different areas will be observed with different frequency and will have 
more observations than other regions in the sky.
The fields of view (FoVs) of the two telescopes on board overlap onto the focal plane of all charge-coupled devices (CCDs) used for the photometry. 
The light is integrated while it crosses a CCD in the along-scan (AL) direction.
Gates have been implemented on the \Gaia\ CCDs to reduce the effective exposure time of an observation in the case of bright sources by limiting the area of the CCD in which integration takes place. Twelve different gate configurations (including the no-gate case) can be activated in configurable magnitude ranges. 
Observations are limited to a small area (window) centred on the detected sources. Different window
sizes and shapes are used on board, depending on the on-board estimated magnitude. These are referred to as window classes.
Different gate and window class configurations effectively constitute different instruments that need to be calibrated to the same system.

The photometric data in the three bands \G, \BP, and \RP\ come from different CCDs on the focal plane: measurements in the \G\ broad band are taken in the astrometric field (AF) CCDs, while the
\BP\ and \RP\ fluxes are integrated over the low-resolution spectra collected in the BP and RP CCDs.

The rest of this section describes the input data, focussing on the differences with respect to \GDR1. 
Many more details on the input data and on the processing leading to \GDR2 are available in 
\citeads{PhotProcessing}.

The \Gaia\ processing within DPAC is organized in cycles. At each cycle new, recent data are added
to the processing, and improved algorithms and models are deployed in all systems feeding each other,
thus potentially leading to higher quality results.
This applies in particular to the second cycle of operations (leading to \GDR2), where the input data
covered a significantly longer period of operations, results from other systems became available to the
photometric processing, and updated and more sophisticated algorithms were adopted in the data processing
to produce a more accurate and complete catalogue than the one released in \GDR1.

The data entering the process of calibrating the \G-band photometry are the image parameters, that is, flux
and centroid location within the window, that are part of the \Gaia\ intermediate data.
These are first determined on a daily basis as soon as the data are downloaded from the satellite in the 
system; this is known as initial data treatment \citepads{2016A&A...595A...3F}.  
A further, more sophisticated and consistent, image
parameter determination (IPD) runs at the beginning of each cycle of operations and covers all data from the
start of operations \citepads[intermediate data update, IDU, see ][and the online \GDR2 documentation at 
\href{http://gaia.esac.esa.int/documentation/}
{this link}\footnote{\href{http://gaia.esac.esa.int/documentation/}{http://gaia.esac.esa.int/documentation/}}]{gdr2Astrometry}. While for \GDR1 the daily image parameters were used, \GDR2
is purely based on the updated cyclic values.
This also implies a more complete input dataset, eliminating the various interruptions in the daily systems 
particularly in the first months of mission when operations were still being established.
An additional improvement coming from the usage of cyclic image parameters is due to 
the better handling of saturated data for \G\ as compared to what was done in the daily chain.

The \BP\ and \RP\ photometric processing uses the raw data directly (i.e. the unpacked telemetry) and therefore requires
some pre-processing steps to be performed to generate the input data for the calibration process.
The algorithms and procedures used in the pre-processing for \GDR2 are similar to those in
place for \GDR1, the only exception being the usage of the astrophysical coordinates of the sources determined
from \Gaia\ data itself \citepads{gdr2Astrometry}.
The source astrometry, together with the satellite attitude and the geometry of
the instrument, are used to predict the location of each source within the observing window. This is particularly
important when a wavelength calibration of the positions along scan within a spectral window is required. As the
photometric calibration is based on detailed colour information extracted from the low-resolution spectra by
integrating in specific bands within the wavelength range covered by the BP and RP instruments, it is clear that
an accurate wavelength calibration is necessary \citepads{2016A&A...595A...7C}. 
In the processing that led to \GDR1, an extrapolation of the source
position from the astrometric field onto the BP and RP CCDs was adopted
for the geometric and wavelength calibration. For \GDR2, this has been
replaced by a prediction based on the source astrophysical coordinates,
the satellite attitude, and an assumed nominal geometry thanks to the high
accuracy of the AGIS \citepads{gdr2Astrometry} results
available at the start of photometric processing. The resulting geometric calibration
consists of a set of corrections to the adopted
nominal geometry.

A much improved cross-match algorithm \citepads{DR2-DPACP-45}
led to a cleaner source catalogue and a better removal of detection
artefacts. The \Gaia\ cross-match is unusual in the sense that it is not simply assigning observations to a
pre-defined list of sources, but is also creating the source list and continuously adding new sources as required,
starting from the data itself.
This is necessary given the much higher resolution and efficiency at detecting faint sources by \Gaia\ in all sky regions.
This difficult task is complicated by the presence of a significant number of spurious
detections. For \GDR2, the algorithms in place for cleaning the input observations from such artefacts and
for performing the clustering analysis that eventually lead to the generation of the source catalogue have
been improved thanks to the experience gained in the previous processing cycle.

In terms of the photometric processing itself, several improvements were made with respect to the
processing done for \GDR1. Most of these were triggered by the results of the validation and various investigations
based on the results of the first cycle of operations. In the following, we list the most relevant
changes. For more details on the photometric processing improvements in \GDR2 with respect to the first release,
see \citetads{PhotProcessing}.

A significant improvement in the photometric catalogue was achieved by a
more robust accumulation of the epoch photometric measurements to generate the mean source photometry, with outlier
rejection driven by a careful statistical analysis \citepads[see Section 4.5 in][]{PhotProcessing}.

The validation of the \GDR1 photometry showed systematic differences in the internal photometric
system at magnitude $G=13$ and $G=16,$ which correspond to configuration changes in the windowing scheme adopted on board (i.e.
causing differences in window shape and size for different observations). The systematic differences arose because only a small fraction of sources were observed in more than one configuration.
Their magnitude is sufficiently close to the magnitude boundary of each configuration to be observed
with different configurations in different transits. The precision of the on-board magnitude estimate at $G=13$ and $G=16$ means
that the useful range of magnitude is rather small. This implies that the calibration process
may initially converge to different photometric systems for different configurations and that a consistent system over the entire
set of configurations may only be reached with very many iterations. A first
attempt at calibrating out the offsets in the photometric systems defined for each configuration was made in the first
cycle of operations and showed promising results. This was further improved for \GDR2
by introducing time dependency in the calibration of the links between different instrumental
configurations \citepads[see Section 4.3 in][]{PhotProcessing}.

The period of operations covered by \GDR1 was heavily affected by rapid and discontinuous variations in the
satellite throughput due to water-based contamination of the satellite instrument components \citepads{2016A&A...595A...1G}. 
The increased level of systematics in the input data constituted one of the main challenges for the photometric
calibration in the first cycle of processing.
The input data for \GDR2, however, contain a long stretch of more than one year in which the contamination effect
was quite low
and stable. The photometric processing strategy was therefore adapted to take advantage of this period for the
initialisation of the reference system \citepads[see Section 4.2 in][]{PhotProcessing}.

Finally, a full external calibration of the internal system onto the absolute Vega photometric system
allowed reconstructing the true passbands. These constitute an important addition to the
release.

At this stage in the mission, the processing is still unable to handle some non-nominal data. Observations obtained with
more than one gate activated at different times within the window have not been processed.
These situations occur when a bright source is observed at a time sufficiently close to another (fainter) source
to trigger the activation of a gate for a period that only partly overlaps the observation period of the fainter source.
Truncated windows are currently also excluded from the stream of input data. These are caused by the overlap on the CCD of two or more
windows and are likely to happen in particularly dense regions. As described before, the two FoVs of \Gaia\ are both projected onto the same focal plane and therefore non-dense regions may experience a high number
of complex gate or truncated window cases if the other FoV points towards a dense region.
BP and RP observations are also more likely to be collected in these non-nominal configurations given the much longer
windows assigned to them.

At this stage, the data available are in
some cases not sufficient or not of sufficient quality
to produce a reliable calibration. This is especially true for some combination of gate and window configurations that
are in principle not expected but occur nevertheless because sources of different magnitudes
are observed simultaneously.
In some cases, this also affects the bright end of the data, where the very limited number of sources makes the calibration
process more challenging. This will affect the completeness of the epoch data, but is not expected to significantly affect the mean
photometry, except possibly for very bright sources ($G<5$).

Some short periods of observations have been excluded from the input data because the scientific quality of the corresponding
data is poor. These periods correspond to the activities of refocus and of decontamination, when several components of the instruments where heated to vaporise
the water-based contaminant, thereby compromising the thermal stability of the instrument. The exact ranges are given in \citetads{PhotProcessing}.
These gaps do not affect the completeness of the mean photometric catalogue, but will cause short gaps in the epoch photometry.

\section{Study of LS and SS calibration coefficients}\label{Sect:LsSsCoefficients}

\begin{figure} \resizebox{\hsize}{!}{\includegraphics{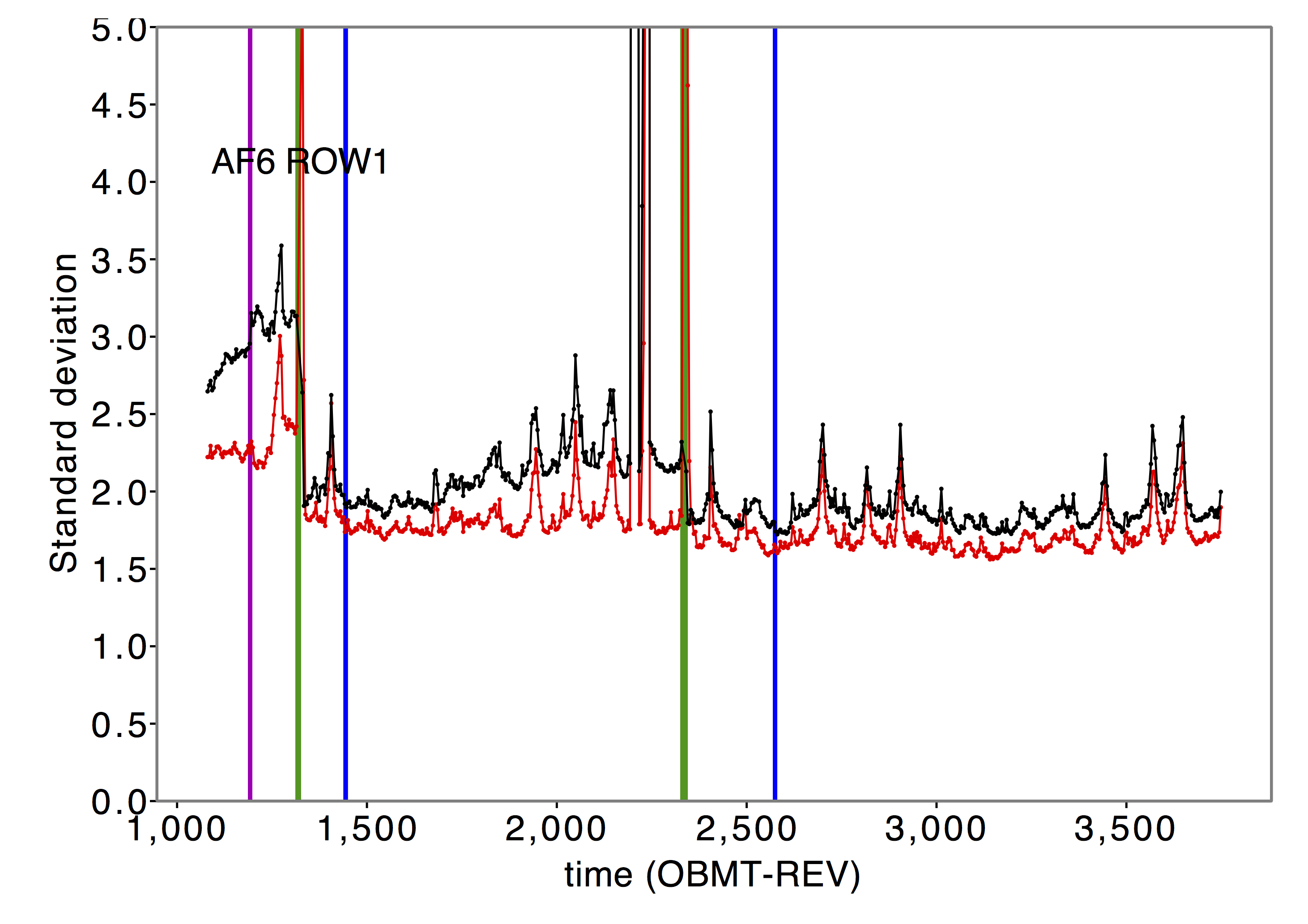}} \caption{Unit-weight standard deviation of the large-scale calibration as a function of
time (in satellite revolutions) for an example calibration unit. In this case, AF6, Row 1, Window Class 1, No Gate. This is the same as shown 
in \citetads{2017A&A...600A..51E} for \GDR1. 
The black lines are for the preceding and red for the following FoV calibration units. The vertical lines represent significant satellite events: scanning law change (magenta), 
decontamination (green), and refocussing  (blue). The approximate time range covered by this plot is July 2014 to May 2016.}
\label{Fig:SdLs}
\end{figure}

\begin{figure} \resizebox{\hsize}{!}{\includegraphics{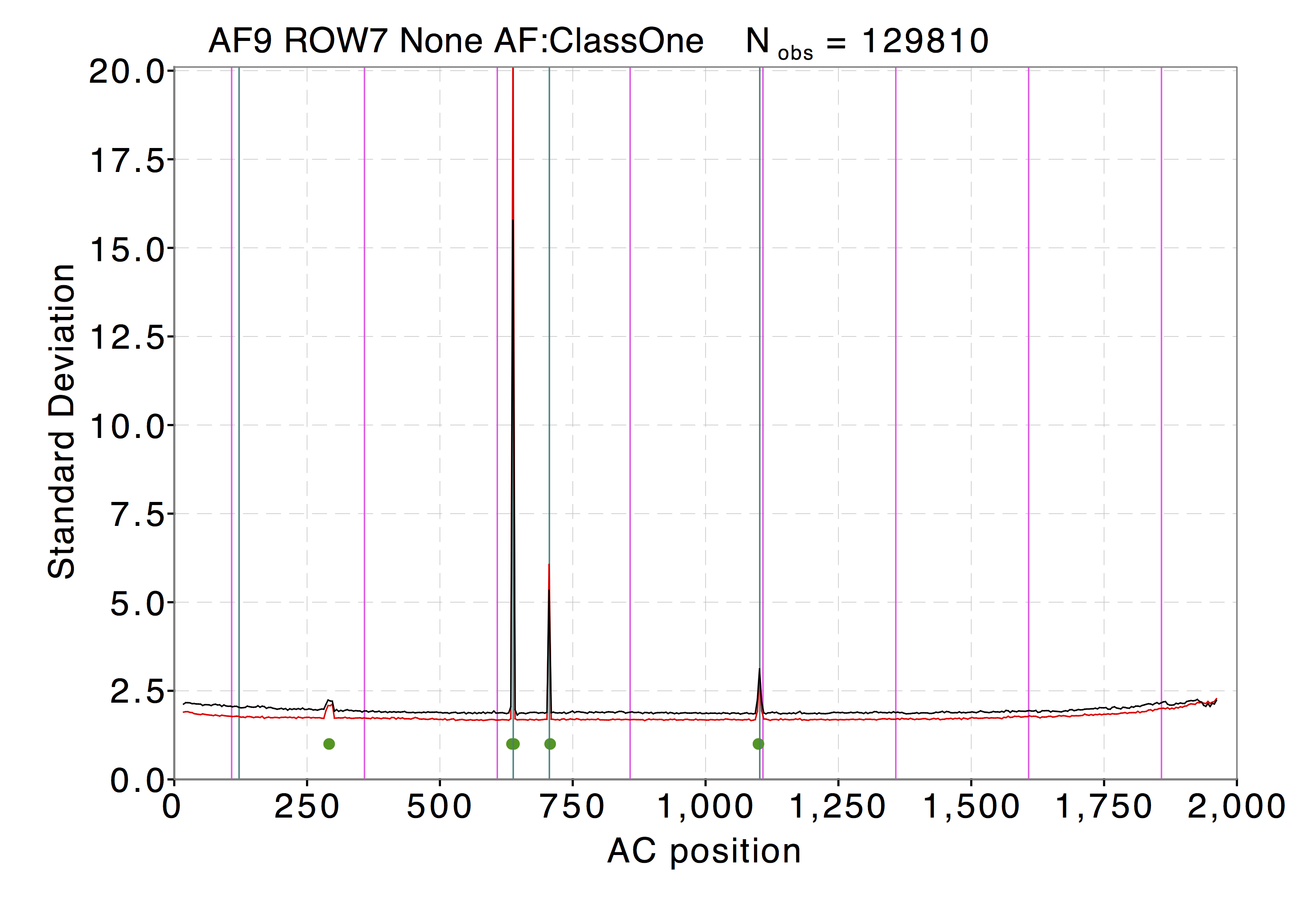}} \caption{Unit-weight standard deviation of the small-scale calibration as a function of
across-scan position on the CCD for an example calibration unit. In this case, AF9, Row 7, Window Class 1, No Gate. This is the same as 
shown in \citetads{2017A&A...600A..51E} for \GDR1. 
The black lines are for the preceding and red for the following FoV calibration units. 
The magenta lines show the locations of the CCD stitch blocks, and the green dots
show the location of detected bad columns (from these calibrations). The cyan lines show the bad columns detected from the low-level CCD calibrations.}
\label{Fig:SdSs}
\end{figure}

\begin{figure} \resizebox{\hsize}{!}{\includegraphics{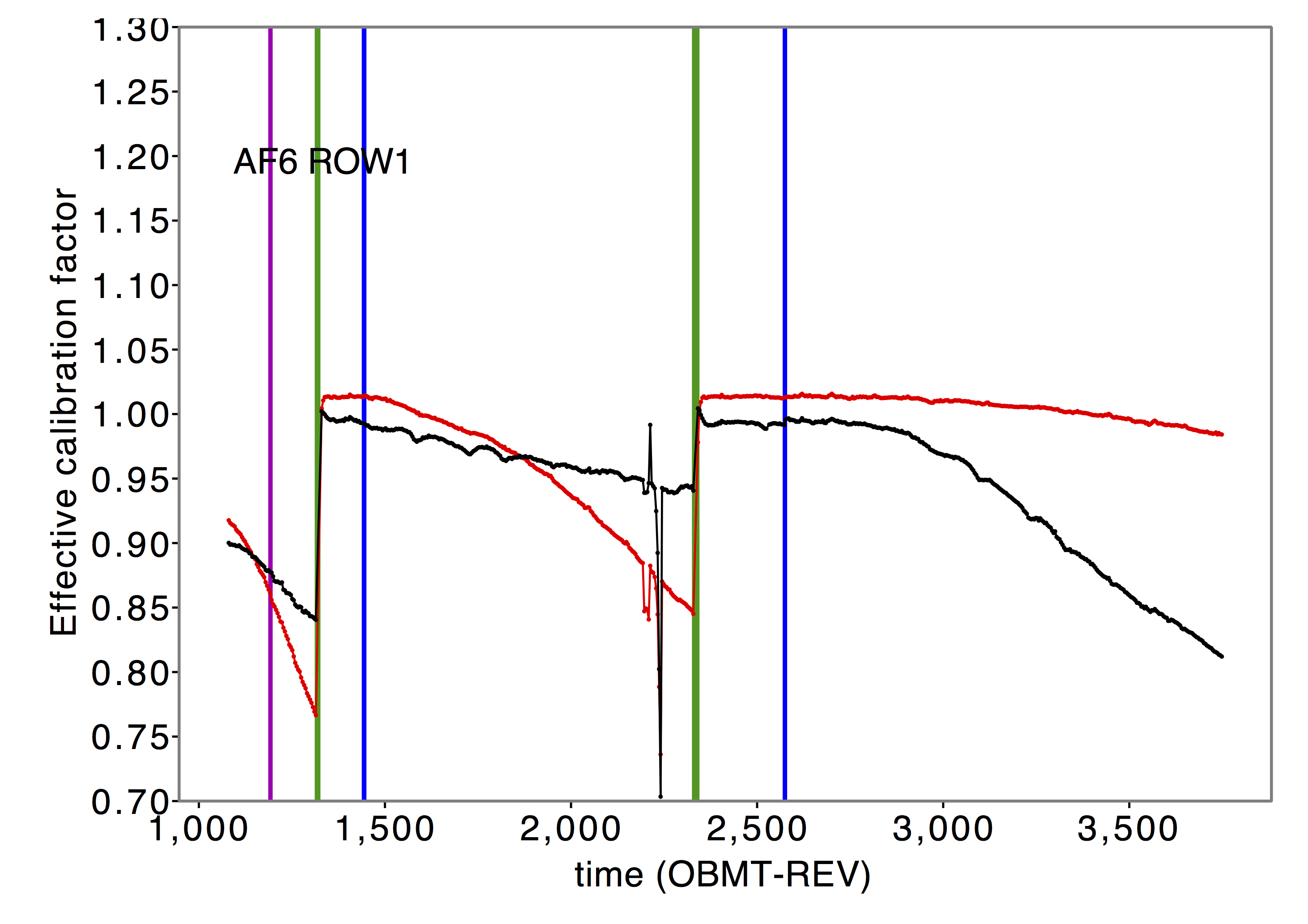}} \caption{Effective calibration factor of the large-scale sensitivity calibration as a function of
time for an example calibration unit. The calibration unit is the same as in Fig.~\ref{Fig:SdLs}, as are the vertical lines.
For this plot, the detailed colour terms of the calibration model have been combined to form an effective calibration factor using default colours.
This is necessary since there is no constant term in the calibration model equivalent to a zeropoint.
See \citetads{2016A&A...595A...7C} 
for more details.}
\label{Fig:ZpLs}
\end{figure}

\begin{figure} \resizebox{\hsize}{!}{\includegraphics{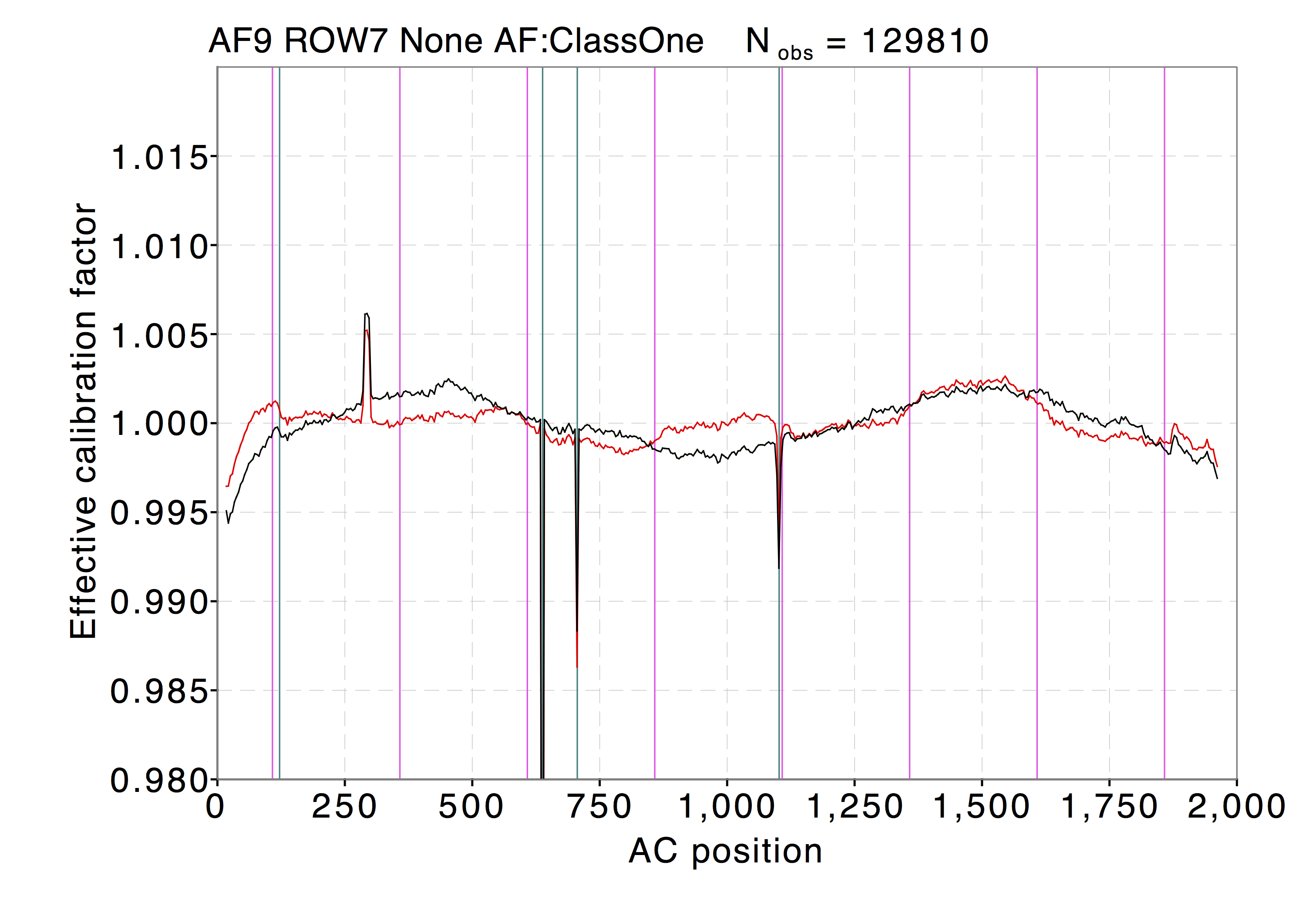}} \caption{Effective calibration factor of the small-scale sensitivity calibration as a function of
across-scan position on the CCD for an example calibration unit. 
The calibration unit is the same as in Fig.~\ref{Fig:SdSs}, as are the vertical lines.
}
\label{Fig:ZpSs}
\end{figure}

The photometric calibration is defined by two sets of calibrations
that describe different instrumental effects. These are referred to as large- and
small-scale calibrations: the large-scale (LS) calibration describes features that vary smoothly across the focal plane but vary rapidly,
while the small-scale (SS) calibration takes into account effects that change slowly but may only affect a small section of the
CCD, even a single pixel column. It is convenient to separate these effects because many more data need to be accumulated
to calibrate out the response of a single pixel column than what is needed to determine the overall response of a CCD with respect to others. On the
other hand, the overall response can change quite rapidly (and at times in a discontinuous fashion, as in the case of decontamination or refocus
events), and therefore our LS calibration needs to be able to quickly adjust to these changes.

Similar to the processing carried out for the first data release, the calibration coefficients and quality statistics for the LS and SS
calibrations were
analysed to assess the overall quality of the photometry and to indicate where improvements might be necessary. Updated versions of the plots in
\citetads{2017A&A...600A..51E} 
are shown in Figs. \ref{Fig:SdLs} to \ref{Fig:ZpSs}.
The unit-weight standard deviation is defined as the square root of the normalised chi-square 
\citepads{2007ASSL..350.....V}  
and gives an indication of how good the solution model is. When the systematic effects have all been removed and the quoted errors are
representative of the underlying distribution, this value should be around 1.0.

The significant change that has occurred in the standard deviation plots is that the overall levels have improved. This is a consequence of the better handling of the formal uncertainties
 on the image parameters and within the calibrations. There is also consistency between the median standard deviation values of the LS and SS calibrations for comparable
magnitude ranges. This is to be expected since the most recent iteration of the SS calibration is applied to the observations used by the LS calibrations, and vice versa
\citepads{2016A&A...595A...7C}. 
For Window Class 1\footnote{see \citetads{2016A&A...595A...1G} 
and \citetads{2016A&A...595A...2G} 
for more details on the windowing and gating strategy used for the \Gaia\ observations.}, 
which has a magnitude range of $13<G<16$, the median standard deviation is 1.8 compared to an average of 5.0 for the LS calibrations in
\GDR1. The value being higher than the expected value of 1.0 indicates that either the formal uncertainties are underestimated or that
additional terms are needed in the calibration models. This does not affect the uncertainty on the weighted mean fluxes since the measured scatter is taken into account in the 
calculation, as described in \citetads{2016A&A...595A...7C} 
and \citetads{PhotProcessing}.

For the LS calibration, more features can be seen in Fig. \ref{Fig:SdLs} compared to the previous release because the standard deviation
values are improved \citepads{2017A&A...600A..51E}. 
While some of
the peaks correspond to anomalous calibrations that are still causing problems, the majority of the peaks seen are caused by major planets in the FoV. By comparing the
standard deviation in Fig. \ref{Fig:SdLs} with the calibration factor in
Fig. \ref{Fig:ZpLs}, it can be seen that the planet crossings that caused the three large peaks in the standard deviation around OBMT-REV 2000 have not affected the derived calibration 
factors.

In Fig. \ref{Fig:SdSs}, the only difference compared to \GDR1 is the small peak at about an across-scan (AC) position of 300. 
At this position, a feature is seen in the dark signal calibrations
and might indicate that some slight improvement could be made. However, this is a very small effect and has been mainly accounted for by the SS calibration, see Fig. \ref{Fig:ZpSs}.

Figure \ref{Fig:ZpLs} shows an example of the time evolution of the effective calibration factor from the LS calibrations. 
This is defined as the multiplicative factor to be applied to the observed flux to obtain its calibrated value. 
The factor is different for sources with different colour and for different AC coordinates; the effective value shown here corresponds to a 
default set of colour terms and the centre of the CCD.
The main features seen here are the effect of the change in response
that is due to the changing levels of contamination on the mirrors and CCDs, see \citetads{2016A&A...595A...1G}.\ 
Just before the second decontamination (green line), the contamination is worse in the following FoV, while at the end of the period used for
\GDR2, the contamination is worse in the preceding FoV.

As pointed out in \citetads{2017A&A...600A..51E}, 
the effective calibration factor from the SS calibrations, see Fig. \ref{Fig:ZpSs}, is equivalent to a 1D flat field. While there is an overall slowly changing difference between the
two FoVs, many of the detailed low-level peaks are in common, indicating that the SS calibration measures the CCD response variation at the sub-millimagnitude level
and that these peaks are not due to noise.

\begin{figure} \resizebox{\hsize}{!}{\includegraphics{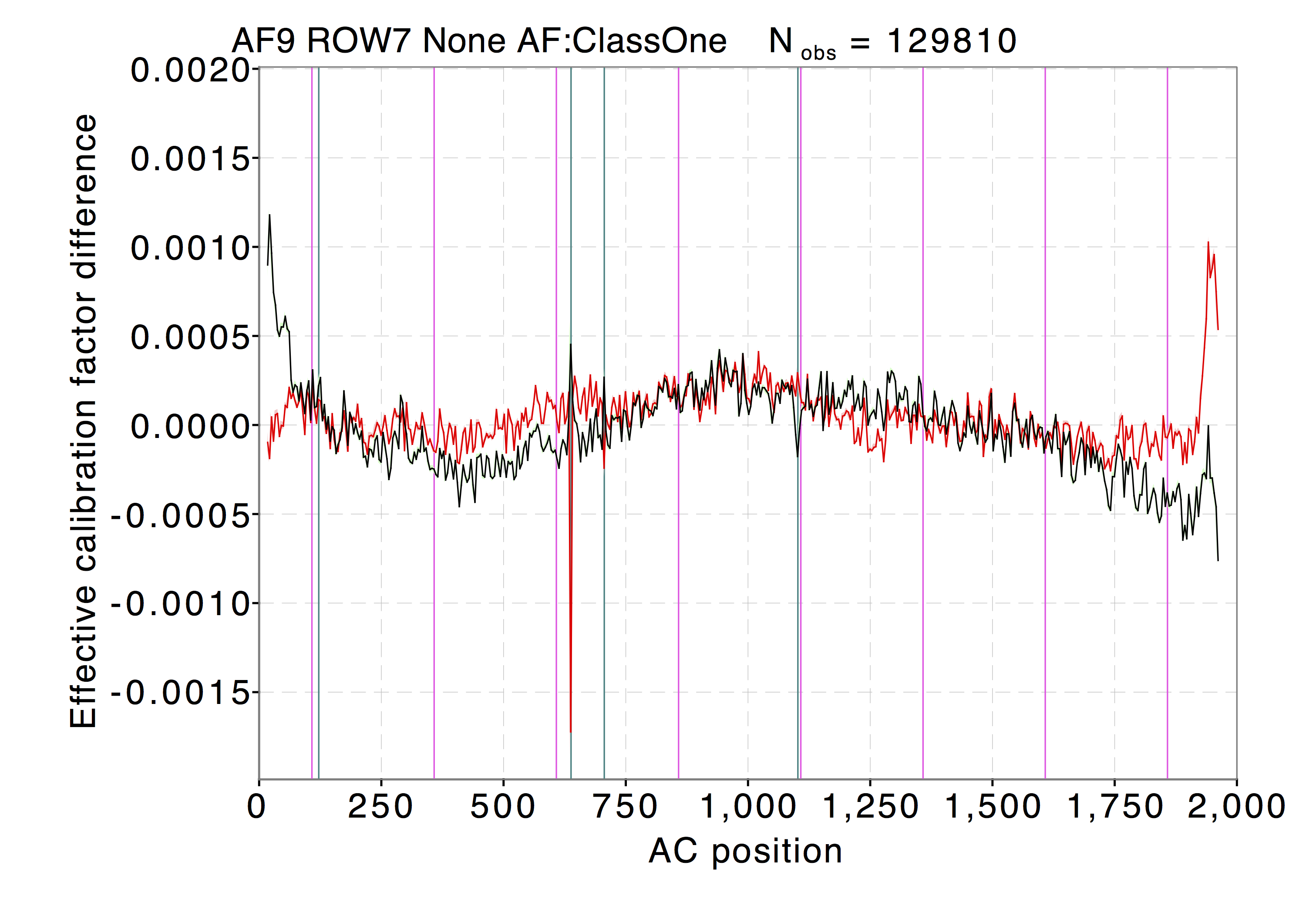}} \caption{ Difference in effective calibration factor of the SS calibrations as a function of
across-scan position on the CCD (same calibration configuration as Fig.~\ref{Fig:SdSs}) between the two time periods used for the SS calibrations.
We note the change in the ordinate scale. The colour scheme for the lines is the same as for Figs. \ref{Fig:SdSs} and \ref{Fig:ZpSs}.}
\label{Fig:SsVar}
\end{figure}

For the SS calibrations, the observations were split into two periods, see \citetads{PhotProcessing}. Figures \ref{Fig:SdSs} and \ref{Fig:ZpSs} refer to calibrations from the initialisation
period following the second decontamination.  Figure \ref{Fig:SsVar} shows the difference between two sets of calibrations from the two periods and shows that the magnitude of the
differences is very small. Some features between the two FoVs appear to be in common, indicating that these are real changes in the CCD response on the timescale of one year. However,
this is approaching the noise level of these calibrations.
The standard deviation of the difference between the two sets of calibrations is less than 0.5 mmag for all Window Classes and thus
does not raise concerns about the frequency of the SS calibrations.

\section{Convergence of the large-scale calibrations}\label{Sect:LsConvergence}

\begin{figure} \resizebox{\hsize}{!}{\includegraphics{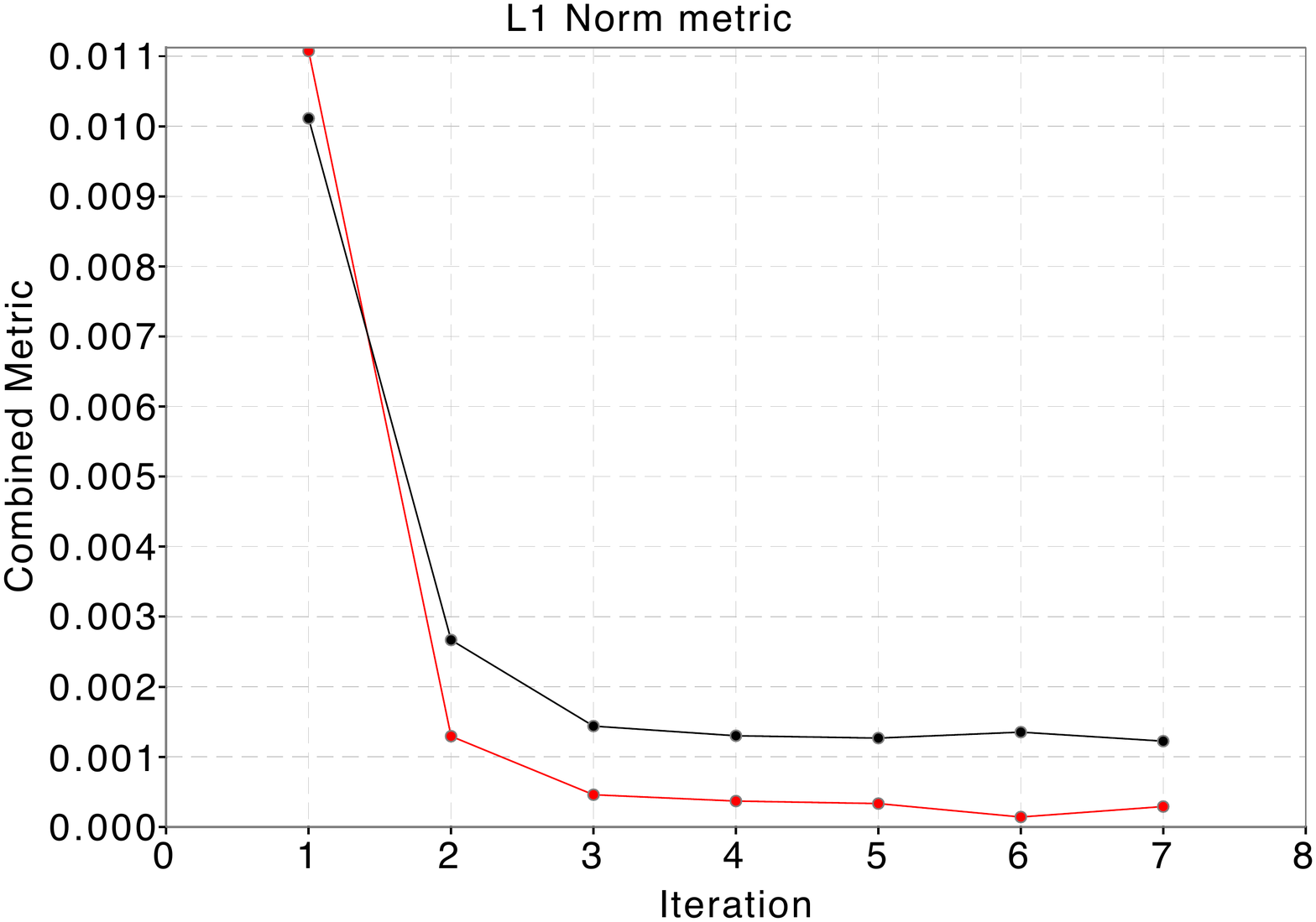}} \caption{Convergence metric as a function of iteration for the \G-band Window Class 2 large-scale
calibrations. The black line shows the L1 norm metric which compares the large-scale
calibrations between two iterations. The last two points compare successive large-scale calibrations at the end of the initial set of iterations
where the small-scale calibrations have also been carried out (see \citetads{2016A&A...595A...7C} 
and \citetads{PhotProcessing} for more details). The red line shows a modification of the metric where a default colour system has been used rather than a representative
sample.}
\label{Fig:Convergence}
\end{figure}

As described in \citetads{2016A&A...595A...7C} 
and \citetads{PhotProcessing}, a series of iterations of the LS and SS calibrations are carried out in order to establish the internal photometric system. To check the convergence of the
system, an L1 norm metric is used, as described in \citetads{2017A&A...600A..51E}. 
Figure \ref{Fig:Convergence} shows this convergence metric as a function of iteration number for \G-band Window Class 2 calibrations, covering the fainter magnitudes ($G>16$).
The black line shows the usual metric using a representative range of colour parameters. Although this drops to a low value (1.2 mmag), it does not
converge to zero as might be expected. This is probably due to the noise level in determining the calibrations and represents a limit in the achievable calibration
precision. To test this
idea further, the L1 norm metric was also calculated using a default colour system formed from an average set of colour parameters (red line in Fig. \ref{Fig:Convergence}).
In this case, the noise associated with the colour terms of the LS calibrations does not contribute to the metric, and thus it drops to a lower level (0.3 mmag) as expected.
The convergence results for Window Class configurations brighter than Window Class 2 show marginally better results.

In future development cycles, it is intended to develop a metric based on the changes in the reference fluxes between iterations since this would directly measure the stability of
the photometric system. The current system shown here effectively measures this by proxy.

\section{Analysis of source photometry statistics}\label{Sect:Accumulations}

\begin{figure} \resizebox{\hsize}{!}{\includegraphics{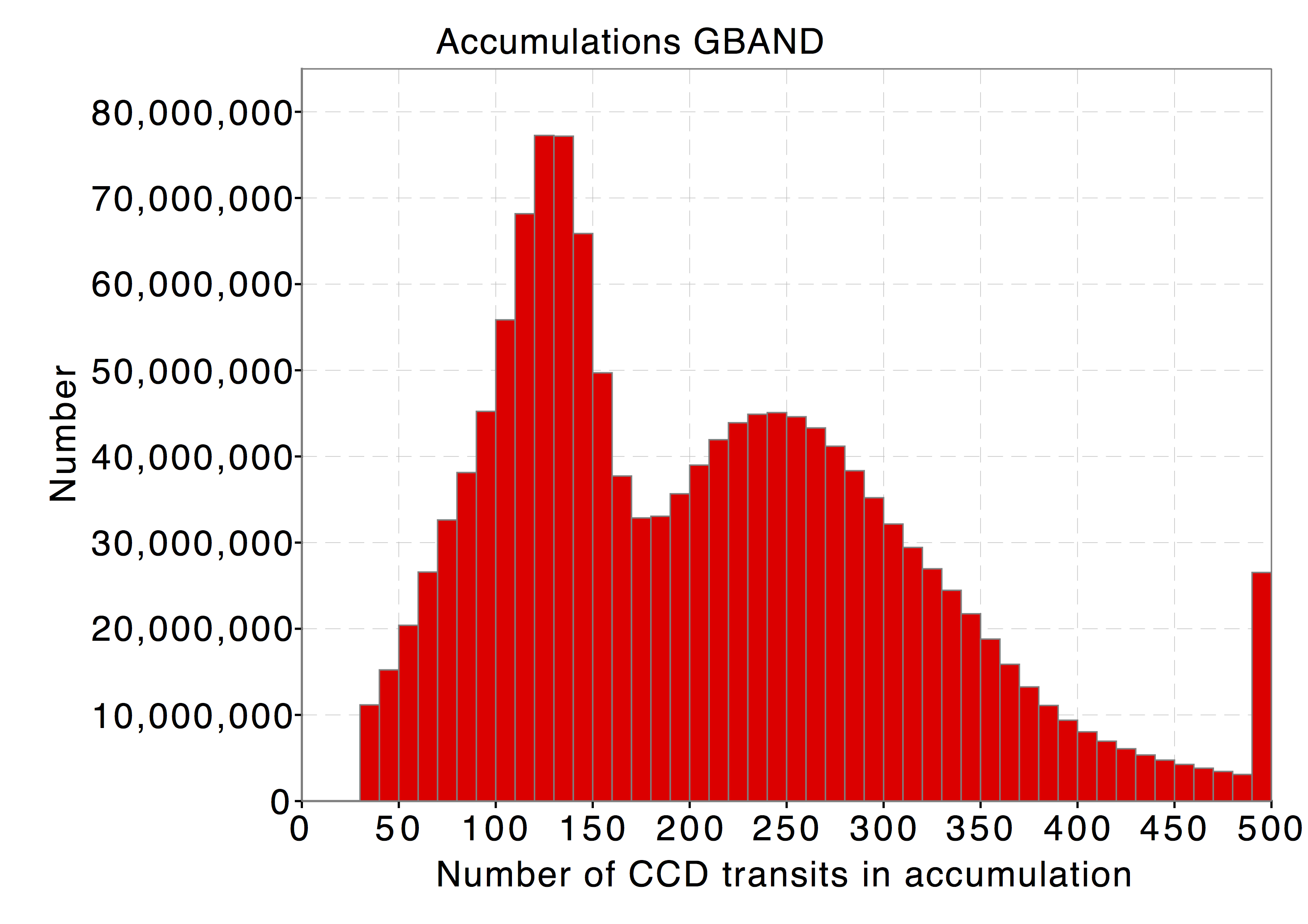}} \caption{Distribution of the number of \G-band CCD transits for each source. The last bin in the
distribution reports the number of sources with more than 500 \G-band CCD transits.}
\label{Fig:AccumNobs}
\end{figure}

As in \citetads{2017A&A...600A..51E}, 
the analysis of the accumulation data for \GDR2 has a restriction on the minimum number of observations. For \G\ observations, this is 30, while for \BP\ and \RP, it is 3. The
reason for this restriction is to minimise the effect of spurious detections, which will tend not to be cross-matched with other observations and therefore not have many
observations.
Figure \ref{Fig:AccumNobs} shows the distribution of the number of \G\ observations for each source. The mean and median values are 209 and 196, respectively, which is approximately
twice the number of observations contributing to the first data release. As can be seen from Fig. \ref{Fig:AccumNobs}, the distribution is broad and bimodal, which is a consequence of the
scanning law. The distribution of the number of \BP\ and \RP\ observations is similar, but reduced by a factor of approximately 9.

\begin{figure} \resizebox{\hsize}{!}{\includegraphics{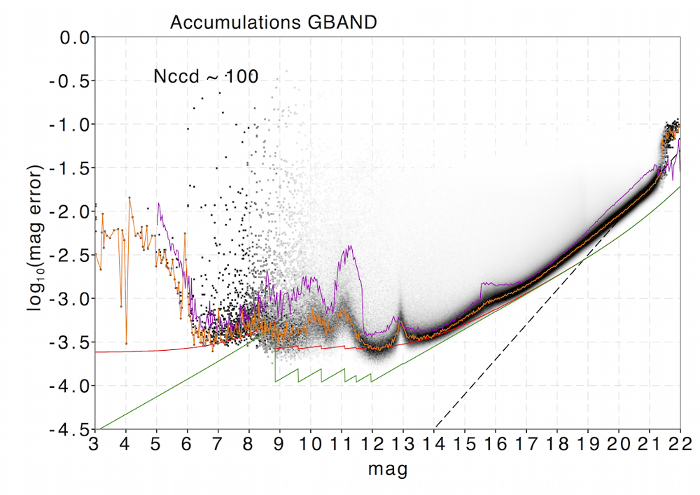}} \caption{ Distribution of uncertainty on the weighted mean \G\text{}
     value as a function of the same
 \G\ magnitude. The orange line shows the mode of
the distribution.
This plot is restricted to all sources with between
90 and 110 \G\ observations. The magenta line shows the equivalent results from \GDR1 as shown in Fig. 17 of \citetads{2017A&A...600A..51E}. 
The green line shows the expected uncertainties for sources with 100 \G-band CCD transits and for a nominal mission with perfect calibrations. 
The red line shows the same uncertainty function, but with a calibration uncertainty of 2 mmag added in quadrature to the individual observations.
The dashed black line has a slope of 0.4 and indicates that the faint end is sky dominated.
The distribution has been normalized along the magnitude axis, i.e. scaled so that each magnitude bin has the same number of sources in order to show features along the whole magnitude range.}
\label{Fig:Accum2dG}
\end{figure}

\begin{figure} \resizebox{\hsize}{!}{\includegraphics{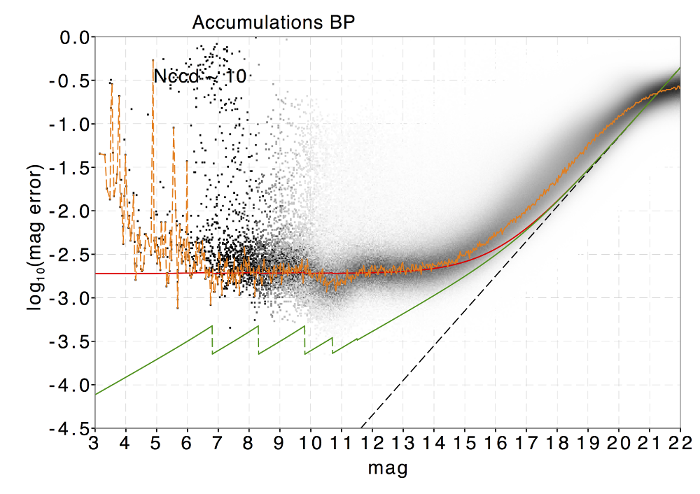}} \caption{ Distribution of uncertainty on the weighted mean \text{}\BP\
       value as a function of the same \G\ magnitude. 
The lines shown are equivalent to those of Fig. \ref{Fig:Accum2dG}. In this case, the plot is restricted to all sources with 10 \BP\ observations. 
The green line shows the expected uncertainties for sources with 10 \BP\ transits and for a nominal mission with perfect calibrations. 
The red line shows the same uncertainty function, but with a calibration uncertainty of
5 mmag added in quadrature to the individual observations.
The dashed black line has a slope of 0.4 and indicates that the faint end is sky dominated.}
\label{Fig:Accum2dBp}
\end{figure}

\begin{figure} \resizebox{\hsize}{!}{\includegraphics{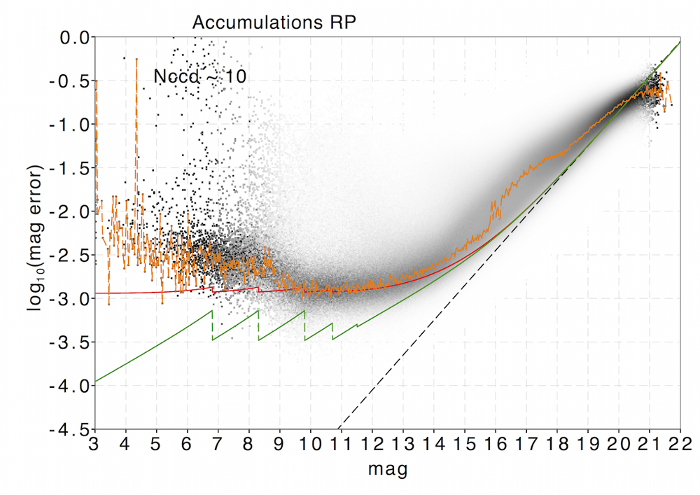}} \caption{ Same as Fig. \ref{Fig:Accum2dBp}, but for \RP.
In this case, the red line shows nominal uncertainty function with a calibration uncertainty of 3 mmag added in quadrature to the individual observations.}
\label{Fig:Accum2dRp}
\end{figure}

Figures \ref{Fig:Accum2dG} to \ref{Fig:Accum2dRp} show the distribution of the standard uncertainty on the weighted mean as a function of \G\ magnitude. These plots are restricted to
sources with approximately 100 CCD transits in \G\ (equivalent to about 10 \BP\ and \RP\ transits) so that they can be compared to predictions of the uncertainty using the 
formulation that can be found
in \citetads{2010A&A...523A..48J}. 
The value of 100 was chosen so that the results from \GDR1 for the \G\ band could be compared with these results. For future releases, a higher value will be chosen.

The green lines in these plots show the predicted uncertainties using nominal mission parameters for sources with 100 CCD observations. 
The red lines show the same predictions, but with a calibration floor added in quadrature.
The estimated calibration floors are 2, 5, and 3 mmag for \G, \BP, and \RP, respectively. For the \G\ band, this represents an overall improvement in the calibrations from the
3 mmag level attained in the first data release.
No \BP\ and \RP\ photometry was released for \GDR1 to be compared with the current results.
We also show in Fig. \ref{Fig:Accum2dG} the results from \GDR1\ (magenta line), which indicate an overall improvement in the calibrations and image parameter determination
(IPD) in this data release over all magnitude ranges. A number of features in these plots require further comment, however, since many of these features were present at a 
higher level in \GDR1; more detail about their causes can be found in \citetads{2017A&A...600A..51E}. 

At the faint end, $G>18$, the performance has improved as a result
of changes in the IPD. This is not as good as the predicted values because stray light levels
are higher than expected. The faintest sources, $G>21.3$, are probably spurious detections or have unreliable photometry. These number a few million, but are visible in 
Fig. \ref{Fig:Accum2dG} because  the distribution is normalised along the magnitude axis.

The features in Fig. \ref{Fig:Accum2dG} seen at around $G=13$ and 16 are related to changes in the window class of the observations. Improvements in the Gate/Window Class link calibrations
\citepads{PhotProcessing} have reduced the size of these features with respect to the first data release. Although the peak at $G=16$ has almost gone, the peak at $G=13$ remains, 
but is narrower. Another reason for the performance gain at $G=16$ is that the determination of the line spread function (LSF)
has been improved and thus does not provide a possible calibration floor at the bright end of this window class configuration.

There are considerable improvements in the performance for sources brighter than  $G=12$, where saturation becomes important. This is attributed to an 
improved masking of saturated samples as part of the cyclic IPD process. A further improvement in this data release for the
brighter sources comes from a better calibration of the point spread function (PSF).

The data for the \BP\ and \RP\ photometry shown in Figs. \ref{Fig:Accum2dBp} and \ref{Fig:Accum2dRp} are very different and show far fewer features. There is only
one window class configuration change for these observations;
this occurs at $G=11.5$. Even though the windows for the brighter sources are transmitted to the
ground as 2D windows, the photometric processing treats them as aperture photometry in the same way as for the 1D windows. This means that it is unlikely that
a jump is introduced into the photometric system during the initialization process. This is confirmed by the external catalogue comparisons, see Fig. \ref{Fig:ExtCompApass}, and by there
being no peak at $G=11.5$ in Figs. \ref{Fig:Accum2dBp} and \ref{Fig:Accum2dRp}. Moreover, because the light is spread by the prisms, saturation is much less of a problem in \BP\ and \RP.
Simulations have shown that saturation only occurs for \RP\ for the brightest and reddest sources.

The most noticeable feature seen in these plots is the broadening of the distributions in the range between $G=15$ and 19. This change in the distribution also moves the mode
(orange line) away from the expected uncertainties (green line). The explanation for this is that the sample of sources used for this analysis is restricted to certain areas of the
sky, which in turn is due to the combination of the selection on number of observations and the scanning law. One of these areas contains the Galactic centre and has strong crowding effects,
where the observations are contaminated by near neighbours and thus will have a larger scatter than normal.
Because of this, some of the sources will have a larger uncertainty on the weighted mean through this increased scatter than sources in other non-crowded areas.
These effects are more prominent in \BP\ and \RP\ because of the larger window size. 
That at the faint end ($G>20$) the uncertainties in \BP\ and \RP\ seem to be smaller than expected might be due to an underestimated background, implying a higher signal-to-noise ratio (S/N) than what would be expected for the \G\ magnitude.

Figures \ref{Fig:Accum2dG} to \ref{Fig:Accum2dRp} also show black dashed lines with a 0.4 slope that match the distributions at the faint end, thus indicating that the observations
there are sky dominated. While this was expected for the BP and RP instruments, the \G\ observations were intended to be source dominated (green line in 
Fig. \ref{Fig:Accum2dG}). The \G\ observations are sky dominated because additional stray light enters the telescope. This is described in detail in
\citetads{2016A&A...595A...1G}. 

Another potentially useful statistic derived from the source-based photometry are the P-values, which can be used to identify variable sources. 
The P-value is the probability that a set of observations is consistent with the null hypothesis. In this case, the null hypothesis is that the
source is constant and that the data are well calibrated, including the errors.
The performance of the \GDR2 P-values  is similar to that reported in \citetads{2017A&A...600A..51E} 
for \GDR1, even though the estimation of the fluxes and formal uncertainties on individual \G\ CCD transits has improved, as have the calibrations.
The P-value test is extremely sensitive to underestimation or overestimation of uncertainties or calibration problems. For \GDR2, it is likely that the formal uncertainties
on the \G\ CCD transits
are slightly underestimated. However, the calculation of the weighted mean accounts for this and the estimation of the uncertainty on this quantity is realistic.
The \BP\ and \RP\ photometry is not affected by this. It should be pointed out that the uncertainties quoted on the mean values are showing precision from a random 
perspective:  systematics that are a function of source-based parameters are significantly larger, as described in other sections.

\begin{figure*}
\centering
\includegraphics[width=\hsize*3/10]{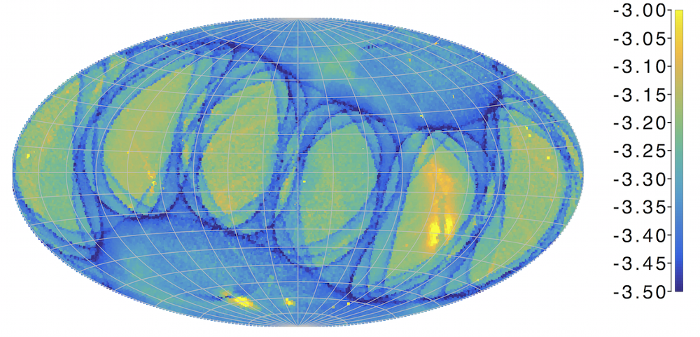} 
\includegraphics[width=\hsize*3/10]{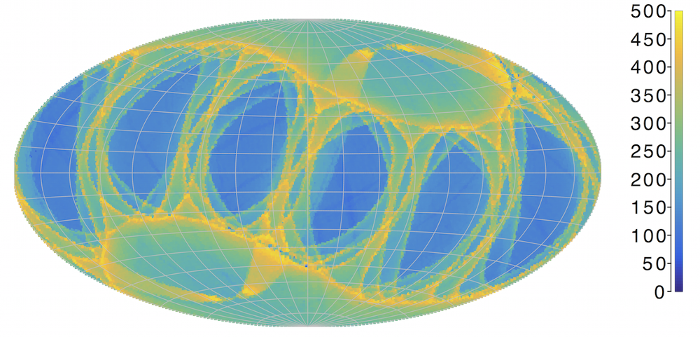} 
\includegraphics[width=\hsize*3/10]{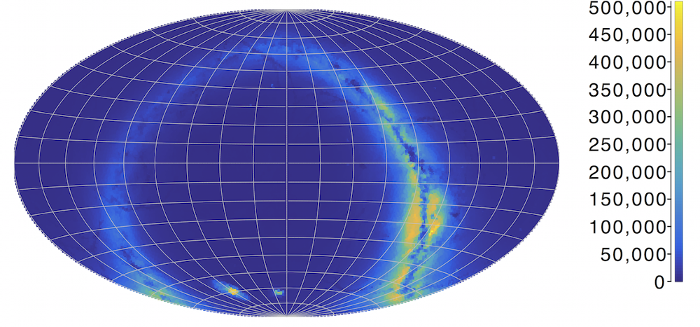} 
\caption{Sky distribution in equatorial coordinates of the uncertainty on the weighted mean in \G\ for sources in the magnitude range 13 to 16 (left plot). 
The middle plot shows the number of observations per source and the right plot the number of sources with $G>16$ per level 6 Healpixel.} 
\label{Fig:AccumSky}
\end{figure*}

The left plot in Fig. \ref{Fig:AccumSky} shows the variation with sky position of the uncertainty on the weighted mean for \G\ for sources that generally are observed in Window Class 1,
that is, a
magnitude range of $13<G<16$. The other plots in this figure show the distribution of the number of observations per source and the source density. As can be seen in these plots, the uncertainty improves with more observations and degrades in areas of high density through crowding
effects. The density plot shown in Fig. \ref{Fig:AccumSky} covers the faintest magnitude range that is analysed since this is the most relevant for the assessment of crowding.
The variation seen in the other passbands and Window Classes is very similar.

\section{Comparisons with external catalogues}\label{Sect:ExternalComparisons}

\begin{figure*}
\centering
\includegraphics[width=\hsize*3/10]{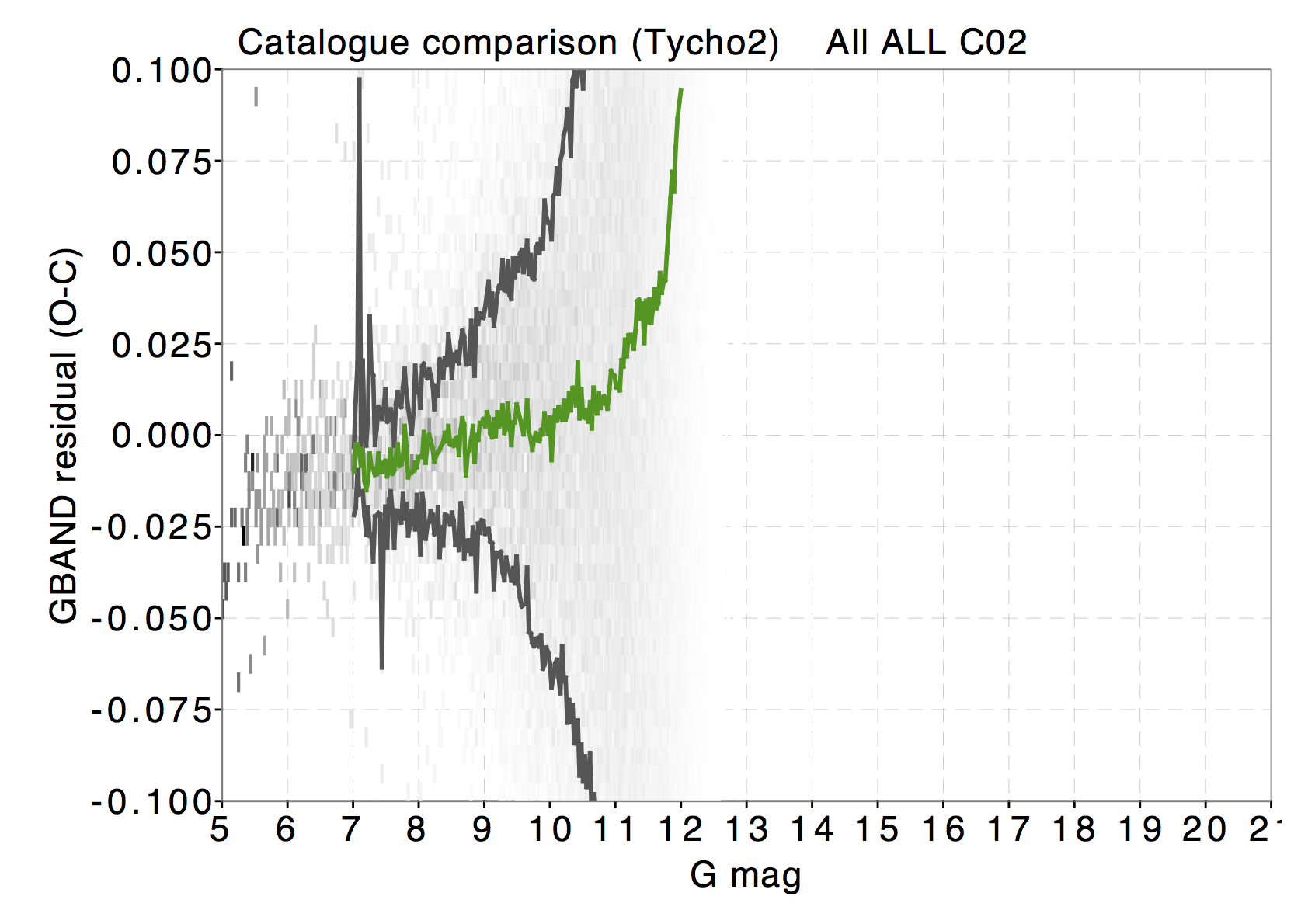} 
\includegraphics[width=\hsize*3/10]{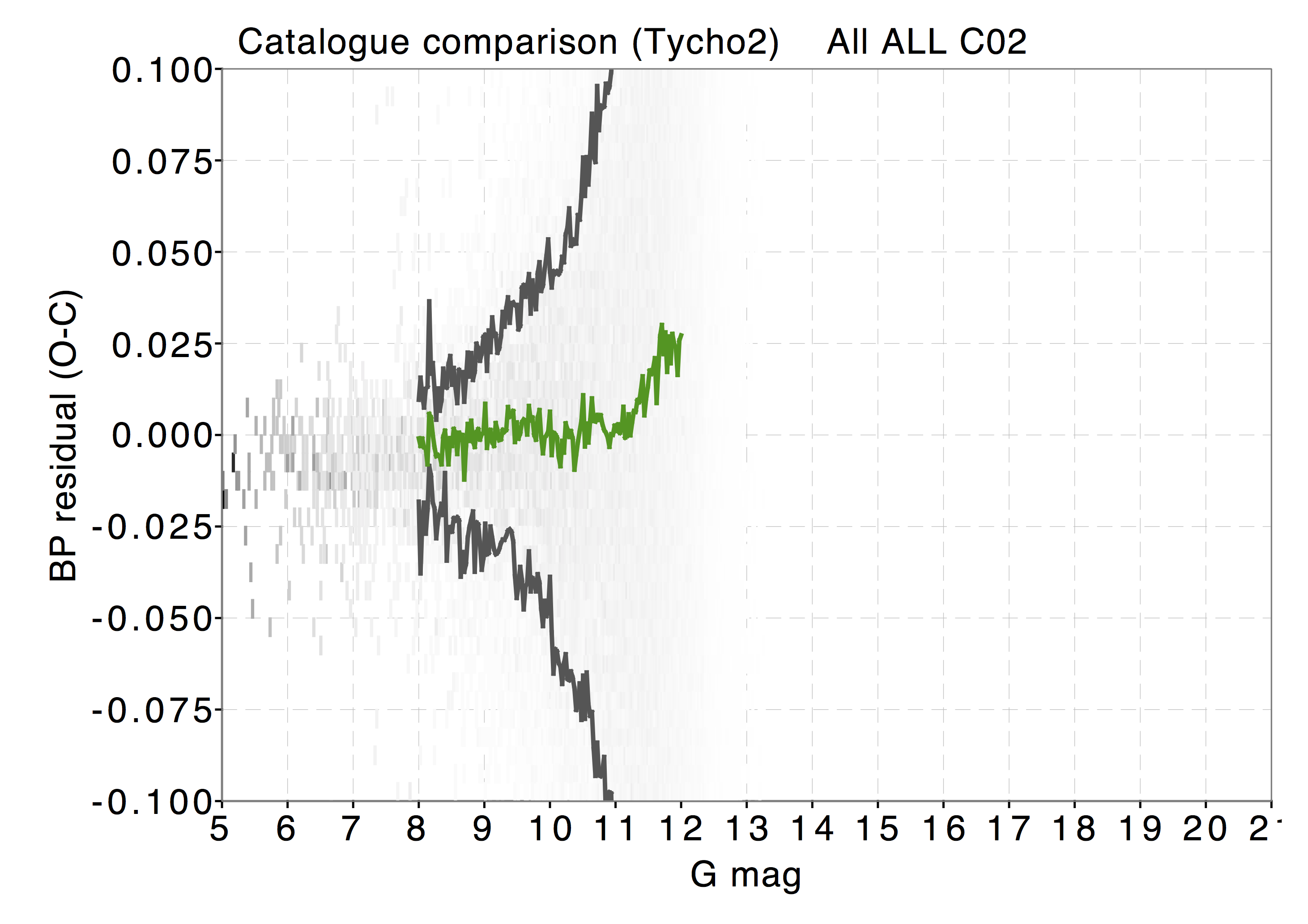} 
\includegraphics[width=\hsize*3/10]{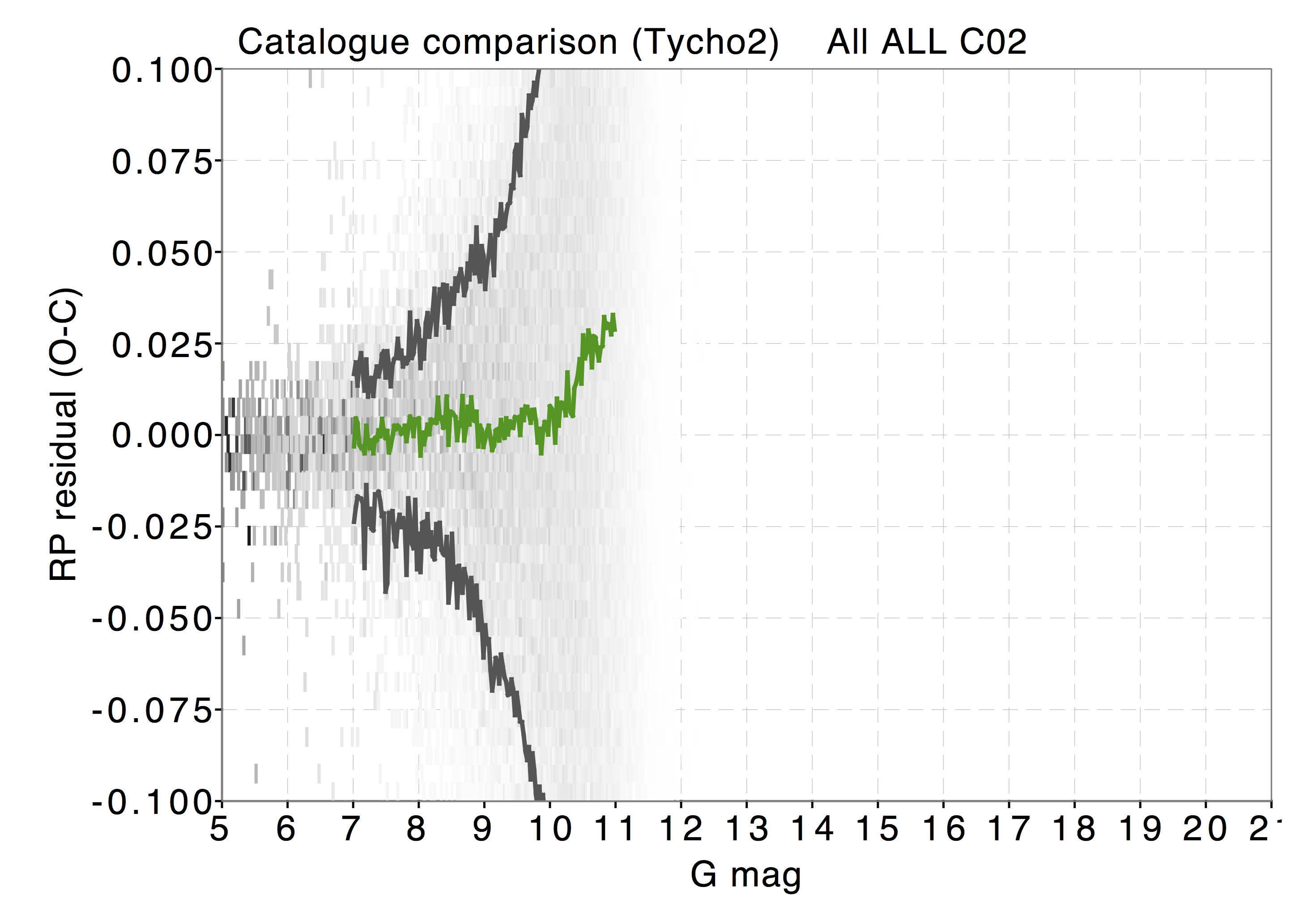} 
\caption{Comparisons of the three \Gaia\ passbands, \G, \BP, and \RP, with respect to the transformed $V_{T}$ photometry from the external photometric catalogue \TychoTwo.  
The green and black lines show the median and one sigma points of the residual distributions, respectively.} 
\label{Fig:ExtCompTycho2}
\end{figure*}

\begin{figure*}
\centering
\includegraphics[width=\hsize*3/10]{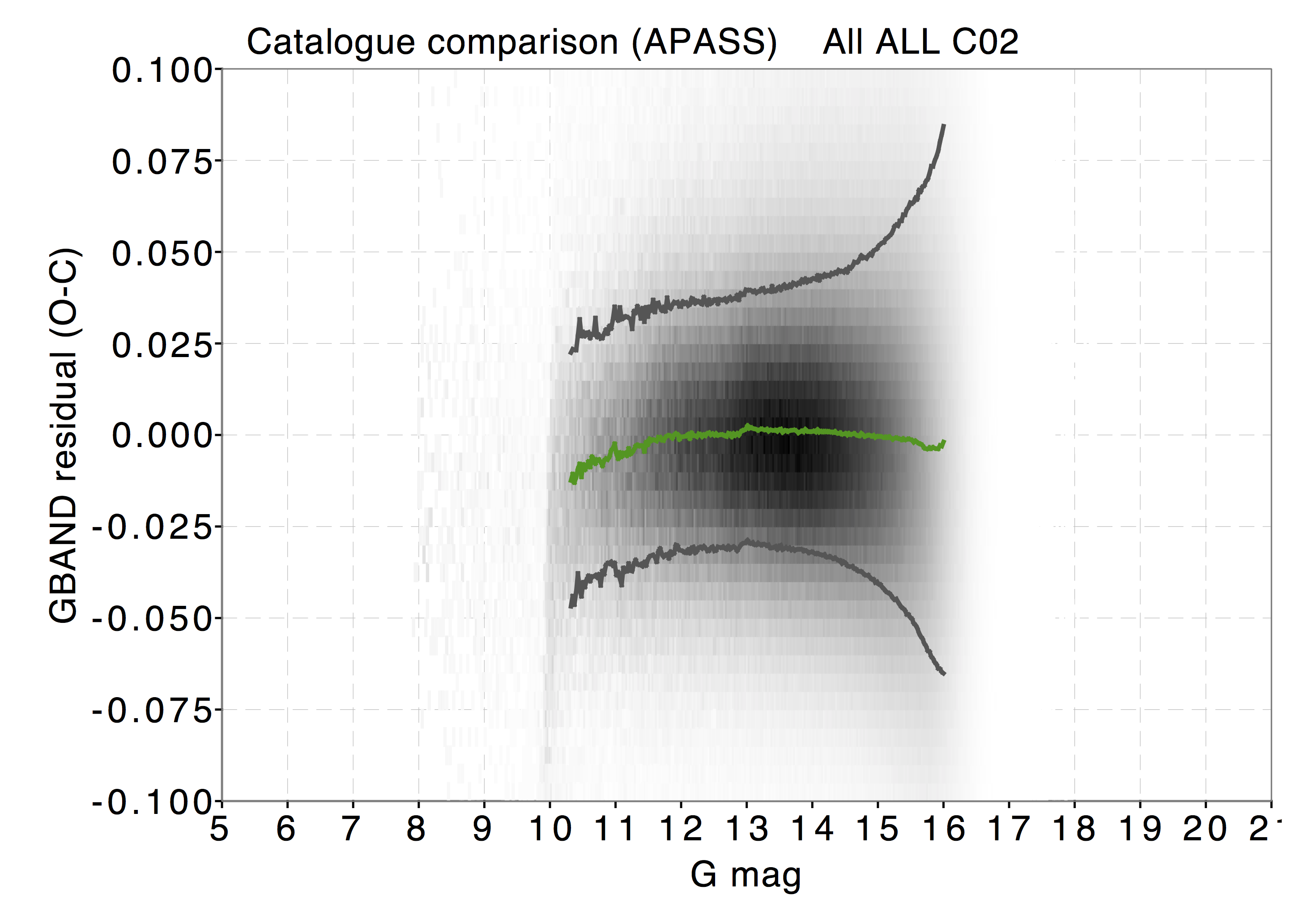} 
\includegraphics[width=\hsize*3/10]{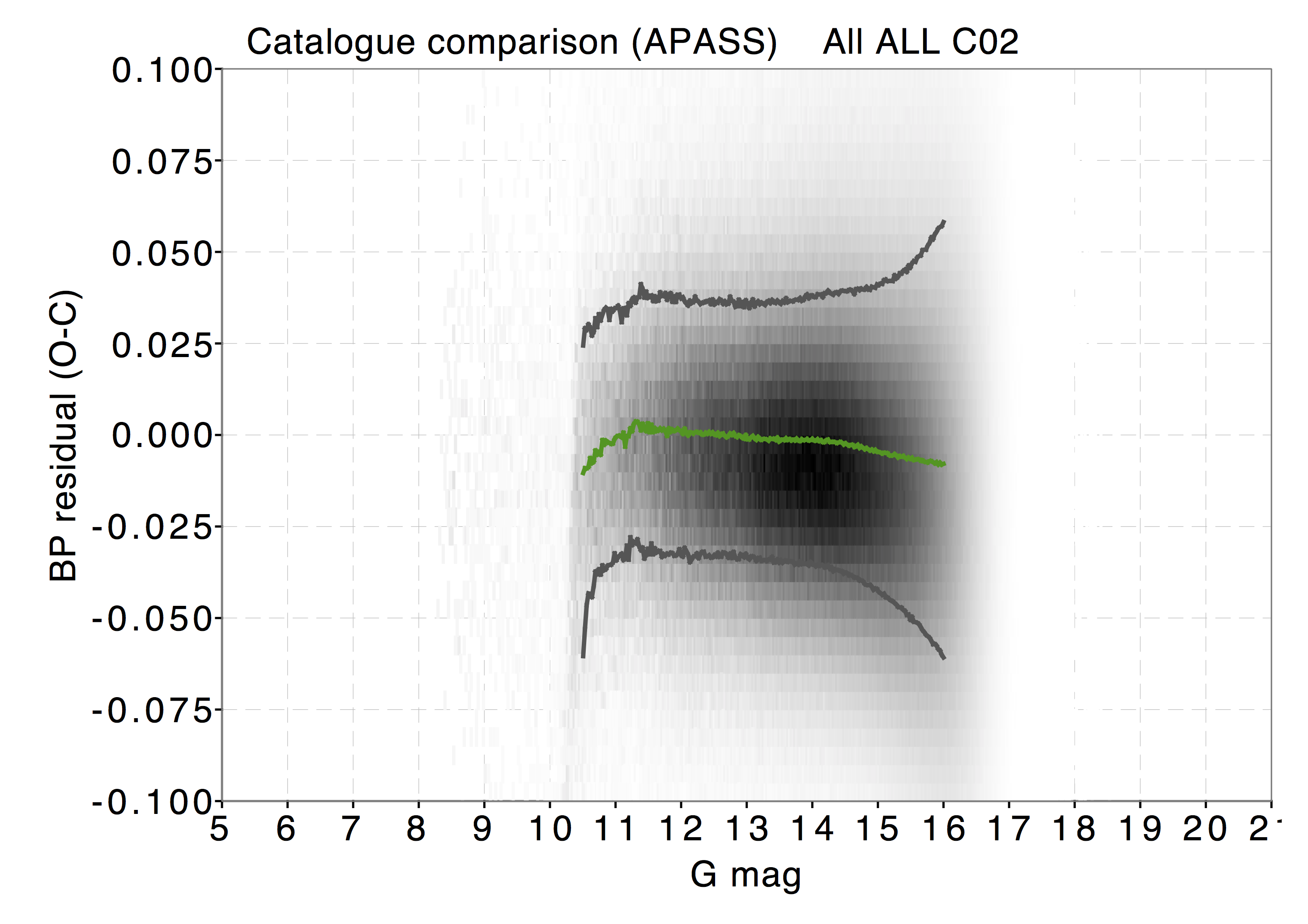} 
\includegraphics[width=\hsize*3/10]{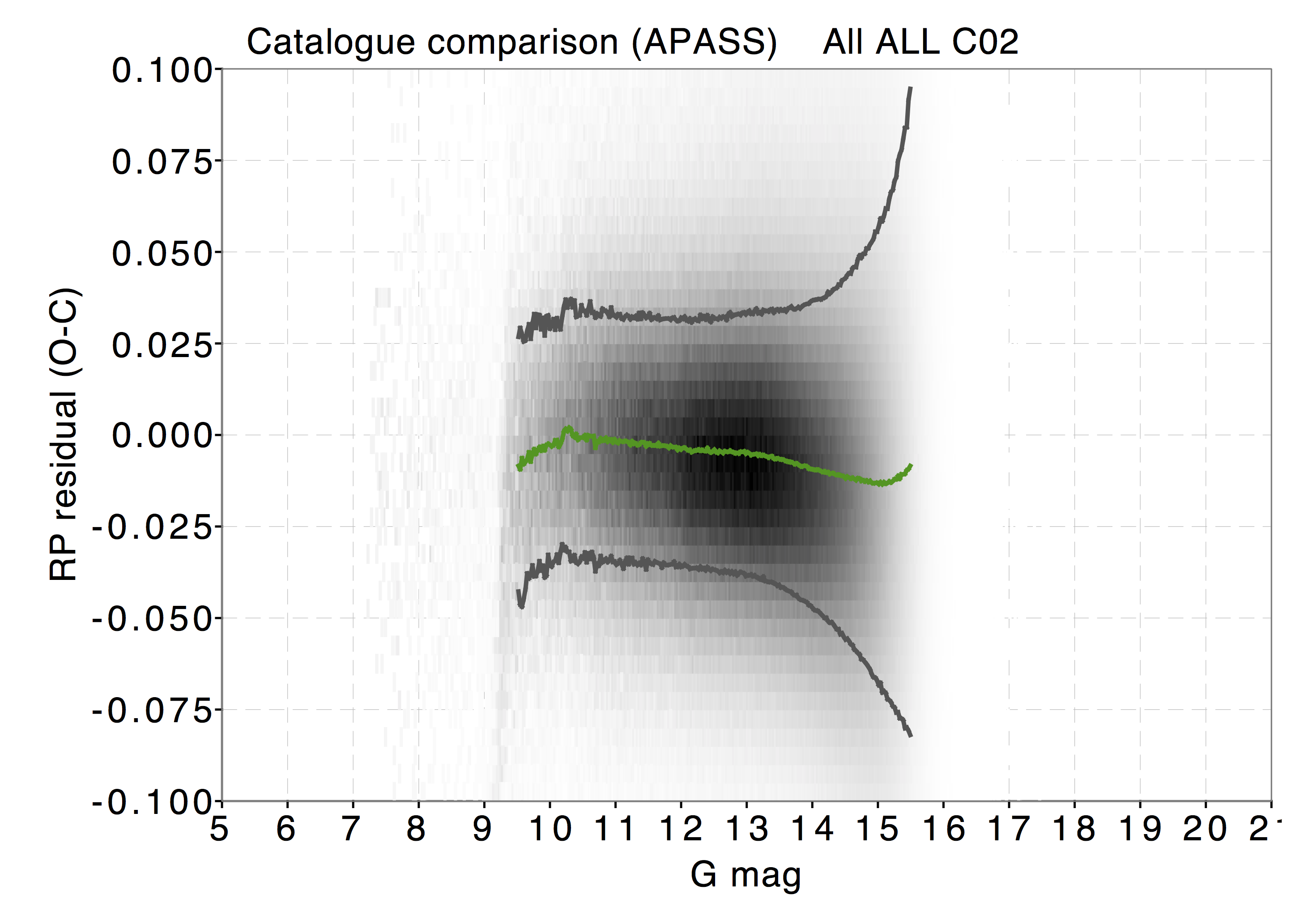} 
\caption{Comparisons of the three \Gaia\ passbands, \G, \BP, and \RP, with respect to the transformed $r'$ photometry from the external photometric catalogue APASS.  
The green and black lines show the median and one sigma points of the residual distributions, respectively.}
\label{Fig:ExtCompApass}
\end{figure*}

\begin{figure*}
\centering
\includegraphics[width=\hsize*3/10]{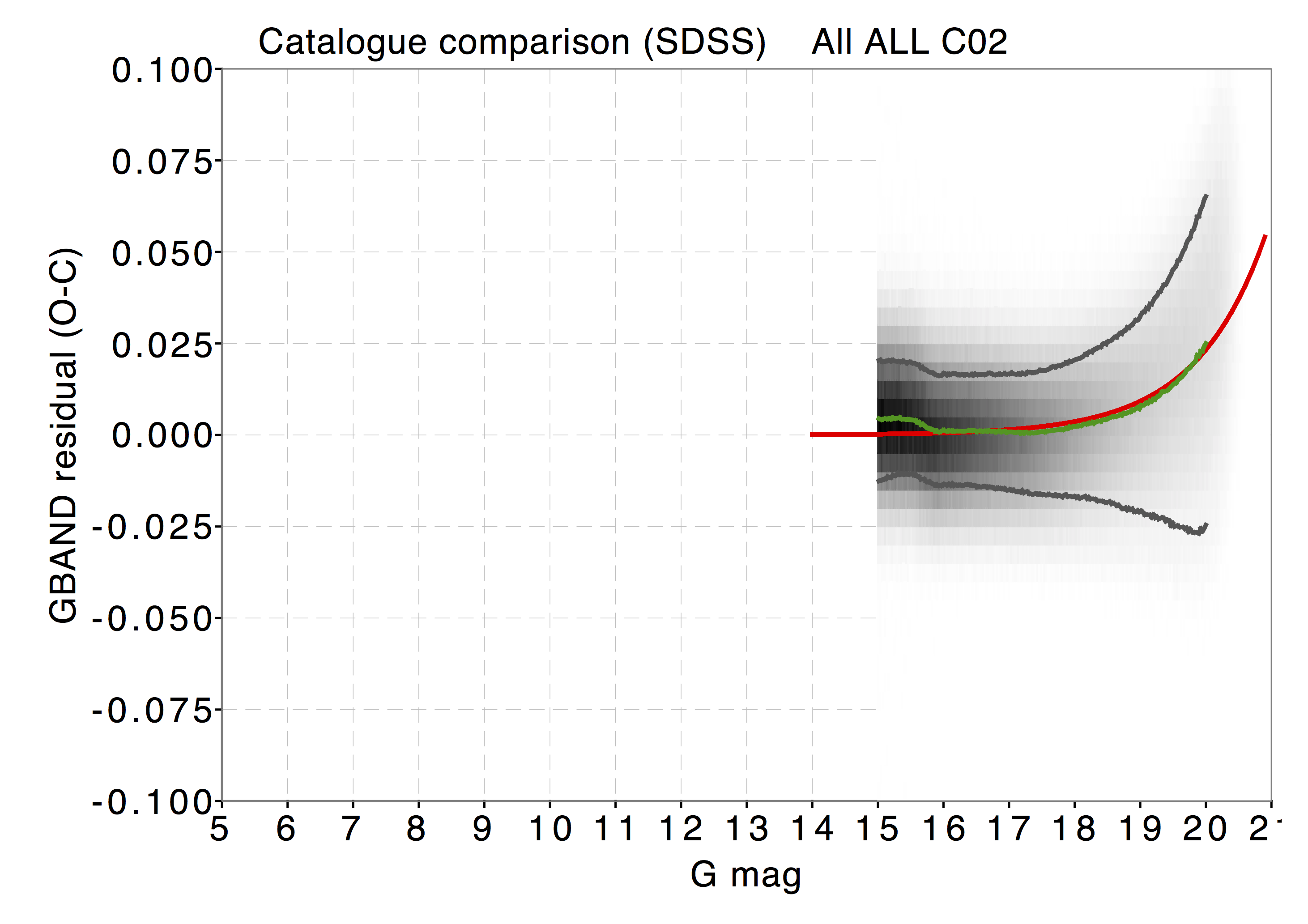} 
\includegraphics[width=\hsize*3/10]{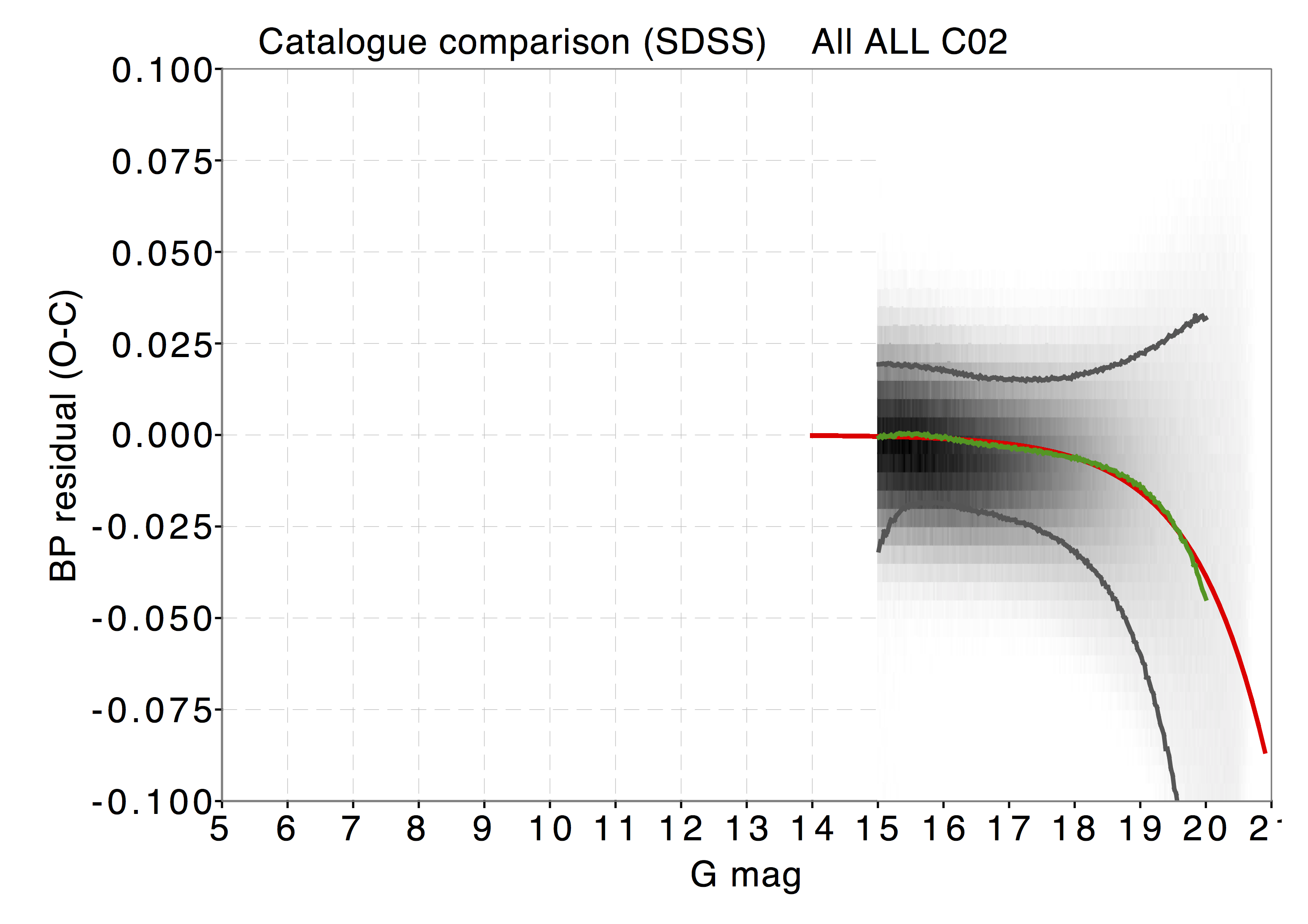} 
\includegraphics[width=\hsize*3/10]{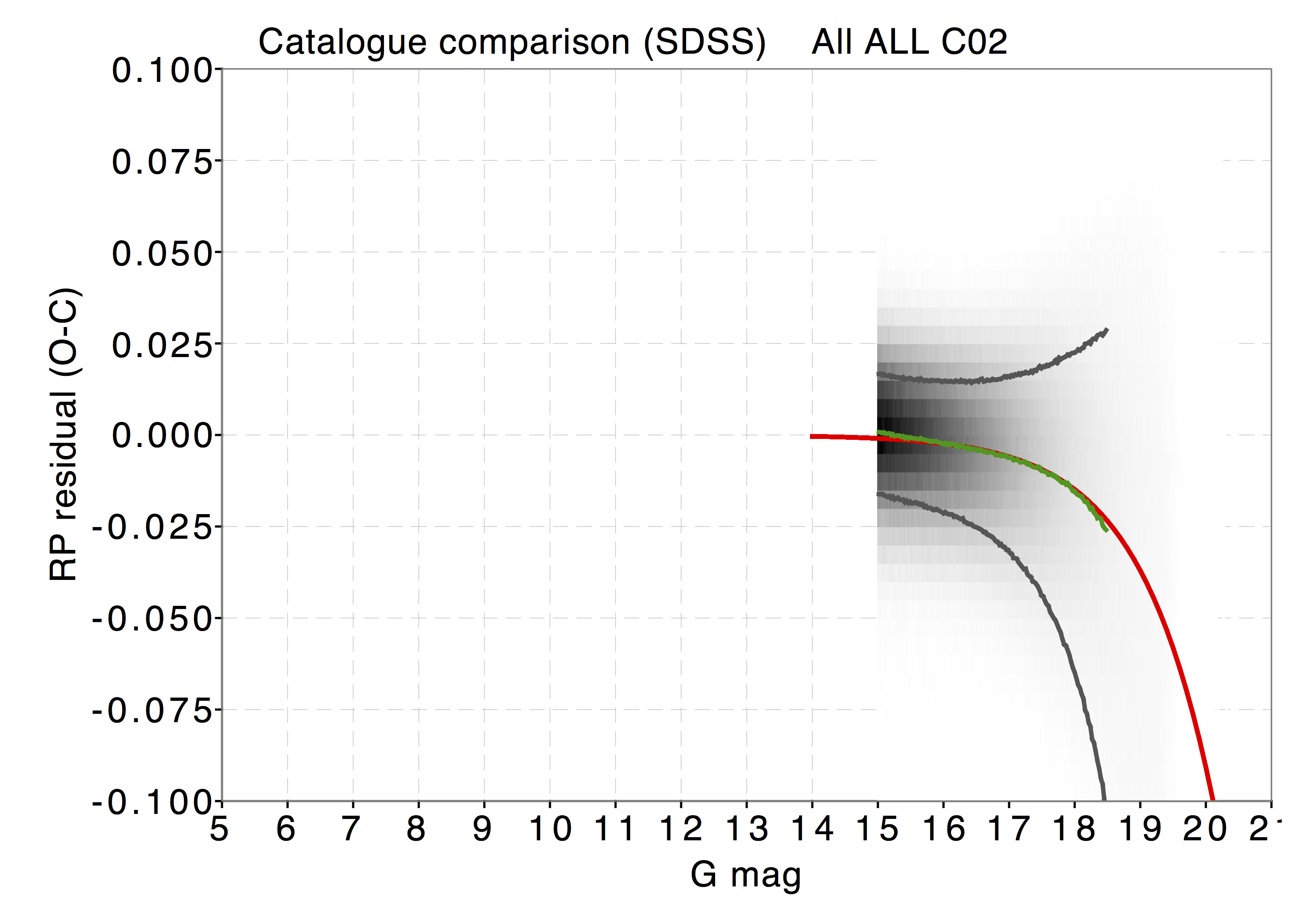} 
\caption{Comparisons of the three \Gaia\ passbands, \G, \BP, and \RP, with respect to the transformed $r'$ photometry from the external photometric catalogue SDSS.  
The green and black lines show the median and one sigma points of the residual distributions, respectively. The red lines show possible flux offsets to the \Gaia\ photometry that match the systematics seen at the faint end. These correspond to
values of -4, 5, and 6 e$^{-}$s$^{-1}$ for \G, \BP, and \RP, respectively.
} 
\label{Fig:ExtCompSdss}
\end{figure*}

As pointed out in \citetads{2017A&A...600A..51E}, 
comparisons with external photometric catalogues can prove useful, but care must be taken that any differences are attributed to the correct source.
The photometric
precision of \Gaia\ is usually much better than that of the external catalogues,
so that the width of the distribution of differences between \Gaia\ magnitudes and those coming from other external catalogues will be dominated by the uncertainties in the external catalogue photometry.
The average difference and its variation with magnitude are the most useful indicators when validating \Gaia\ photometry versus other catalogues.
It should also be noted that the angular resolution of \Gaia\ is better than that of
ground-based catalogues, and in order to avoid high-density regions, where crowding may be a problem, sources close to the Galactic plane ($|b|<10^{\circ}$) have been
excluded from these comparisons. This also reduces any problems linked with reddening in these comparisons.

The main comparisons made for DR2 have been made with respect to 
\TychoTwo\ \citepads{2000A&A...355L..27H}, 
APASS
\citepads[AAVSO Photometric All-Sky Survey, ][]{2015AAS...22533616H} 
and SDSS DR12
\citepads[Sloan Digital Sky Survey, ][]{2015ApJS..219...12A}. 
In the \GDR1 analysis, only a \G-band comparison was required. In order to avoid complications linked to absorption, luminosity class, and population, a simple approach
was used in which the external passbands were compared directly with \G\ and no transformations were used. This method was aided by the comparison being out of the
Galactic plane, thus avoiding most extinction effects, and it
was restricted to the colour range 1.0$<$\BP$-$\RP$<$1.2. In this narrow colour range, the colour term between
the \G\ and $r'$ bands is very small and contributes very little to the overall width of the comparison. However, for \GDR2, comparisons must be made with the \BP\ and \RP\
photometry, where the assumption of a small colour term is invalid.
Using transformations given in Appendix \ref{Sect:Transformations},
the $r'$ and $V_{T}$ photometry from the external catalogues was converted into the three \Gaia\ passbands. The narrow colour restriction was maintained, thus avoiding any
problems with the transformations at extreme colour. A zeropoint offset was also applied to aid the comparisons.

Figure \ref{Fig:ExtCompTycho2} shows the comparison with the \TychoTwo\ catalogue. In the \G\ comparison, a small gradient is seen between $G=7$ and 11, resulting in a
systematic of up to 10 mmag. This matches the consistency results in 
\citetads{DR2-DPACP-39}. 
The \BP\ and \RP\ comparisons in the same magnitude range are consistent with no significant differences between the two catalogues.
The upturn seen at the faint end in all three comparisons is probably due to a bias in the \TychoTwo\ data, especially since this upturn is not seen in the APASS comparison.

Figure \ref{Fig:ExtCompApass} shows the comparison with the APASS catalogue.
The small jump of 2 mmag seen at \G=13 in the \G\ band comparison is probably
linked to a Window Class change in \Gaia. Although the Gate/Window Class calibration is specifically run to solve this problem, see \citetads{2016A&A...595A...7C} 
for more details, it does not remove this effect entirely, and the performance is similar to that of \GDR1 at this magnitude. For \BP\ and \RP, the corresponding Window Class change occurs at \G=11.5, and 
the above comparisons show that the calibration has worked very well in this case.

The systematic differences seen at the bright end are likely to originate in the APASS photometry since they are seen in all three comparisons. Comparisons with the \GDR1 performance
shown in Sect. 10 of \citetads{2017A&A...600A..51E}, 
where a bump was seen at \G=11, indicate that the handling of saturation effects have been improved.

Figure \ref{Fig:ExtCompSdss} shows the comparison with the SDSS catalogue. The 3--4 mmag jump seen at \G=16 is associated with another Window Class change in 
\Gaia\ and indicates that the Gate/Window Class calibration has not fully succeeded in this case. The size of the jump is about half that seen in \GDR1.

At the faint end, ``hockey stick'' features are seen in all three passband comparisons. Assuming that this has a \Gaia\ rather than an SDSS origin, 
a possible explanation for these trends is that the background subtraction has not been
fully successful. In order to estimate the size of these effects on the three passbands, lines have been drawn on the plots in Fig. \ref{Fig:ExtCompSdss} that represent the
effect of a single flux offset that might be representative of an erroneous background determination. These flux offsets are not fits to the data, but representative values. It is still unclear whether a single value can be used for all sources.
Additionally, the trends seen at the faint end could also be affected by catalogue incompleteness caused by the magnitude detection threshold in the \Gaia\ data.

\begin{figure}
\centering
\includegraphics[width=\hsize]{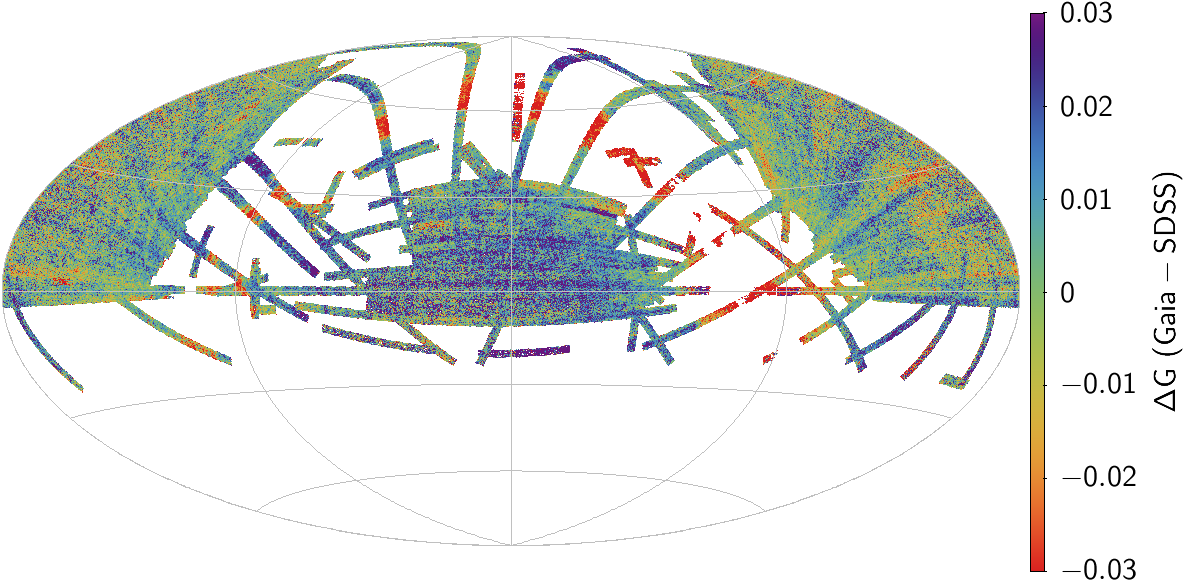}  
\caption{Sky distribution of the median difference between the \Gaia\ \G\ magnitude and a \G-band magnitude estimated from the SDSS $r$ and
  $i$ photometry using the photometric transformations listed in Appendix \ref{Sect:Transformations}. The comparison is limited to sources
  with $13<G<21$, $0.6<r-i<1.2$, \Gaia\ photometric errors lower than 0.03, and SDSS photometric errors lower than 0.05. The median value for
  each HEALPix pixel is computed over all sources falling in the same pixel of level 8.
} 
\label{Fig:SdssSky}
\end{figure}
A sky distribution of the median difference in the \G\ band between \Gaia\ photometry and SDSS is shown in Fig. \ref{Fig:SdssSky}. 
All SDSS sources with a cross-match to the \GDR2 catalogue and with photometric errors in the $r$ and $i$ bands lower than $0.05$ were considered for this plot. 
The \G-band photometry for the SDSS catalogue was derived from $r$ and $i$ photometry using the photometric transformations presented in the Appendix \ref{Sect:Transformations}. An additional zeropoint offset of $0.16$ mag was removed to centre the distribution of residuals around zero.
Only sources with a $r-i$ colour within the range $[0.6, 1.2]$, well within the validity range of the photometric transformation, were selected.
The Galactic plane stands out with larger differences between \Gaia\ and SDSS magnitudes.
Beyond the Galactic plane, the systematics are dominated by the SDSS photometry, as proven by the fact that the most prominent features are aligned with the SDSS scans through the sky.

\section{BP/RP flux excess}\label{Sect:BpRpExcess}

\begin{figure} 
\resizebox{\hsize}{!}{\includegraphics{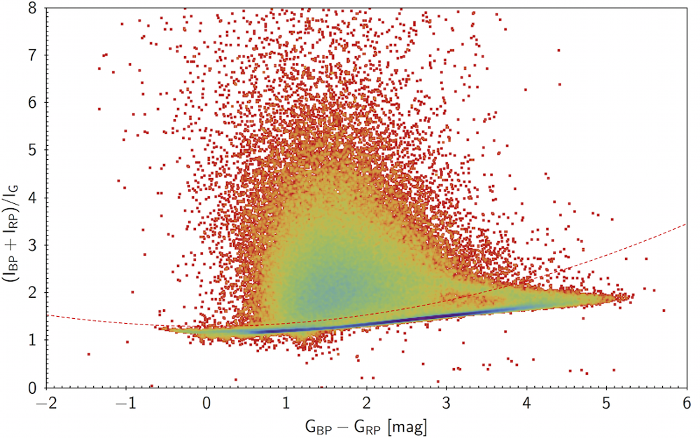}} 
\resizebox{\hsize}{!}{\includegraphics{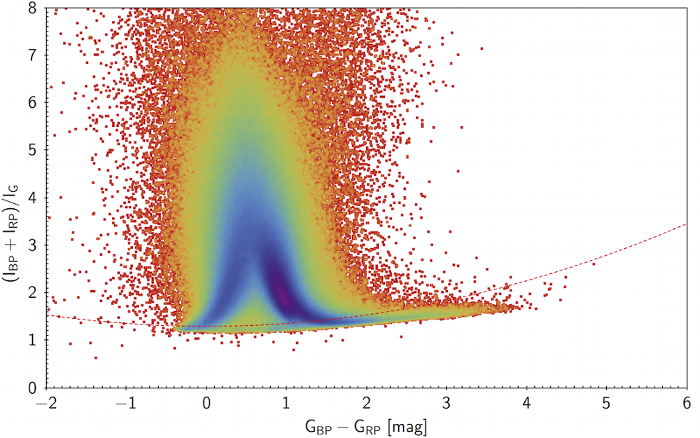}} 
\caption{Flux excess versus colour. {\em Top:} Nearby sources 
($\varpi > 15$\,mas). {\em Bottom:} Sources near the centre of the LMC.
The dotted line corresponds to $1.3+0.06 (G_{BP}-G_{RP})^2$.}
\label{fig:ratioVsCol}
\end{figure}

For most sources we have flux estimations in three bands: \G, \BP, and \RP.
The \G\ flux, $I_G$, is determined from a profile-fitting to a narrow image,
while the $BP$\ and $RP$\ fluxes, $I_{BP}$ and $I_{RP}$, give the total flux in a
field of $3.5\times 2.1$ arcsec$^2$.  These fluxes are therefore much more
susceptible to contamination from nearby sources or an unusually bright sky
background than the \G\ flux.  For \GDR 2, no deblending was applied, and we may
expect that the colour information for a source often suffers from
contamination. It is therefore recommended to check that the fluxes in the three
bands are consistent with the assumption of a source being isolated if accurate
colour information is required.

The \BP\ and \RP\ passbands overlap slightly and have a somewhat better
response in their respective wavelength ranges than \G, 
so to a first approximation, we expect the sum of the $BP$\ and $RP$\
fluxes to exceed the \G\ flux by only a small factor. We define the {\tt
phot\_bp\_rp\_excess\_factor} in the \GDR2 archive 
as the simple flux ratio $C=(I_{BP}+I_{RP})/I_G$.
Figure~\ref{fig:ratioVsCol} shows this factor versus the observed colour for
nearby sources in the top panel and for sources towards the centre of the Large Magellanic Cloud (LMC)
in the bottom panel. Most of the nearby sources are confined to a narrow band
slightly below the dotted line ($1.3+0.06 (G_{BP}-G_{RP})^2$). We interpret
this band as the well-behaved single sources. We also note a cloud
of points with excess around 2--3, however, which must be heavily affected, plus
points with very low or very high values. The full range extends from 0.02
to more than 700. For the very crowded LMC area, the narrow band of well-behaved
sources is still visible, but now more like a lower envelope. The affected
sources here cluster at much bluer colours than for the general nearby sources. 

\begin{figure} 
\resizebox{\hsize}{!}{\includegraphics{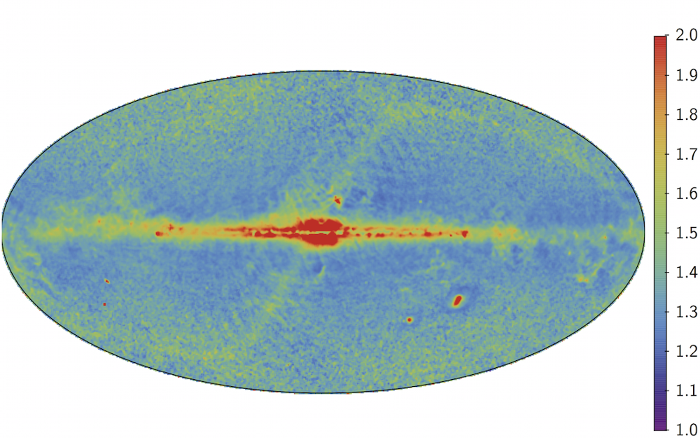}} 
\caption{Median flux excess in galactic coordinates for a random set of sources.}
\label{fig:ratioMap}
\end{figure}

Figure~\ref{fig:ratioMap} shows the median excess factor across the sky.  Very
dense areas like the Galactic centre and the Magellanic clouds stand out with
high excess levels, as can be expected because of the crowding.  The
central Galactic regions resemble near-infrared images. We also note a
narrow band along the ecliptic plane, where the excess is likely due to
insufficient subtraction of zodiacal light. This suggests that the
sky background is not always well modelled and can leave an imprint on the
fluxes, and thereby the colours, for faint sources. 

\begin{figure} 
\resizebox{\hsize}{!}{\includegraphics{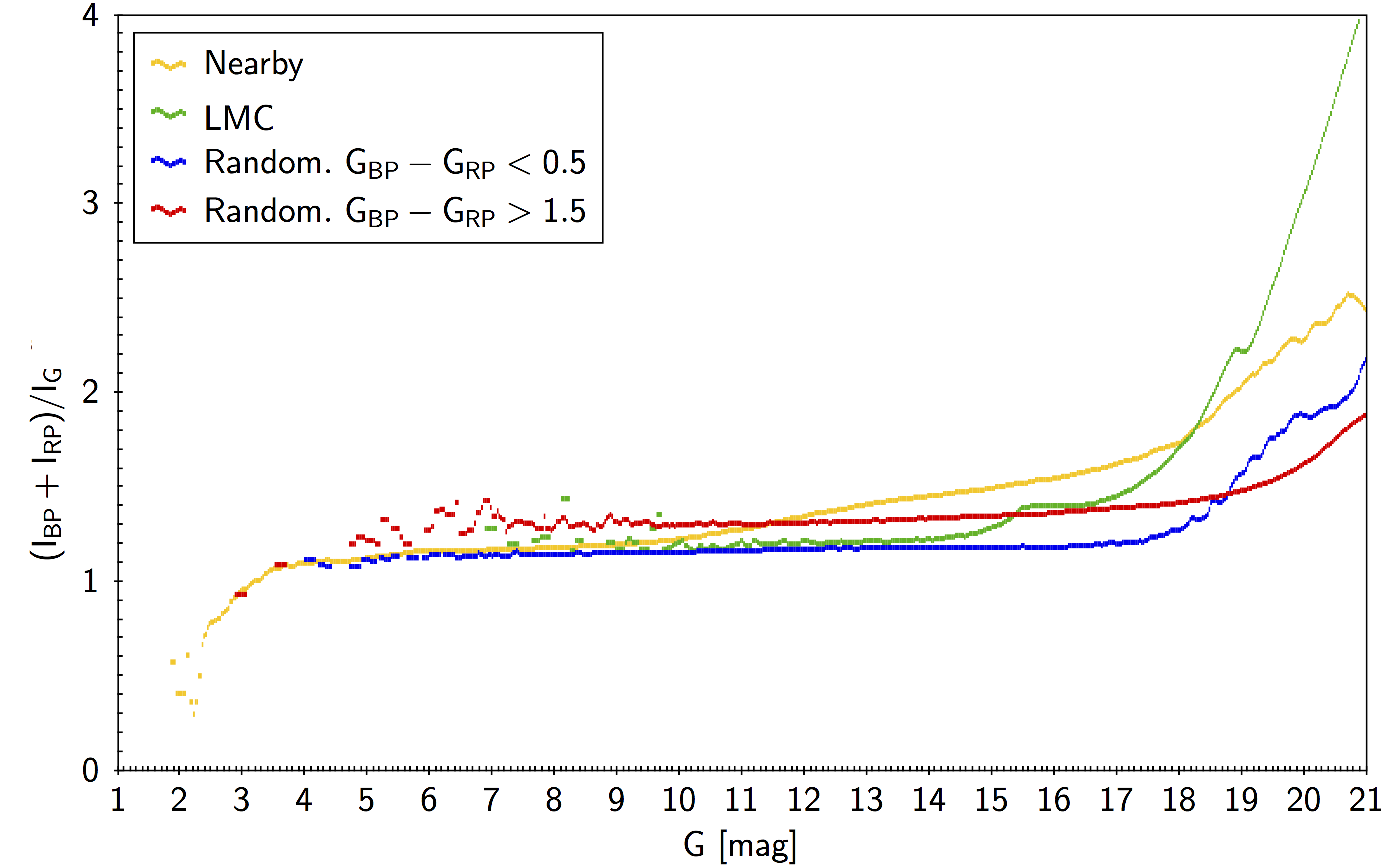}} 
\caption{Median flux excess versus magnitude for nearby sources, for sources
near the LMC centre, and for random sets of blue and red sources.}
\label{fig:ratioVsG}
\end{figure}

The magnitude dependence of the flux excess is shown in
Fig.~\ref{fig:ratioVsG}.  We show median values for the nearby and LMC
datasets, as well as for randomly selected blue and red sources. For sources brighter than 17--18~mag, the behaviour is good, with the
exception of the very bright end, that is,\ \G\ below 3.5~mag, where saturation
issues are very likely at play.  The blue and red sources stay at almost
constant levels, while the nearby selection becomes redder at fainter colours.
Beyond about 18~mag, the median values shoot upwards, indicating that very many sources show one or more of the above issues.

As mentioned, the flux excess factor may occasionally reach extremely high
values that indicate a very serious crowding, processing, or
calibration issue. In these cases, the $BP$\ and $RP$\ fluxes have little
relation to the source in question, and as a precaution, they have been removed
from the data release. The limit applied in \GDR 2 is a flux excess of five.
The diagrams discussed above have all been produced before applying this
filter.

Sources can show high flux excess for many reasons. The dominating reasons are
binarity, crowding, and incomplete background modelling, but extended
objects or galaxies can cause problems. Obviously, the surroundings of very bright
sources are a particular challenge. All these problems will mostly affect the
\BP\ and \RP\ measurements because of the longer windows, but they can also disturb
the \G-band observations.

We therefore recommend that a filter be applied in order to remove the
more problematic sources. This can be done with a requirement such as

\begin{equation}
\label{eq:excessflux}
C < a + b (G_{BP}-G_{RP})^2,
\end{equation}

\noindent where $C$ is the flux excess, and $a$ and $b$ must be chosen depending on
the characteristics of the sample under study and on how clean a subset is
needed.

At the present stage of \Gaia\ data processing, many sources
show one or more of these problems, but the number of \Gaia\ sources is
very large, and with suitable filtering, many clean sources
still remain.

\section{Validation of the external flux calibration and passband determination}\label{Sect:Passbands}

\begin{figure*}
\centering
\includegraphics[width=\hsize*3/10]{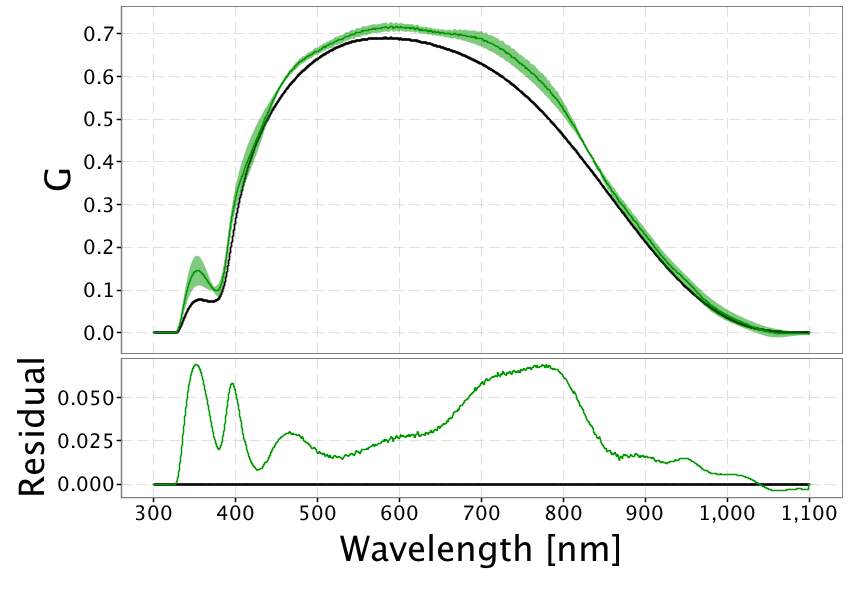} 
\includegraphics[width=\hsize*3/10]{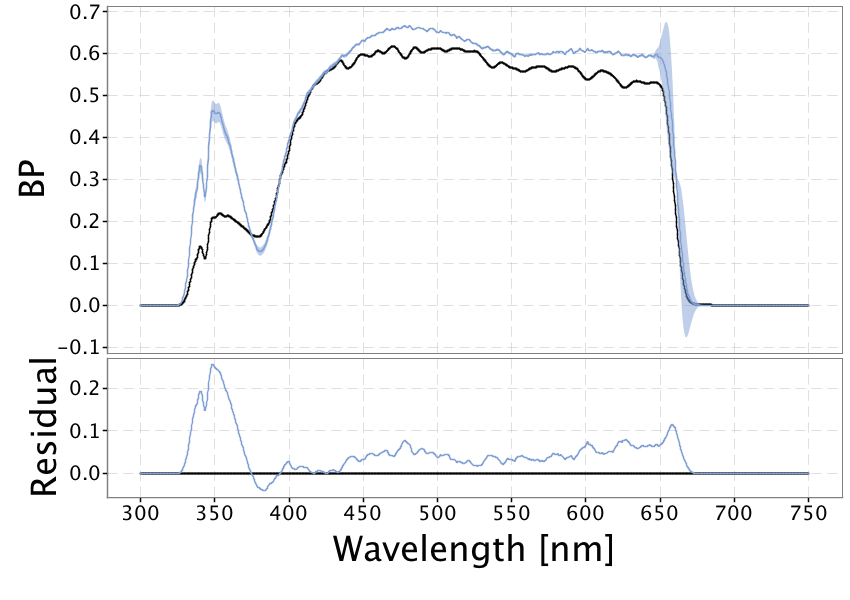} 
\includegraphics[width=\hsize*3/10]{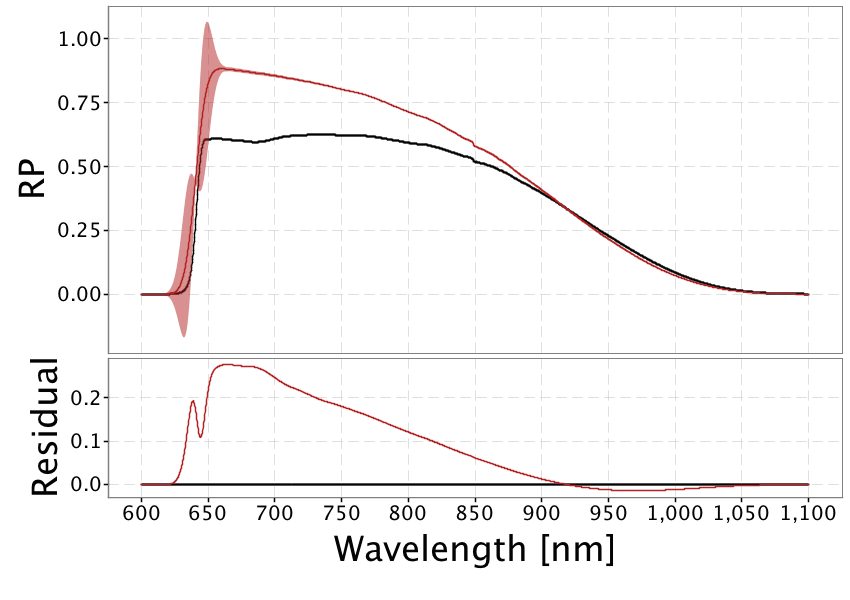} 
\caption{Calibrated three \Gaia\ passbands, \G\ (\emph{left}), \BP\ (\emph{centre}), and \RP\ (\emph{right}) used for \GDR2 processing. The calibration has been achieved using  a set of 93 SPSS. 
Black lines represent nominal response curves, while the coloured lines show the calibrated curves. Shaded areas represent the $\pm 1\sigma$ level: the error curve has been computed by standard propagation of the uncertainty using the model covariance matrix obtained in the model-fitting process. The bottom panels show the difference between calibrated and nominal curves.
} 
\label{Fig:ExtPassbandsJuly}
\end{figure*}

\begin{figure*}
\centering
\includegraphics[width=\hsize*3/10]{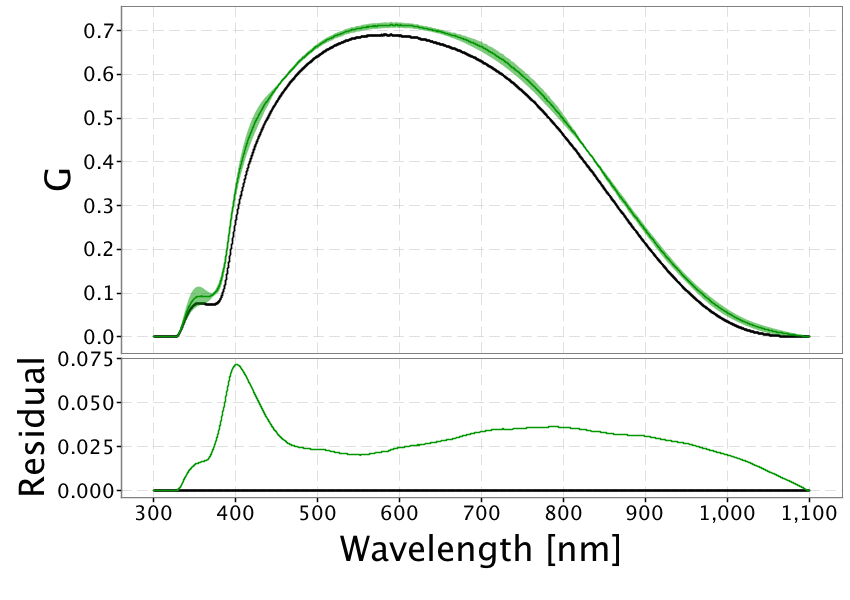} 
\includegraphics[width=\hsize*3/10]{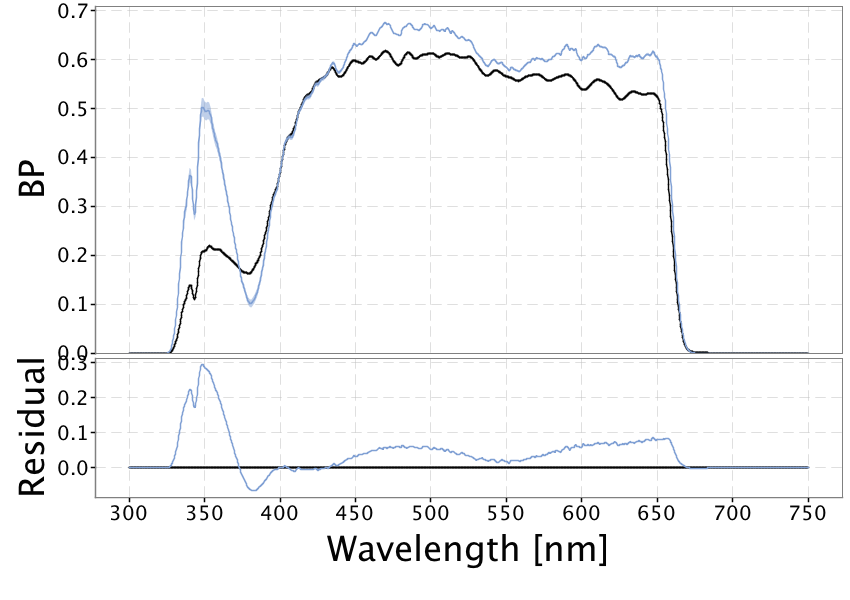} 
\includegraphics[width=\hsize*3/10]{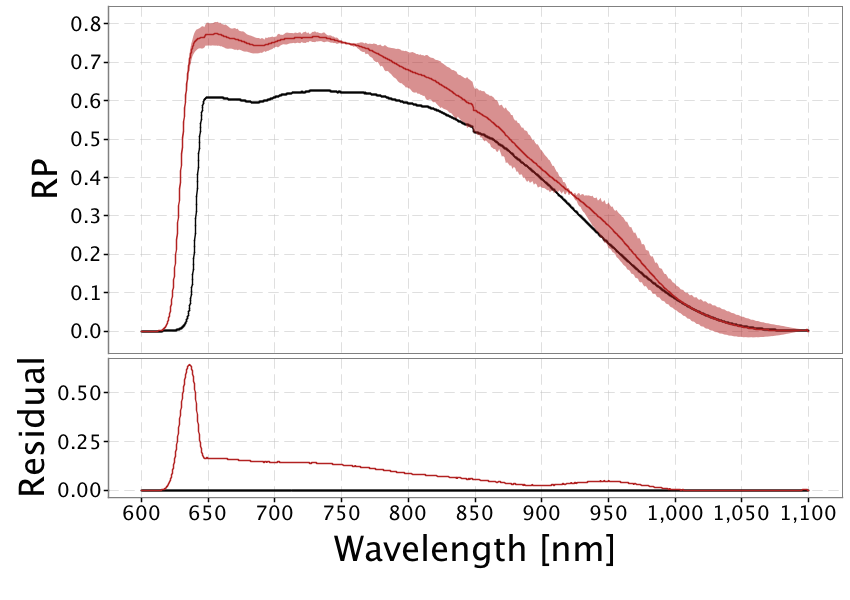} 
\caption{Revised set of calibrated \Gaia\ passbands for \G\ (\emph{left}), \BP\ (\emph{centre}), and \RP\ (\emph{right}). These passbands
represent the \GDR2 photometric system more accurately.  
Black lines represent nominal response curves, while the coloured lines show the calibrated curves. Shaded areas represent the $\pm 1\sigma$ level. 
} 
\label{Fig:ExtPassbandsOctober}
\end{figure*}

The external calibration effectively provides a means to relate the photometric system defined by the internal calibration to other photometric systems and allows for physical interpretation of data. This is achieved by providing a response function and an absolute zeropoint for each passband. The response function model depends on a set of adjustable parameters that modifies its shape and is designed to reproduce the nominal passband curves given in \citetads{2010A&A...523A..48J} 
when all parameters are set to zero.
The optimal set of parameters is obtained by minimising the residuals between synthetic flux predictions and the corresponding internally calibrated value for a set of spectro-photometric standard stars (SPSS) described in 
\citetads{2012MNRAS.426.1767P} 
and  
\citetads{2015AN....336..515A}. 
A complete description of the calibration model and algorithms used for \GDR2 release can be found in the online documentation. 
However, any passband derived in this way has an arbitrary component because the SPSS cannot cover the full space of parameters. For this reason, we have built  the model based on the pre-launch knowledge about the instrument overall response in the form of nominal passband curves. However, this procedure does not fix the passband shapes univocally because of uncertainties in the various elements that make up the response curve, and above all, because the current photometric system is effectively set by the internal calibration through the selection of internal calibrators and by the algorithms adopted to remove spatial and time-varying instrumental  effects. 
The  \BP\ and \RP\ cases offer a concrete example of these complexities because the steep cut-offs cannot be properly constrained  by the fitting process alone:
the assumption made for the \GDR2 processing was that the actual cut-offs could only deviate from the nominal cut-offs by small amounts ($\leqslant3$ nm). The calibrated photonic passbands obtained in this way are shown in Fig. \ref{Fig:ExtPassbandsJuly}. 
However, a subsequent analysis using BP and RP spectral data of the same SPSS revealed that the nominal curves for the photometer transmissivity used in the modelling process were not the most up-to-date curves provided by the satellite manufacturer, and at least in the RP case, a significant deviation with respect to the correct curve was present in the cut-off position. Thus, the redefinition of the cut-off properties along with some other minor fixes in the procedures led to a revised set of passbands that is shown in Fig.  \ref{Fig:ExtPassbandsOctober}.
To distinguish between the two sets, we will hereafter use DR2  for the first set and REV for the revised set.
An important consequence of this revision is that the photometric zeropoints slightly changed from those derived for the \GDR2 processing. Since a reprocessing of the whole catalogue was not feasible in the \GDR2 schedule, it has been decided to proceed with the former zeropoints and passbands to ensure internal consistency between different DPAC subsystems, and to publish the revised passbands and zeropoints as a service to those users who wish to make precise prediction of \Gaia\ photometry for isochrones etc.
The comparison of the photometric properties of the two sets (by computing synthetic magnitudes on extended library spectral
energy distributions, SEDs, with a large coverage in astrophysical parameters) shows that while the predicted magnitudes are basically equivalent in the colour range 
 \BP$-$\RP\  $\simeq [ -0.5, 2.5]$, 
a linear trend arises at redder colours that produces differences of up to 0.1 mag and  0.08 mag at  \BP\ $-$\RP\ $\simeq 6$ in \G\ and \RP, respectively.
\begin{table}[htp]
\caption{Photometric zeropoints in the VEGAMAG  system.}
\label{Tab:ext_zp}
\begin{center}
\begin{tabular}{|l|c|c|c|c|c|}
\hline Band & ZP$_{DR2}$ & $\sigma_{DR2}$ & ZP$_{REV}$ & $\sigma_{REV}$ & $\Delta_{ZP}$ \\
\hline
 \G\      &   25.6884   &   0.0018    &   25.6914   &   0.0011   &   0.0030 \\
 \BP\   &   25.3514   &   0.0014    &   25.3488   &   0.0005   &   -0.0026 \\
 \RP\   &   24.7619   &   0.0019    &   24.7627   &   0.0035   &   0.0008 \\
\hline
\end{tabular} 
\end{center}
\end{table}
Table \ref{Tab:ext_zp} provides a summary of the zeropoints for the two sets of passbands: ZP$_{DR2}$ are the zeropoints used for the \GDR2 processing, that is, the zeropoints used to convert mean integrated fluxes into the final \G, \BP, and \RP\ magnitudes. These zeropoints must be used together with the DR2 passband set to compare for example synthetic magnitudes computed on SEDs with the corresponding published magnitudes.

\begin{figure*}
\centering
\includegraphics[width=\hsize*3/10]{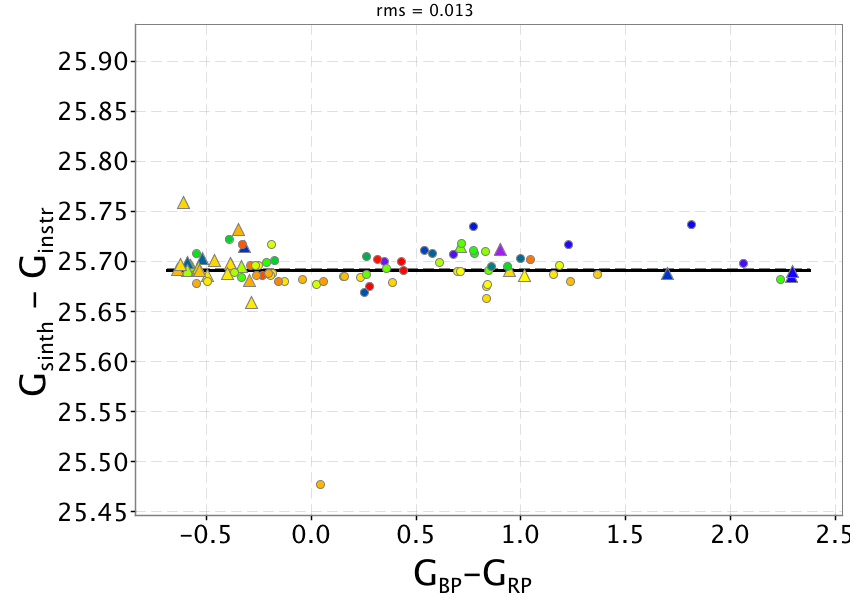} 
\includegraphics[width=\hsize*3/10]{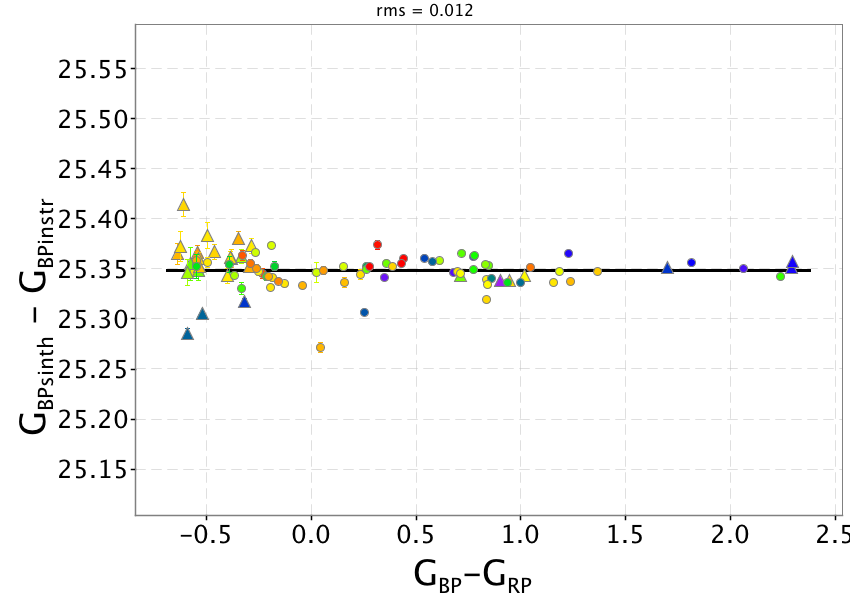} 
\includegraphics[width=\hsize*3/10]{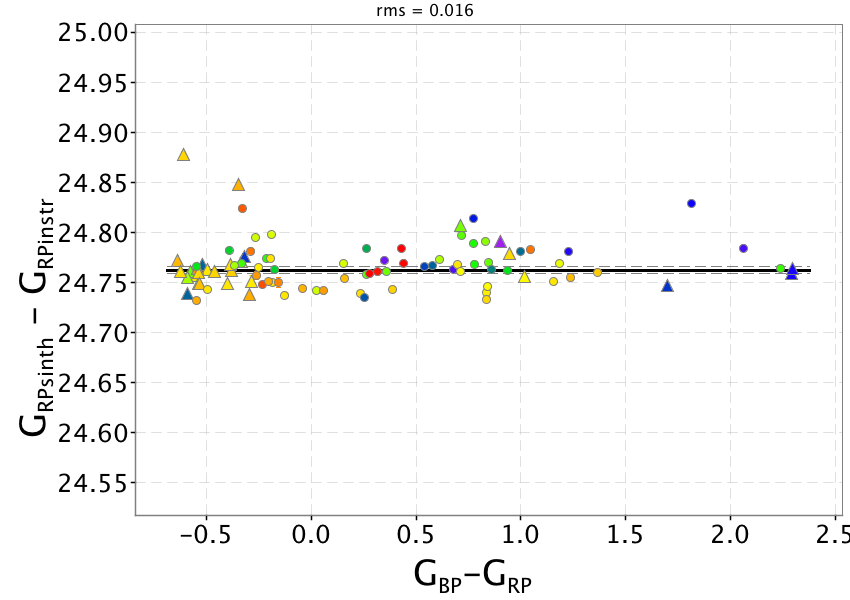} 
\caption{Comparison between synthetic and instrumental magnitudes as function of source {\BP\ $-$\RP\ }  colour for  \G\  (\emph{left}), \BP\ (\emph{centre}), and \RP\ (\emph{right}). Synthetic magnitudes have been computed using the REV passband set. The data point colours encode the \G\ magnitude of the sources (red dots show fainter sources); 
triangles represent the silver sources. The black horizontal line represents the corresponding magnitude zeropoint.
} 
\label{Fig:ExtDeltamagsOctober}
\end{figure*}
\begin{figure*}
\centering
\includegraphics[width=\hsize*3/10]{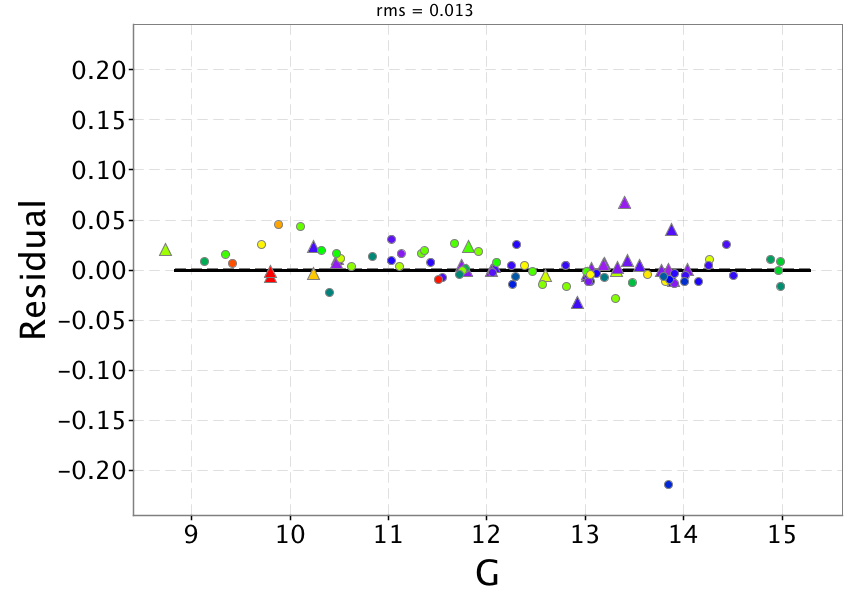} 
\includegraphics[width=\hsize*3/10]{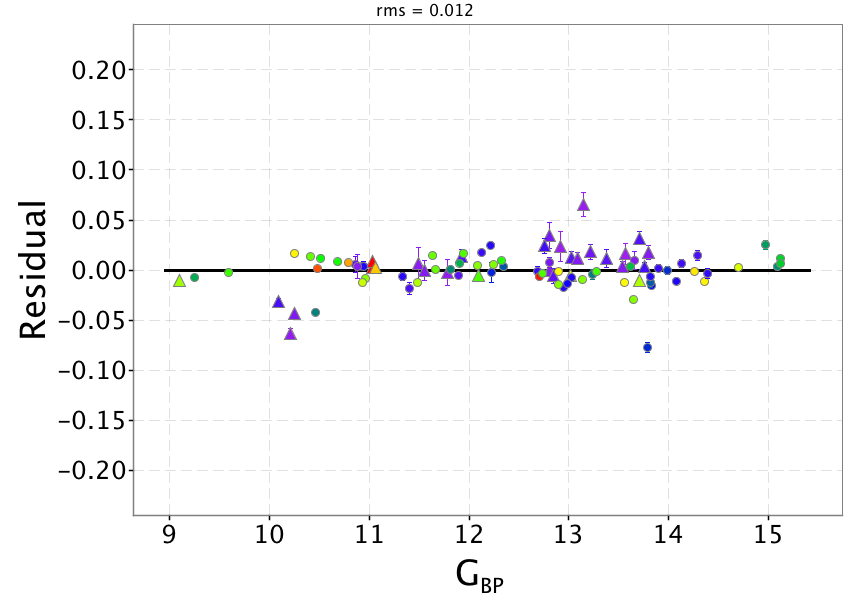} 
\includegraphics[width=\hsize*3/10]{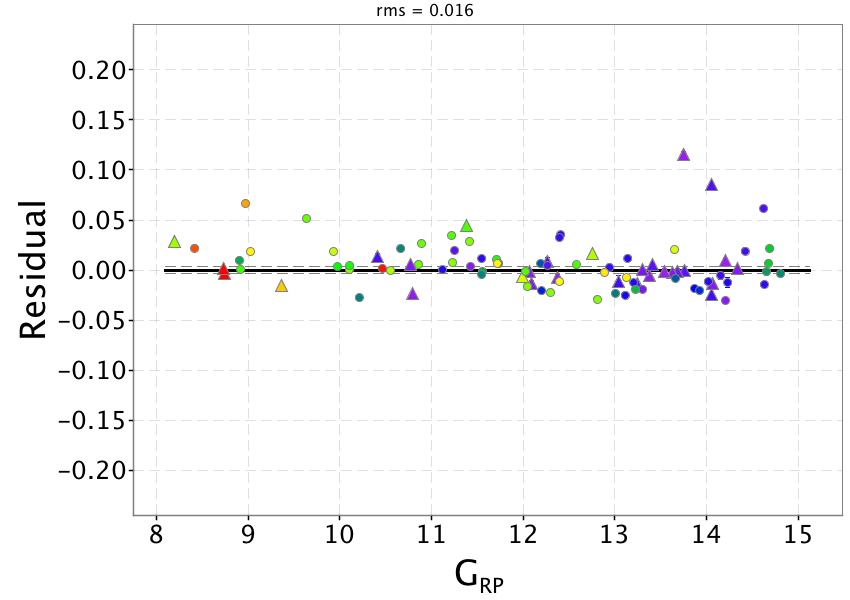} 
\caption{
Residuals with respect to the zeropoints as a function of source magnitude for  \G\  (\emph{left}), \BP\ (\emph{centre}), and \RP\ (\emph{right}); the data point colours encode the 
{\BP$-$\RP}  colour of the sources (red dots show the reddest sources).
} 
\label{Fig:ExtResidualsOctober}
\end{figure*}

Once the photometric system passbands are calibrated, the direct comparison between synthetic and instrumental magnitudes  shown in Fig. \ref{Fig:ExtDeltamagsOctober} exhibits a flat distribution. In this figure the comparison is made by assuming the REV passband set: the corresponding comparison made on the DR2 set is substantially equivalent and is not shown here, but can be found in the online documentation. Notably, the SPSS photometry alone is not sufficient to distinguish  between the two sets of passbands.
The system zeropoints, shown in the plots as horizontal black lines, are not computed as the mean values of the displayed data, but instead as
\begin{equation}
\label{equ:cu5phot_ext_5}
ZP = +2.5~\log~  \left( P_A ~ \int \left( \frac{f_{\lambda}^{\rm Vega}(\lambda)\ \lambda}{10^9~hc} \right) \  S(\lambda)\   d\lambda \   \right) 
,\end{equation}
where $P_A = 0.7278~ \textrm{m}^2$ is the \Gaia\ telescope pupil area, wavelengths $\lambda$ are expressed in nm, $S(\lambda)$ is the calibrated passband, and $f_{\lambda}^{\rm Vega}(\lambda)$ is the energy flux distribution per wavelength units of the reference Vega spectrum expressed in units of $\textrm{W} ~\textrm{nm}^{-1} \textrm{m}^{-2}$. For the \GDR2 release, we assumed as reference distribution the unreddened A0V star Kurucz/ATLAS9 Vega spectrum (CDROM 19) with $T_{eff} = 9550$~ K, log g $= 3.95$~ dex, $[Fe/H] = -0.5$~ dex, and $v_{micro} = 2~ \textrm{km}~ \textrm{s}^{-1}$;  this model has been normalized by imposing the condition that the energy flux at 
 $\lambda = 550~\textrm{nm}$ is $f_{550} = 3.660~10^{-11}~\textrm{W} \textrm{m}^{-2} \textrm{nm}^{-1}$ according to 
\citetads{1992S&T....84..171S}. 

The above formula implies that  the calibrated passbands are normalised so that a source-integrated flux can be estimated by convolving them with the corresponding SED expressed in units of $\textrm{photons}~\textrm{s}^{-1}\textrm{nm}^{-1}\textrm{m}^{-2}$ and scaling the result by the \Gaia\ telescope pupil area. 

The residual distributions of the differences with respect to the zeropoints are shown in Fig. \ref{Fig:ExtResidualsOctober} as function of the SPSS \G, \BP, and \RP\ magnitudes and reflect the self-consistency between the observed and predicted synthetic fluxes. The measured rms of the residuals are 0.013, 0.012, and 0.016 mags for \G, \BP, and \RP, respectively: these values are in agreement with the \Gaia\ end-of-mission requirement for the SPSS flux precision, which has been set to $\simeq 1\%$, and are notably higher than the formal zeropoint errors of Tab.~\ref{Tab:ext_zp}  because they also include the contribution from SEDs noise.
\begin{figure*}
\centering
\includegraphics[width=\hsize*3/10]{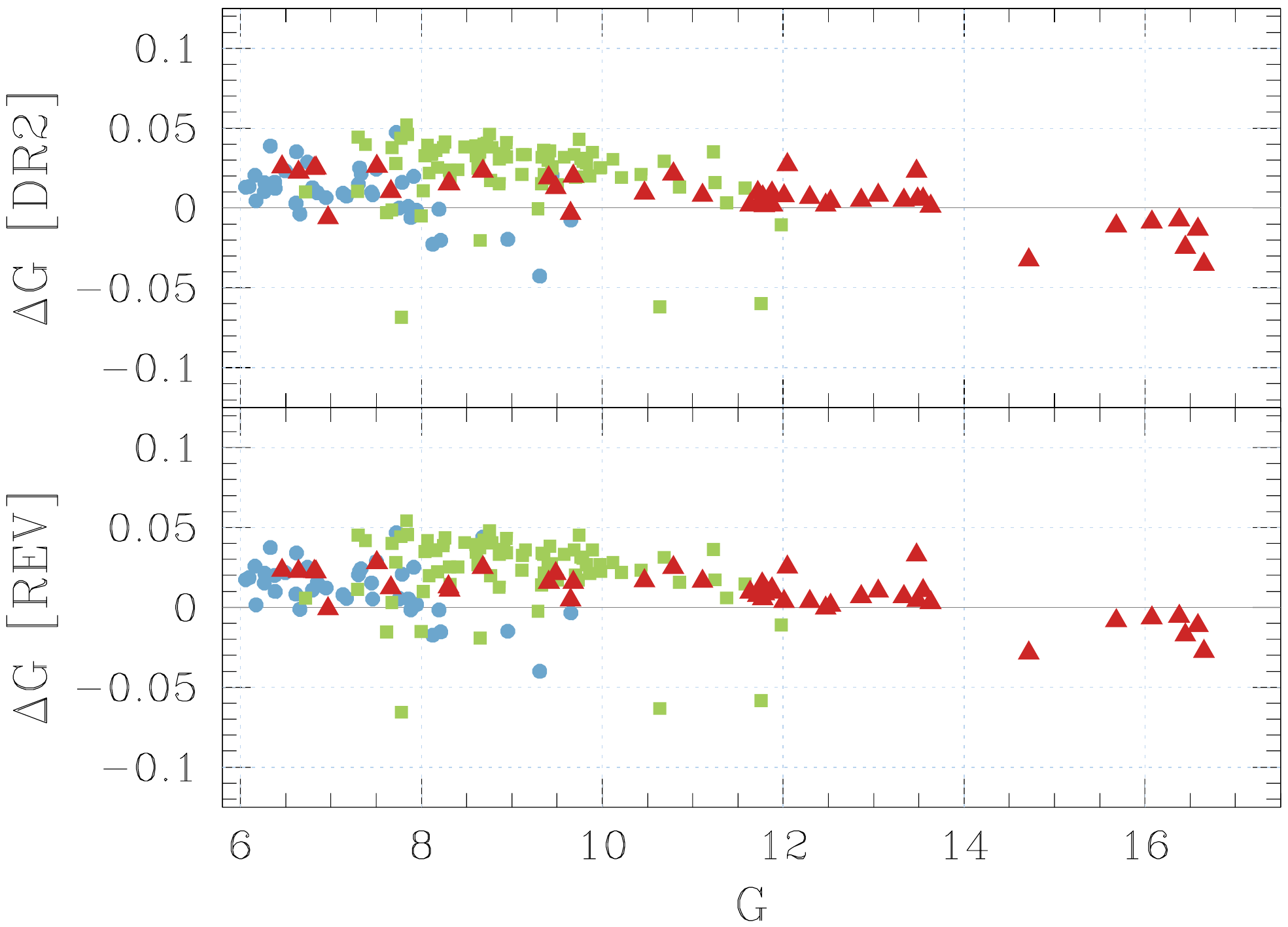} 
\includegraphics[width=\hsize*3/10]{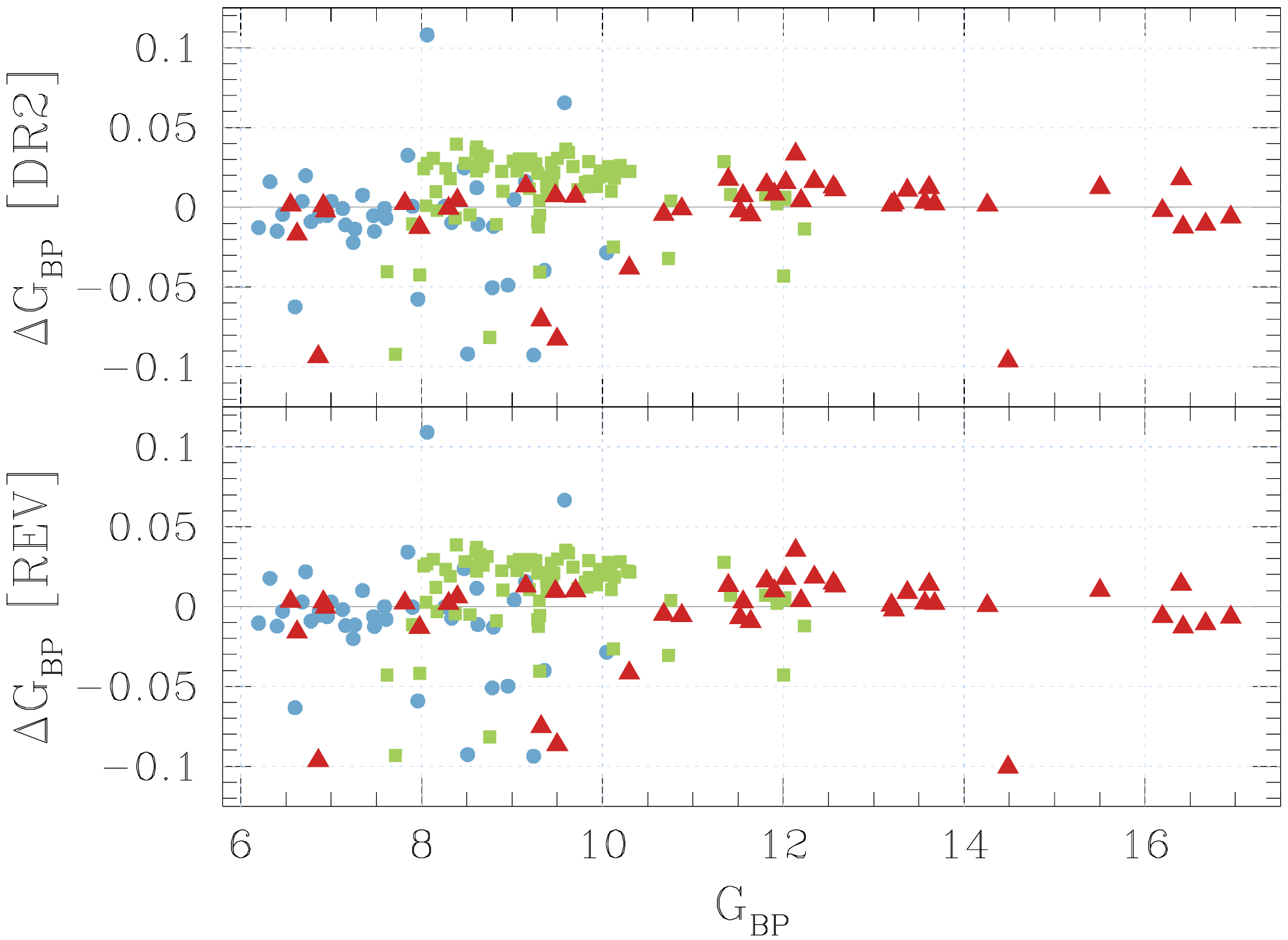} 
\includegraphics[width=\hsize*3/10]{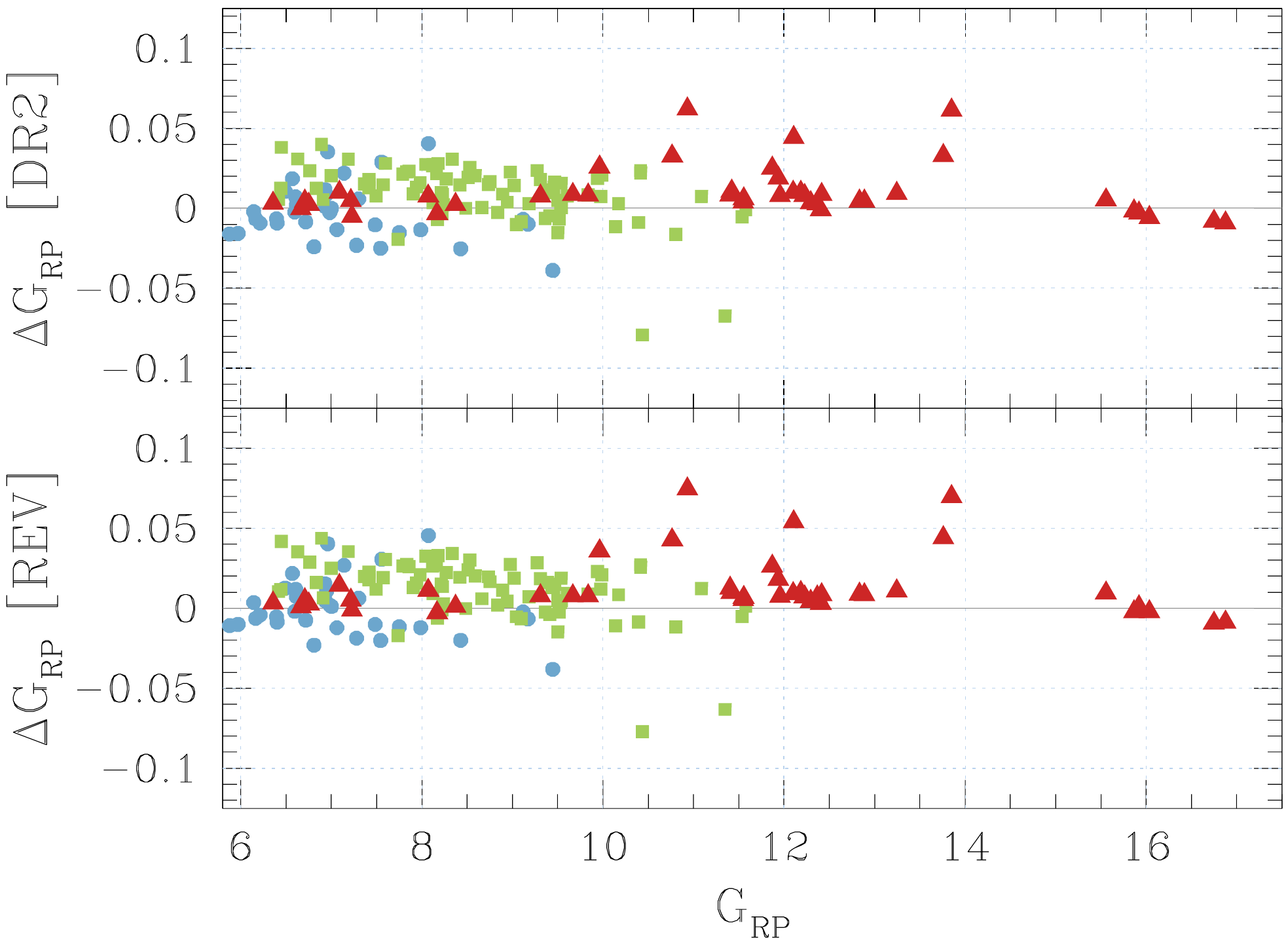} 
\caption{
Residuals between synthetic photometry computed on three spectral libraries and the corresponding \GDR2 magnitudes as a function of the  \G\  (\emph{left}), \BP\ (\emph{centre}), and \RP\ (\emph{right}) magnitude. Red triangles represent 43 sources from the CALSPEC database, green squares represent 81 stars taken from the Stritzinger spectral library, and blue dots show 40 objects from the NGSL library. \emph{Top:}  Synthetic magnitudes computed by convolving the source SEDs with the DR2 passbands and normalising by the photometric zeropoint. \emph{Bottom}: Synthetic photometry obtained with the REV passband set.} 
\label{Fig:ExtValidationMags}
\end{figure*}
\begin{figure*}
\centering
\includegraphics[width=\hsize*3/10]{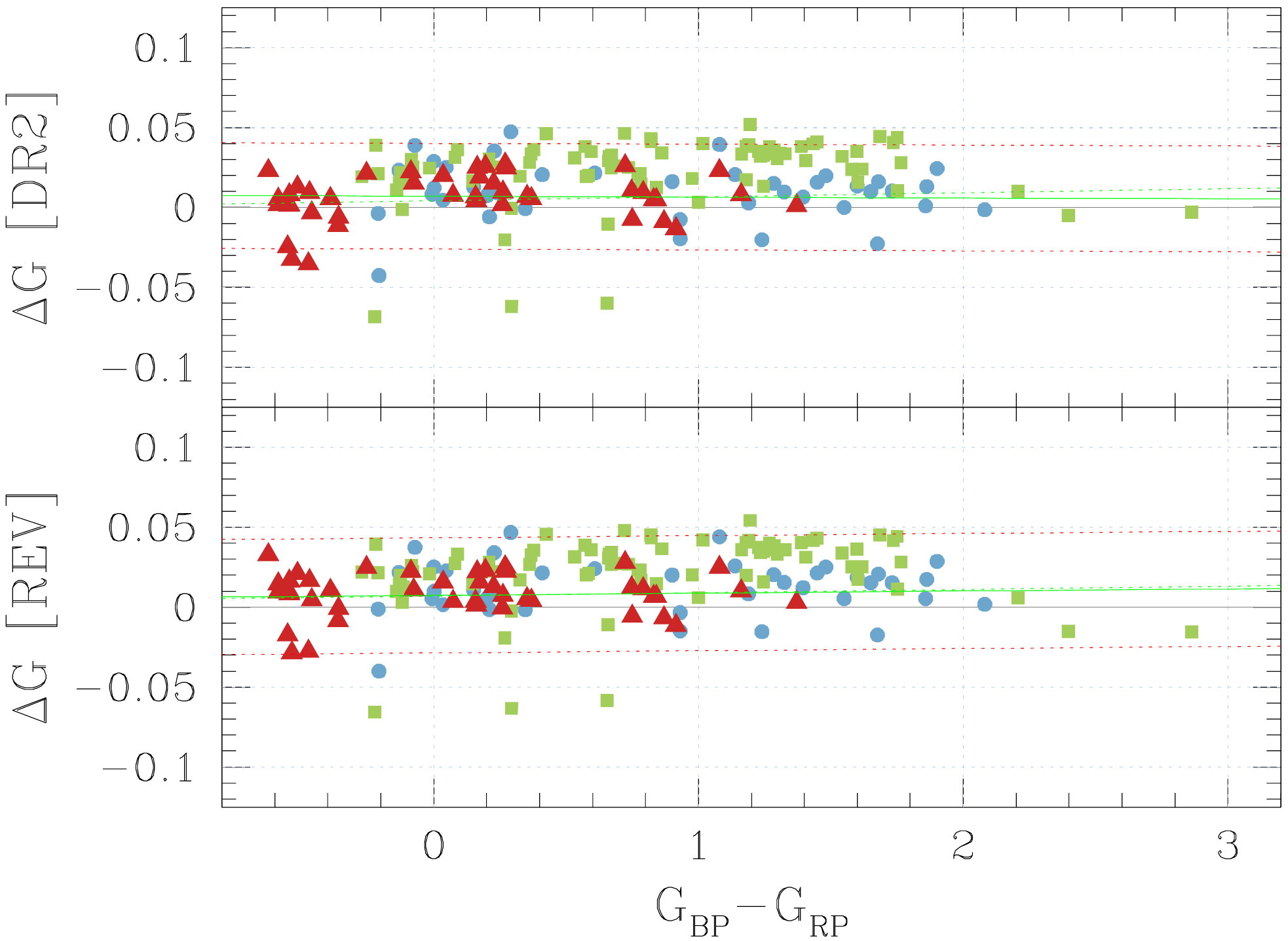} 
\includegraphics[width=\hsize*3/10]{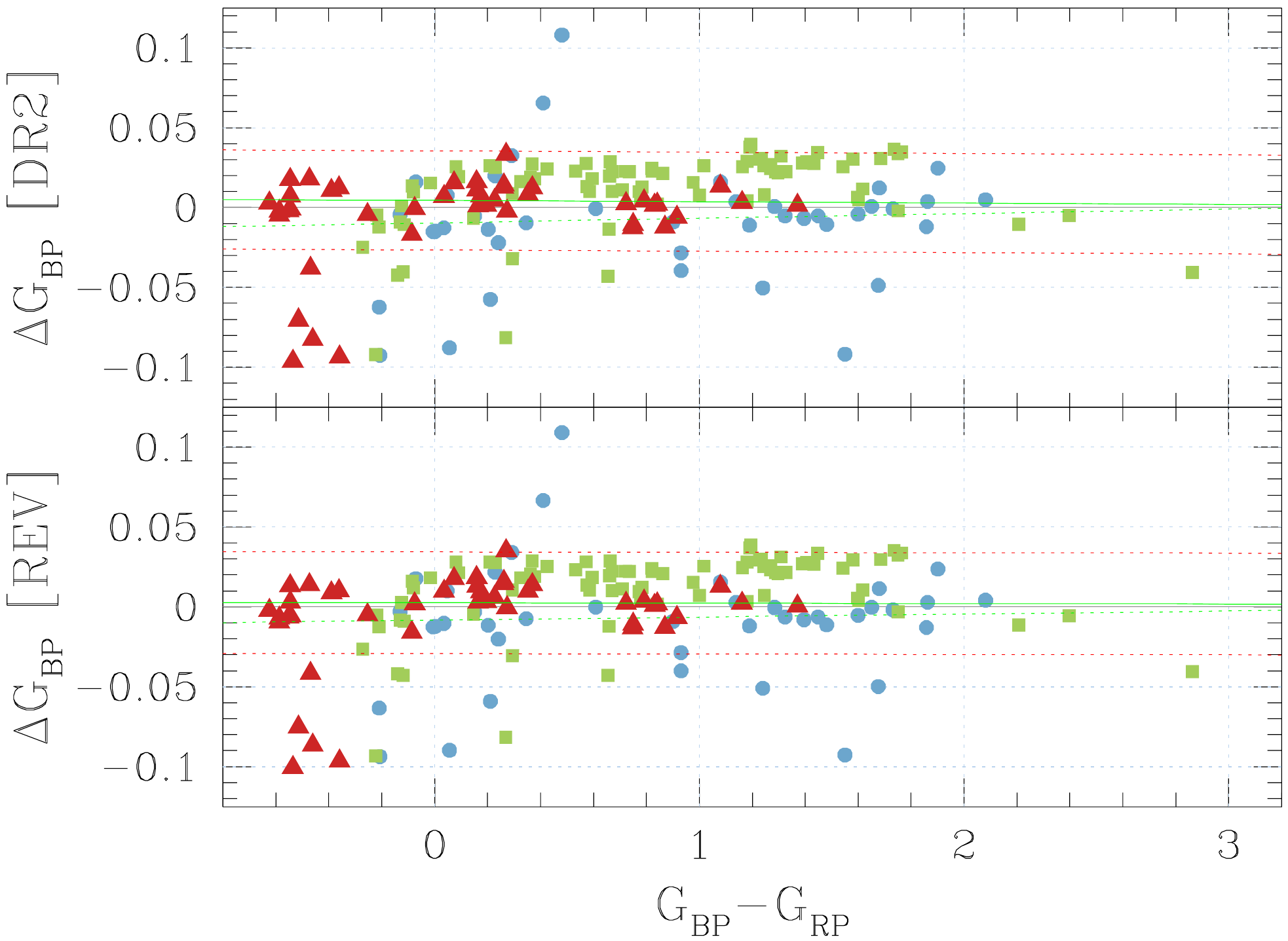} 
\includegraphics[width=\hsize*3/10]{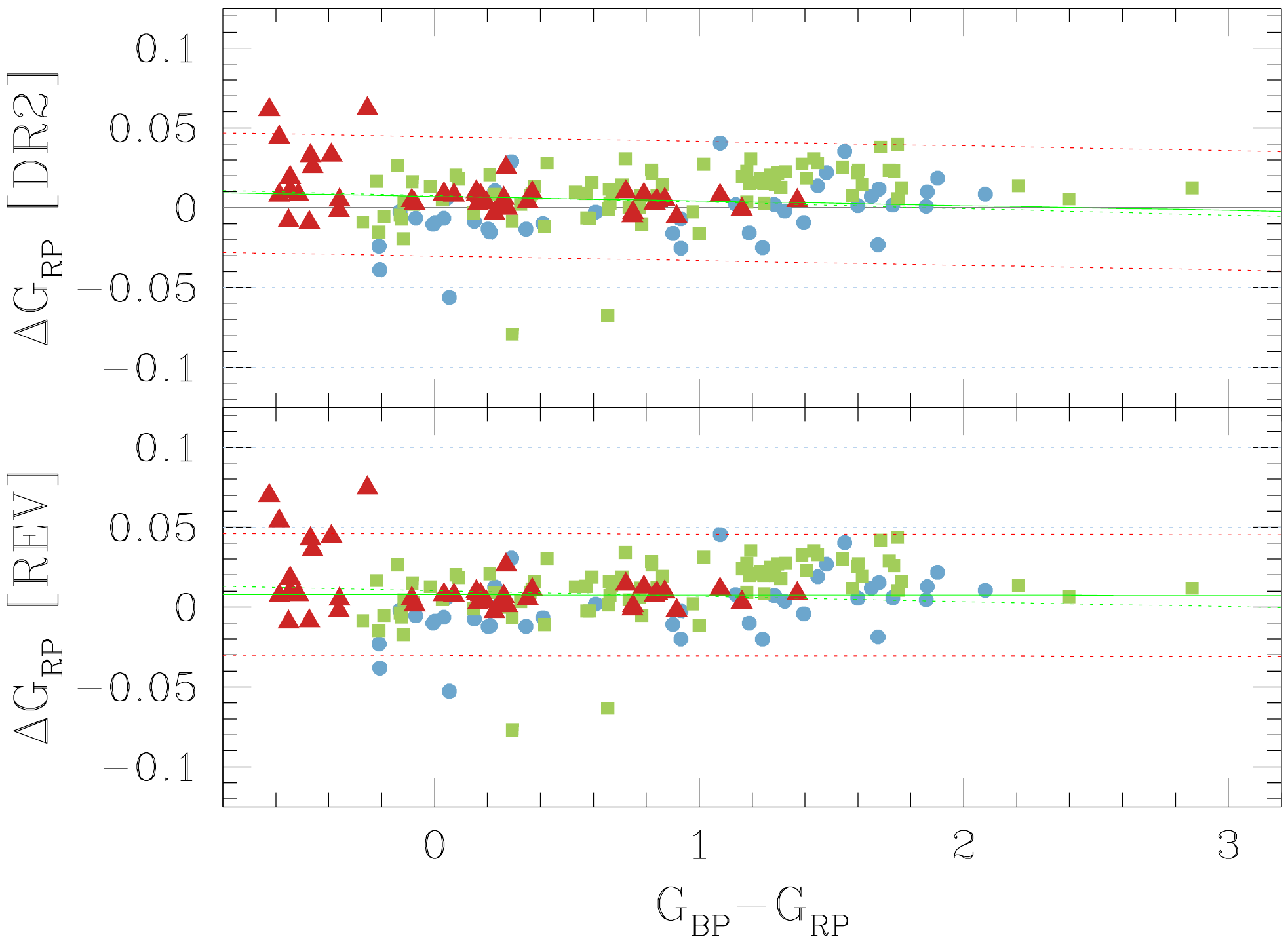} 
\caption{
Residuals between synthetic photometry computed on the NGSL, CALSPEC, and Stritzinger libraries and the corresponding \GDR2 magnitudes as a function of the observed  \BP$-$\RP\   colour. Symbols are the same as Fig. \ref{Fig:ExtValidationMags}. Horizontal green and red lines represent a linear fit (with sigma-rejection clipping) to data with \G $>10$  and the $\pm 1\sigma$ region.
\emph{Top:}  Synthetic magnitudes computed by convolving the source SEDs with the DR2 passbands and normalising by the photometric zeropoint. \emph{Bottom}: Synthetic photometry obtained with the REV passband set.} 
\label{Fig:ExtValidationCol}
\end{figure*}
As a further validation for calibrated passbands, we have used data from three different spectral libraries, the STIS Next Generation Spectral Library (NGSL) 
\citepads{2007ASPC..374..409H},  
the CALSPEC spectral database 
\citepads{2017AJ....153..234B}, 
and the Stritzinger spectral library
\citepads{2005PASP..117..810S} 
to compare synthetic magnitudes computed with the two set of passbands with the corresponding DR2 magnitudes. 
The cross-match between the libraries and \GDR2 release produced an initial list of 118, 50, and 93 objects for the NGSL,  CALSPEC, and Stritzinger library, respectively: selecting sources with $G > 6$ and and |\G$_{synth}-$\G$_{DR2}| < 0.125$ to avoid excessive saturation and possible poor matches left the number of objects equal  to 40, 43, and 81, respectively.
Figure \ref{Fig:ExtValidationMags} shows the residuals for the three bands as a function of the \G, \BP, and \RP\ magnitudes. The top plots refer to the DR2 passbands, and the bottom plots show the residuals obtained with the REV passbands. While the NGSL- and CALSPEC-predicted photometry shows a satisfactory agreement with \GDR2 data (with the only exception of CALSPEC \G\ data, which exhibit a small systematic trend with magnitudes), the Stritzinger data show a residual of up to  $0.03$ mag for sources brighter than $G \simeq 10$. Figure \ref{Fig:ExtValidationCol} shows the same data as the previous plots as a function of the observed  {\BP$-$\RP}  colour. The horizontal green lines represent a linear fit made on data fainter than  $G=10$: the slope is almost zero for all the data sets, and the measured offsets are  0.007, 0.004, and 0.007 and 0.007, 0.003, and 0.008 mag for 
\G, \BP, and \RP\ and the DR2 and REV sets, respectively. The fit was made by applying a sigma-rejection clipping to exclude deviating points, and the red dotted lines represent the final $\pm 1\sigma$ region. 
In conclusion, 
the behaviour shown by the two passband sets is almost indistinguishable, as expected in this colour range, because much redder sources would be needed to reveal possible differences. 
However, the REV passbands are more accurate because they are based on additional constraints from spectral information  and are therefore preferable to the DR2 ones, which were used as a preliminary solution for internal use in the downstream processing.
 
The VEGAMAG  system has been chosen by DPAC as the standard reference photometric system for \Gaia. However, to facilitate the conversion of \Gaia\ photometry between the two systems, the zeropoints for the AB system  \citepads{1983ApJ...266..713O} have also been calculated for both sets of passbands and are available in Tab.~\ref{Tab:ext_zp_ab}. The details of the methods followed for these calculations can be found in the \GDR2 online documentation.

\begin{table}[htp]
\caption{Photometric zeropoints in the AB  system.}
\label{Tab:ext_zp_ab}
\begin{center}
\begin{tabular}{|l|c|c|c|c|}
\hline Band & ZP$_{DR2}$ & $\sigma_{DR2}$ & ZP$_{REV}$ & $\sigma_{REV}$ \\
\hline
 \G\      &   25.7934   &   0.0018    &   25.7916   &   0.0018  \\
 \BP\   &   25.3806   &   0.0014    &   25.3862   &   0.0038   \\
 \RP\   &   25.1161   &   0.0019    &   25.1162   &   0.0020   \\
\hline
\end{tabular} 
\end{center}
\end{table}

\section{Statistical properties of the gold, silver, and bronze samples}\label{Sect:GoldSilverBronze}

\begin{figure} \resizebox{\hsize}{!}{\includegraphics{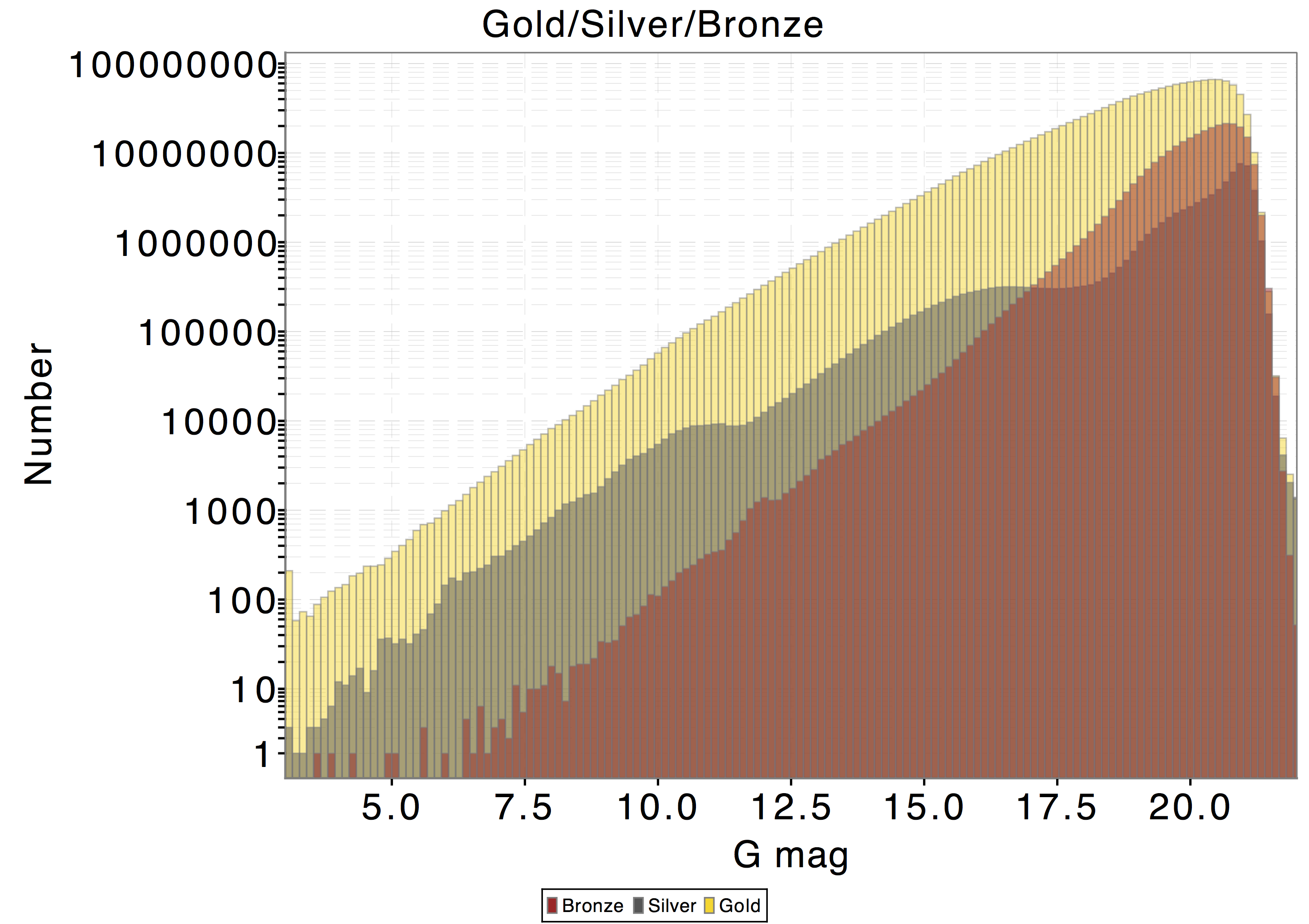}}
\resizebox{\hsize}{!}{\includegraphics{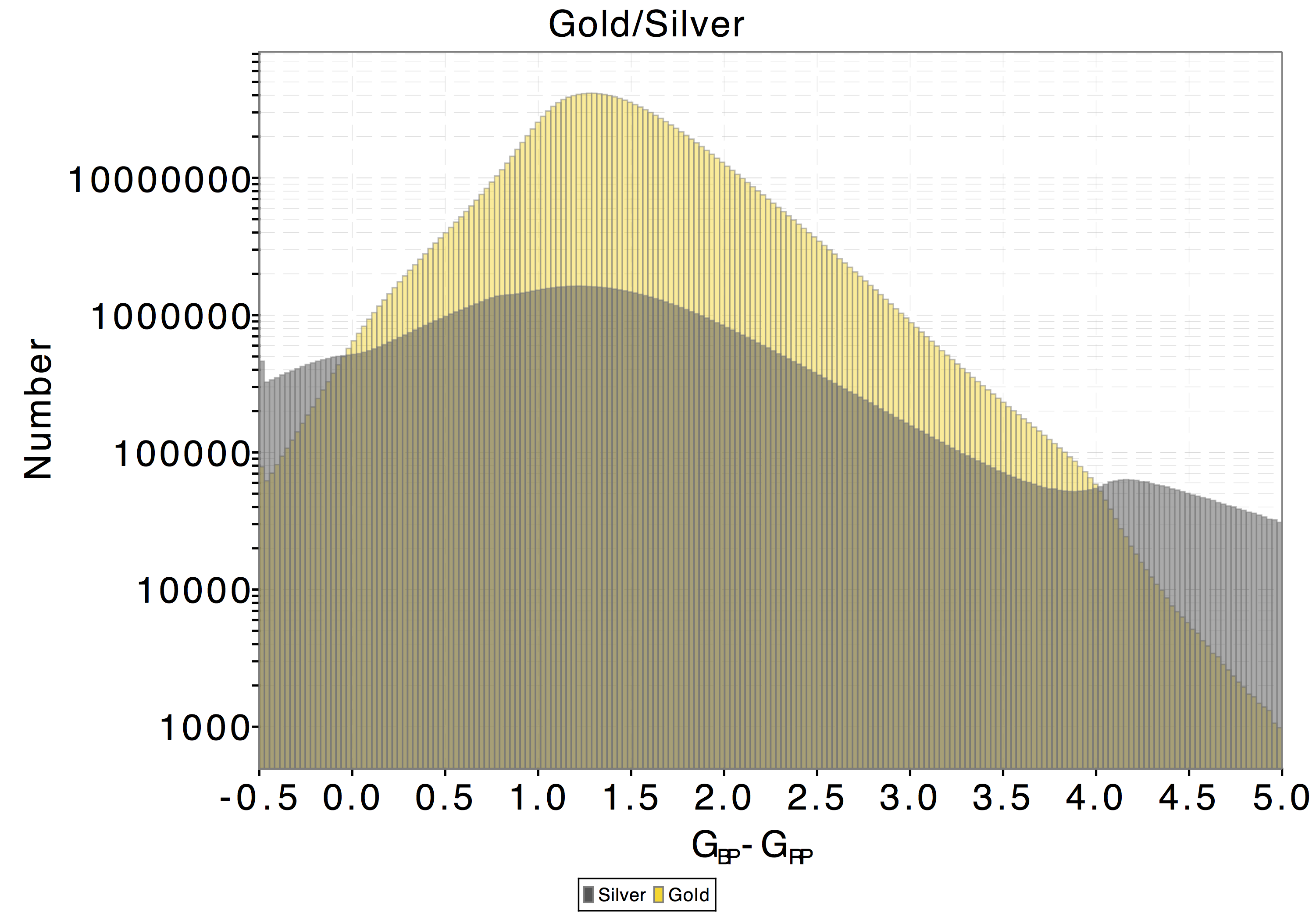}} 
\caption{Distribution of gold, silver, and bronze sources as a function of \G\ magnitude and \BP-\RP\ colour. 
Only sources with at least 30 CCD observations in \G\ are analysed in order to
minimize the effect of spurious detections.}
\label{Fig:GSBMagCol}
\end{figure}

\begin{figure} \resizebox{\hsize}{!}{\includegraphics{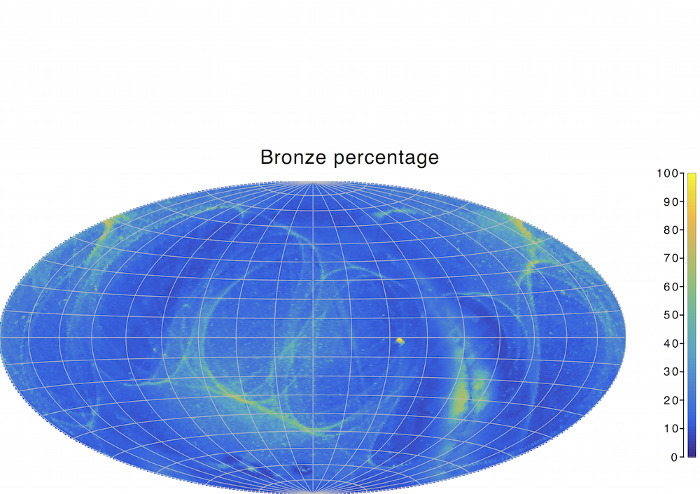}}
\caption{Percentage of bronze sources as a function of sky position in equatorial coordinates.  Only sources with a minimum of 30 \G\ CCD observations are considered.}
\label{Fig:BronzeSky}
\end{figure}

As described in \citeads{PhotProcessing},
the sources are described as having gold, silver, or bronze photometry depending on the processing chain used to generate the photometry. Figure \ref{Fig:GSBMagCol}
shows the magnitude and colour distributions of these samples. The magnitude distribution of the silver sources reflects problems in the processing that have caused
sources at various stages to drop out. They were not calibrated in the normal gold processing chain.
Although the colour distribution of the silver sources shows elevated levels for sources outside the range 0.0 to 4.0, the distribution is dominated by intermediate colours.
Initially, the silver processing chain was developed to calibrate sources with extreme colours that could not be calibrated normally. However, the silver processing chain
also picks up the low fraction of sources that fail to be calibrated at other stages in the calibration process.
The magnitude distribution of the bronze sources is biased towards the faint end and reflects the difficulty of obtaining colour information at fainter colours. Naturally, there is
no colour distribution available for the bronze sources since these, by definition, are sources without \Gaia\ colour information and have been calibrated with a default set
of colour information.

Figure \ref{Fig:BronzeSky} shows the fraction of bronze sources as a function of sky position. In the high-density regions around the Galactic centre, the fraction
of bronze sources becomes very high because the BP/RP windows are likely to overlap with neighbouring sources and are therefore not processed. The patterns caused by the \Gaia\ scanning law
are also visible here. It is unclear why these areas would have a higher bronze fraction, but a possible cause is spurious detections. The bright point at
(RA, dec) = (315$^\circ$, 0$^\circ$), is a result of a processing problem that prevented the colour information from being available in this area and switched the sources to the bronze processing chain.
These sources do not have \BP\ or \RP\ photometry in \GDR2.

\section{Epoch photometry}\label{Sect:EpochPhotometry}

Although this paper focusses on the source photometric catalogue, which is
the main contribution to the release from the photometric point of view, \GDR2
will also contain detailed epoch photometry for a small fraction of sources in
the same \G, \BP, and \RP\ bands. The epoch photometry will be averaged over a FoV
transit for the \G\ band, and the corresponding error will measure the scatter of the 
contributing CCD measurements. For \BP\ and \RP, each epoch flux will come from
one single CCD measurement and its associated error.

Problems affecting single CCD transits or all measurements of a FoV transit
are marginally important when considering mean source photometry as they
will simply slightly increase the scatter of the measurements and therefore
the error associated with the mean value.
However, the same problems cause outliers in the epoch photometry.
For instance, observations affected by hot columns do not currently
receive any special treatment. Cases of poor background estimation have also
been observed in all bands, either due to lack of calibration
data or to a rapidly varying background that could not be captured by the model
resolution. Observations taken in the proximity of very bright
sources may either have underestimated background or may be
spurious detections caused by the diffraction spikes from the bright source.
Nearby sources may also cause outliers through the different scanning
direction in subsequent scans over a region of sky, for instance, two nearby sources falling
in the same window in one scan, but not in others. Furthermore,
sky-related effects may well affect perfectly isolated sources because the two FoVs overlap on the focal plane.

In \GDR2, only a small subset of problematic epoch measurements will be flagged as
rejected in the photometric processing. These are non-nominal cases (such as truncated
windows and observations taken with more than one gate activated during the integration
time), poor results from the image parameter determination, measurements that could
not be calibrated because no appropriate calibration was available. Additional filtering
will be done by the system processing the variability data \citepads{VariabilityAnalysis},
and the corresponding epochs will be correspondingly flagged in the catalogue, but
it is likely that many of the above events will not be reported. Any subsequent analysis
using the epoch photometry in \GDR2 should be robust against outliers.

\section{Conclusions}\label{Sect:Conclusions}

This paper presents the photometric content of the \GDR2. This includes 
mean source photometry in the \G\ passband for 1.7 billion objects and in the integrated \BP\ and \RP\
passbands for about 1.4 billion objects.
An estimate of the three photometric passbands, based on a set of spectro-photometric standards with
high-quality ground-based reference data, will also be published together with the corresponding
zeropoints.
As detailed in Sect. \ref{Sect:Passbands}, the passband definition at this stage is not unique because of
limits both in the information available in the photometric data and in the set of standard sources
available. A better estimate of the three passbands is also presented in this paper. Although this is
also based on the same set of standards and is therefore not fully constrained, it has been determined
taking full advantage of the spectral data as well as the photometry and therefore gives a more accurate
representation of the \Gaia\ average instrument and the internal reference photometric system.

The \GDR2 will also contain epoch photometry in the same three bands for a small fraction of sources
that is classified as variables. More details on these data are given in \citetads{VariabilityAnalysis}.
At this stage in the mission, epoch photometry still shows outliers that are due to problems that affect measurements
at the CCD and FoV level.

The accompanying paper, \citetads{PhotProcessing}, provides a detailed description of the photometric
processing that led to the generation of the \GDR2 catalogue.
Results of the detailed validation campaign have been included in this paper. Validation takes place
at various levels during the photometric processing. Most of the validation is purely internal and is based
on a careful analysis of the calibration parameters, on convergence metrics for the iterative process leading to
the definition of the internal photometric system, on the analysis of the residuals of the calibration
process, and on statistical distribution of the photometric errors at various stages.
Comparisons with external catalogues are also useful, but may be affected by the precision and systematics in the external catalogues.

For the \GDR2 photometric data, our internal validation indicates that a calibration uncertainty of
$2$ mmag has been reached on the individual observations for the \G\ band. Slightly larger uncertainties
of $5$ and $3$ mmag are estimated for \BP\ and \RP, respectively. The expected error distribution
will depend on the number of individual measurements and on other contributions such as read-out noise and
photon noise.
Validation against several external catalogues shows that systematic effects are at the 10 mmag level
except for the brightest sources (\G $<3$ mag).

Section \ref{Sect:BpRpExcess} in this paper shows the result of an investigation on the consistency
between the \G\ and \BP\ and \RP\ photometry. Given the passband definition, the flux in the \G\ band is
expected to be close to the sum of the \BP\ and \RP\ fluxes. We have therefore defined a BP/RP flux excess
factor as the ratio between the sum of the \BP\ and \RP\ fluxes and the \G\ flux. Most sources in the
catalogue follow a distribution of BP/RP flux excess factor close to 1 with a smooth trend with colour.
This quantity is included in the released data, and users are encouraged to use this to select data for
their photometric studies.

A final caveat should be given related to the BP/RP flux excess factor and in general the mean source
photometry for variable sources. Even though the algorithm applied to generate the mean photometry
has been improved with respect to \GDR1, with the addition of robust outlier rejection based on the full
distribution of the individual measurements, this is still not optimal for variable sources.
The larger scatter of the measurements in the case of a variable source implies that outliers may still be
within a few sigma from the median value and therefore are not rejected. Faint outliers, with correspondingly
small uncertainties, may shift our weighted average estimate of the mean source photometry by several
magnitudes. Moreover, outliers may not be present in all bands. This will affect the reliability of the
BP/RP flux excess factor, the value of which is entirely based on mean photometry in the three bands.
Photometric studies of variable sources should take advantage of the provided epoch photometry when
possible.

It is intrinsic in the cyclic nature of the \Gaia\ processing that the quality of the data products
will improve at each release through the increased amount of data available, the better treatment
of the known instrumental effects, and the more sophisticated calibration models and algorithms.
Each new release will therefore allow new challenging research to be pursued, even more so thanks to the
new data products.
Future releases will see the gradual inclusion of epoch photometry and low-resolution spectral data for
all sources, thus creating the most complete and accurate photometric catalogue to date.

\begin{acknowledgements}
  
  This work presents results from the European Space Agency (ESA) space mission \Gaia. \Gaia\ data are
  being processed by the \Gaia\ Data Processing and Analysis Consortium (DPAC). Funding for the DPAC
  is provided by national institutions, in particular the institutions participating in the \Gaia\
  MultiLateral Agreement (MLA). The \Gaia\ mission website is \url{https://www.cosmos.esa.int/gaia}.
  The \Gaia\ Archive website is \url{http://gea.esac.esa.int/archive/}.
  
  This work has been supported by the United Kingdom Rutherford Appleton Laboratory, the United
  Kingdom Science and Technology Facilities Council (STFC) through grant ST/L006553/1, and the United
  Kingdom Space Agency (UKSA) through grant ST/N000641/1.

  This work was supported by the MINECO (Spanish Ministry of Economy) through grant ESP2016-80079-C2-1-R
  (MINECO/FEDER, UE) and ESP2014-55996-C2-1-R (MINECO/FEDER, UE) and MDM-2014-0369 of ICCUB
  (Unidad de Excelencia “Mar\'ia de Maeztu”).

  This work was supported by the Italian funding agencies Agenzia Spaziale Italiana (ASI) through grants 
  I/037/08/0, I/058/10/0, 2014-025-R.0, and 2014- 025-R.1.2015 to INAF and contracts I/008/10/0 and 2013/030/I.0 to ALTEC S.p.A
  and
  Istituto Nazionale di Astrofisica (INAF).

  This research has made use of the APASS database, located at the AAVSO web site. Funding for APASS has
  been provided by the Robert Martin Ayers Sciences Fund.\\

  We thank A. Vallenari for supplying us with spectra for the validation of the external flux calibration
  and passband determination.
\end{acknowledgements}

\bibliographystyle{aa} 
\bibliography{refs} 

\begin{appendix}

\section{Colour-colour transformations}\label{Sect:Transformations}

This section includes colour-colour relationships between the passbands in {\Gaia} and other photometric surveys. These relationships supersede those published before launch using the nominal {\Gaia} passbands \citepads{2010A&A...523A..48J}. 
In addition to $G$ magnitudes, {\GDR2} is the first to also include {\BP} and {\RP} magnitudes. This allows for the first time to provide photometric relationships in both directions: to predict {\Gaia} photometry from external photometry, and vice versa, to predict how {\Gaia} sources would be seen if they were observed with another set of passbands. This could allow the use of {\Gaia} sources as calibrators for other projects using these external passbands. Here we provide photometric transformations between {\GDR2} and 
\Hipparcos\ \citepads{1997A&A...323L..49P}, 
\TychoTwo\ \citepads{2000A&A...355L..27H}, 
SDSS12 \citepads{2015ApJS..219...12A},  
Johnson-Cousins \citepads{2009AJ....137.4186L}, and 
2MASS \citepads{2006AJ....131.1163S}. 

\begin{figure*}[t]
        \begin{center}
        \includegraphics[width=0.3\textwidth]{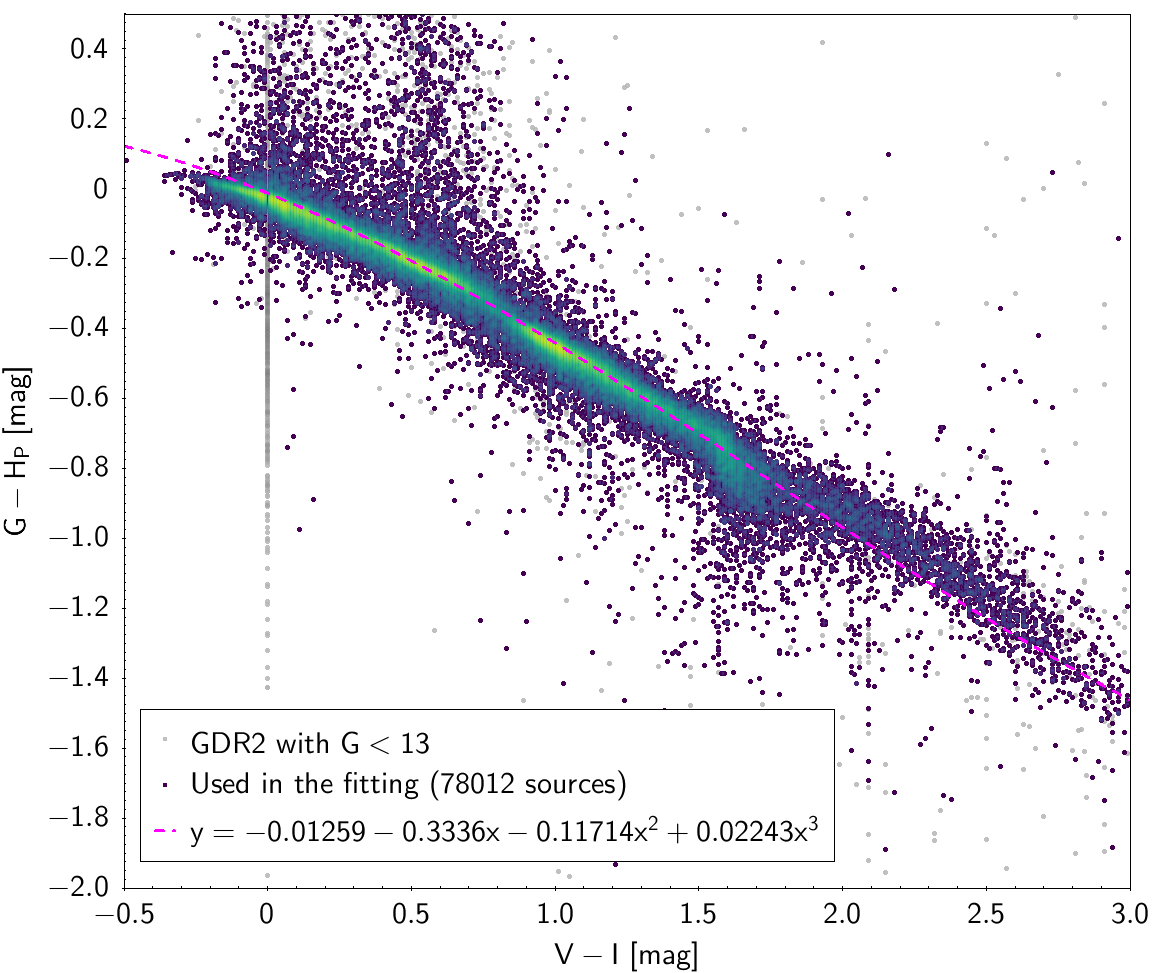}
        \includegraphics[width=0.3\textwidth]{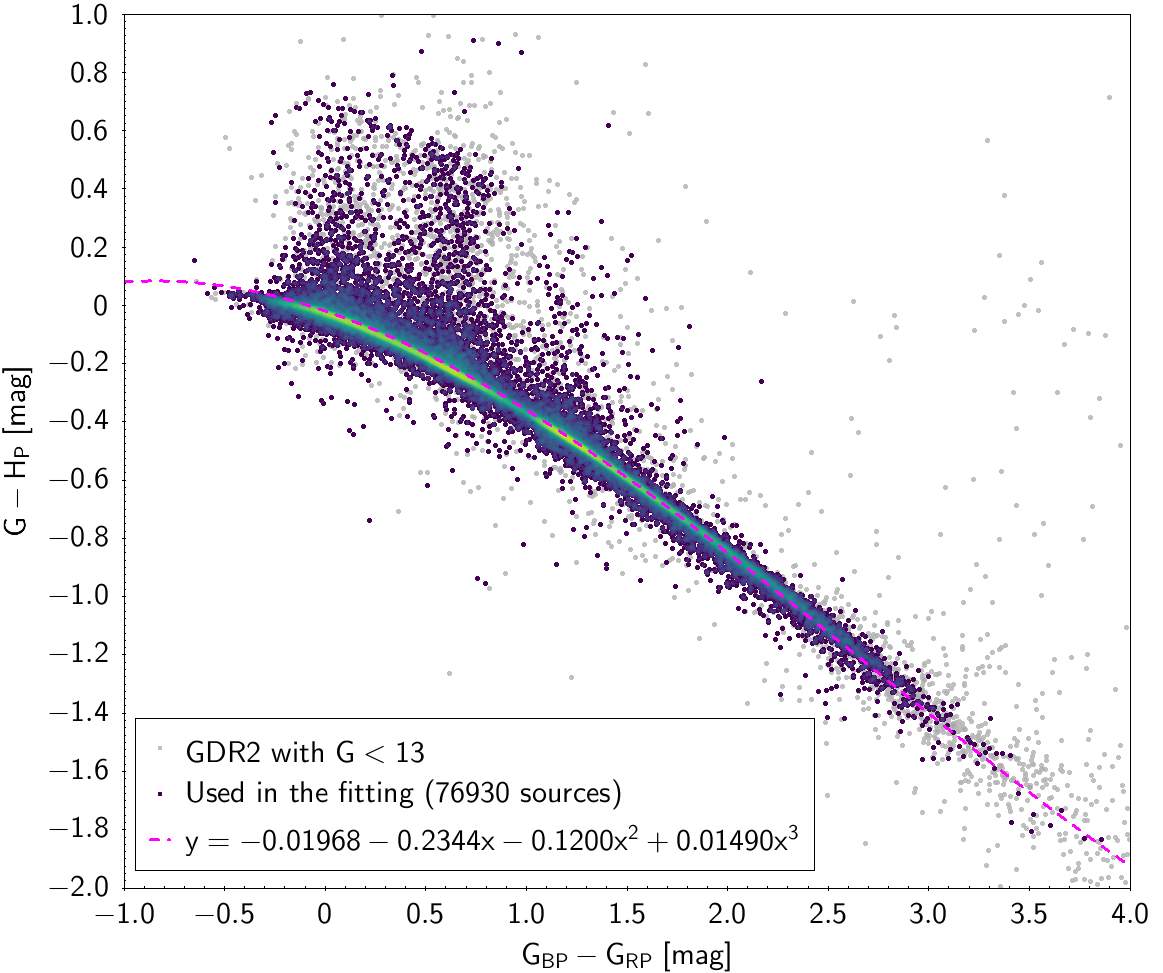}
        \includegraphics[width=0.3\textwidth]{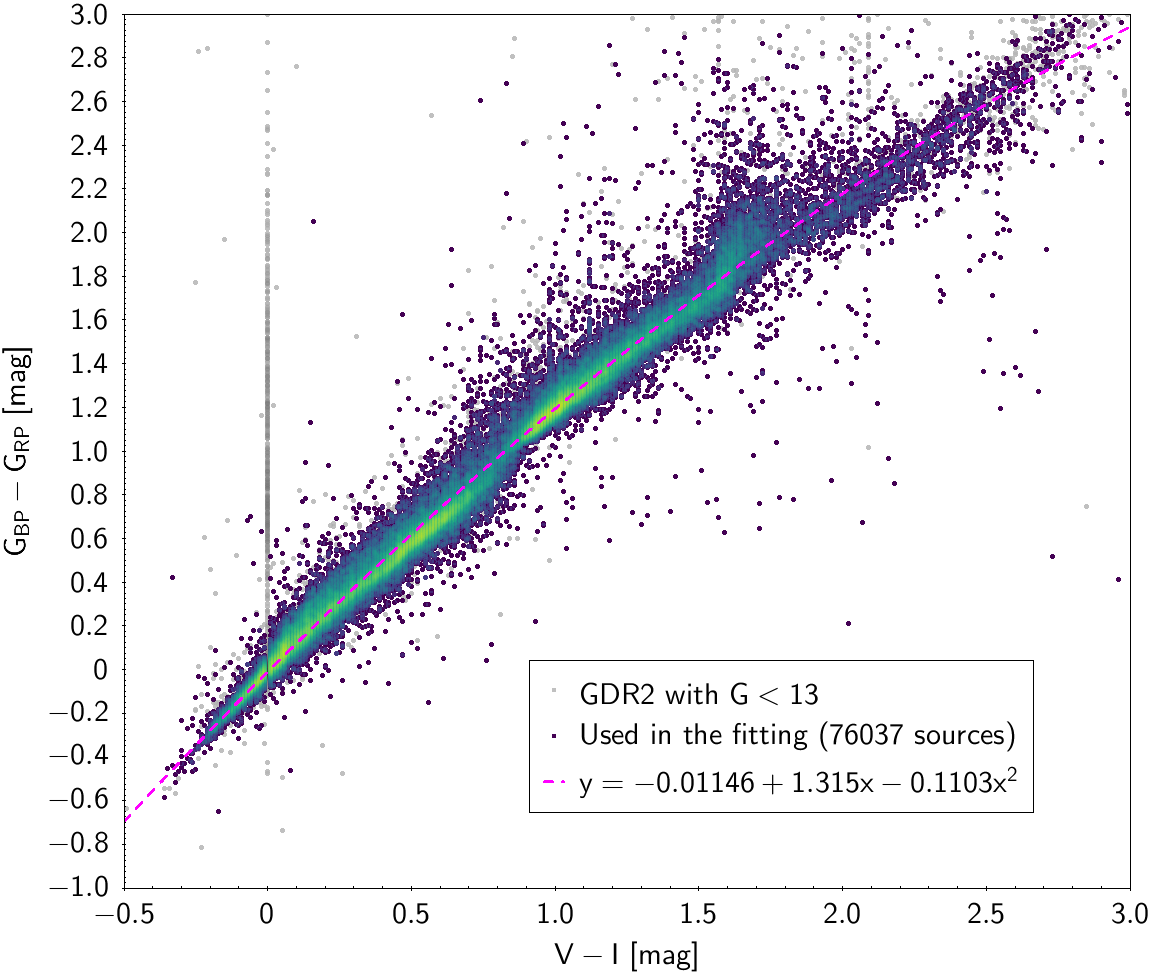}
        \includegraphics[width=0.3\textwidth]{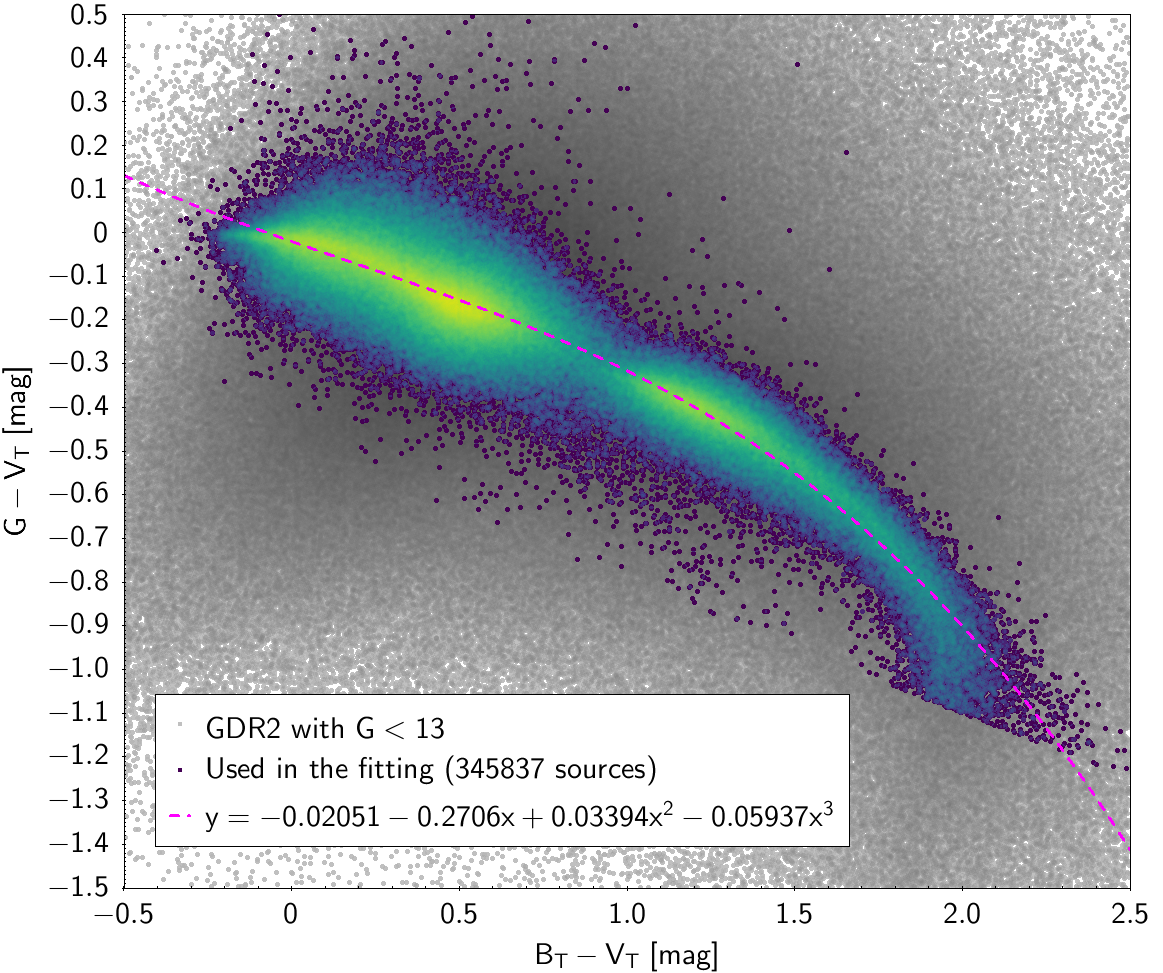}
        \includegraphics[width=0.3\textwidth]{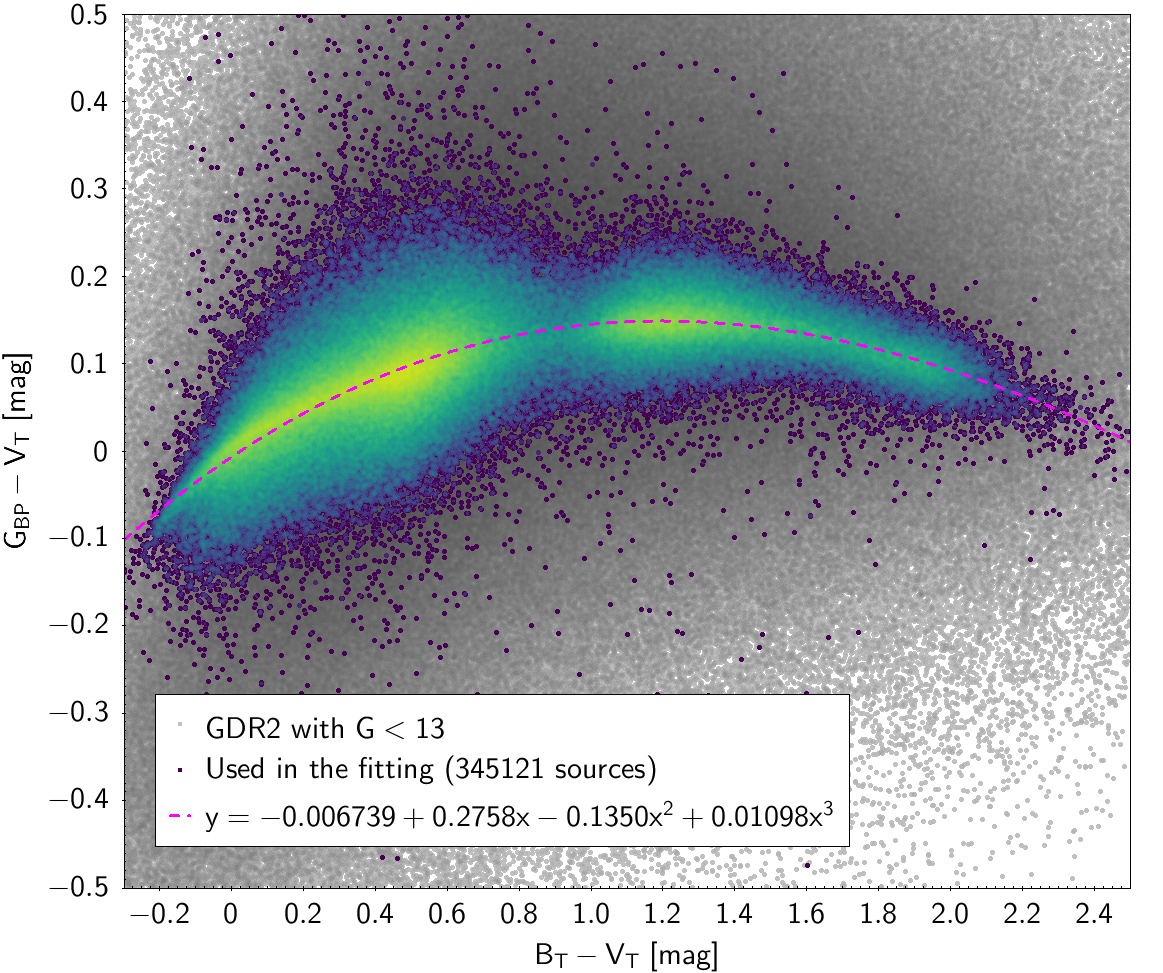}
        \includegraphics[width=0.3\textwidth]{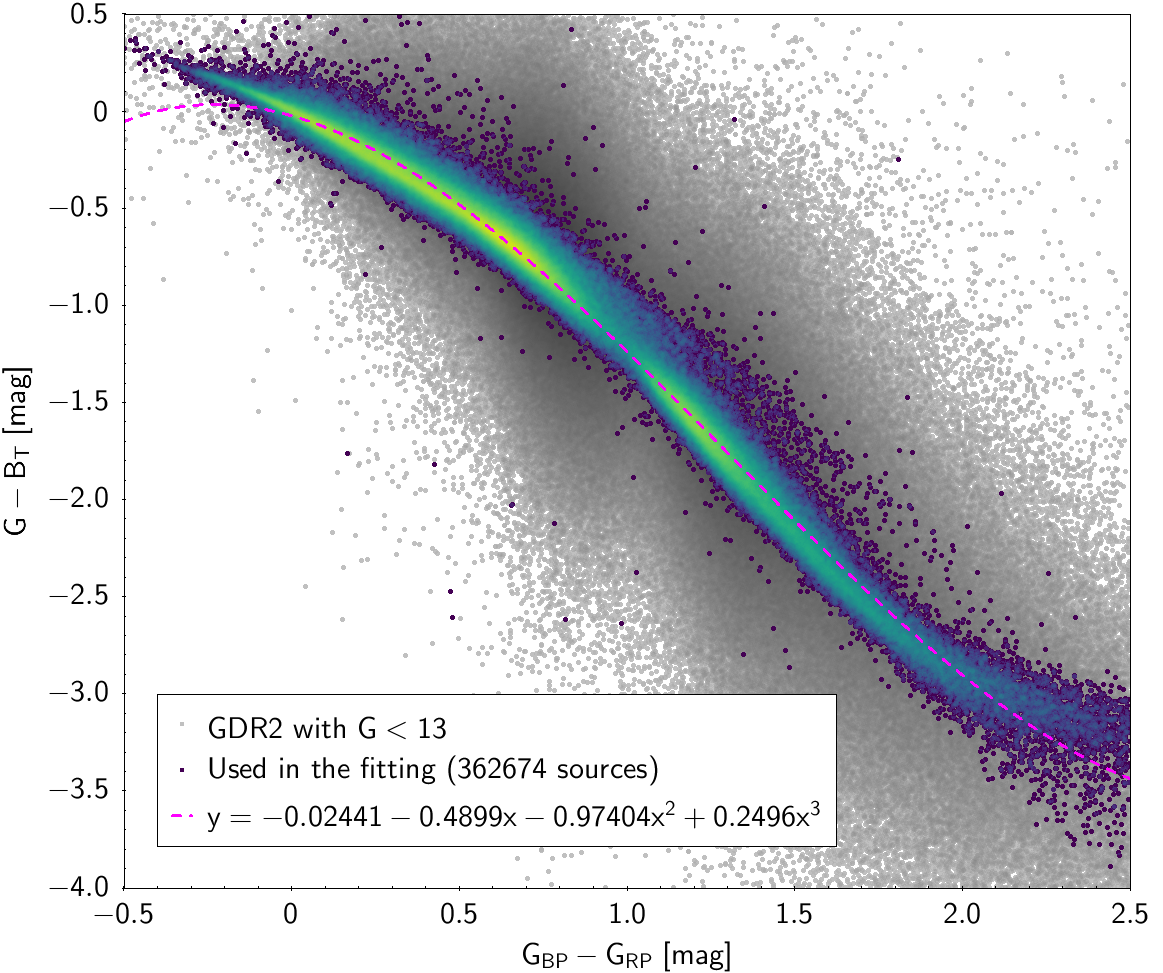}
        \includegraphics[width=0.3\textwidth]{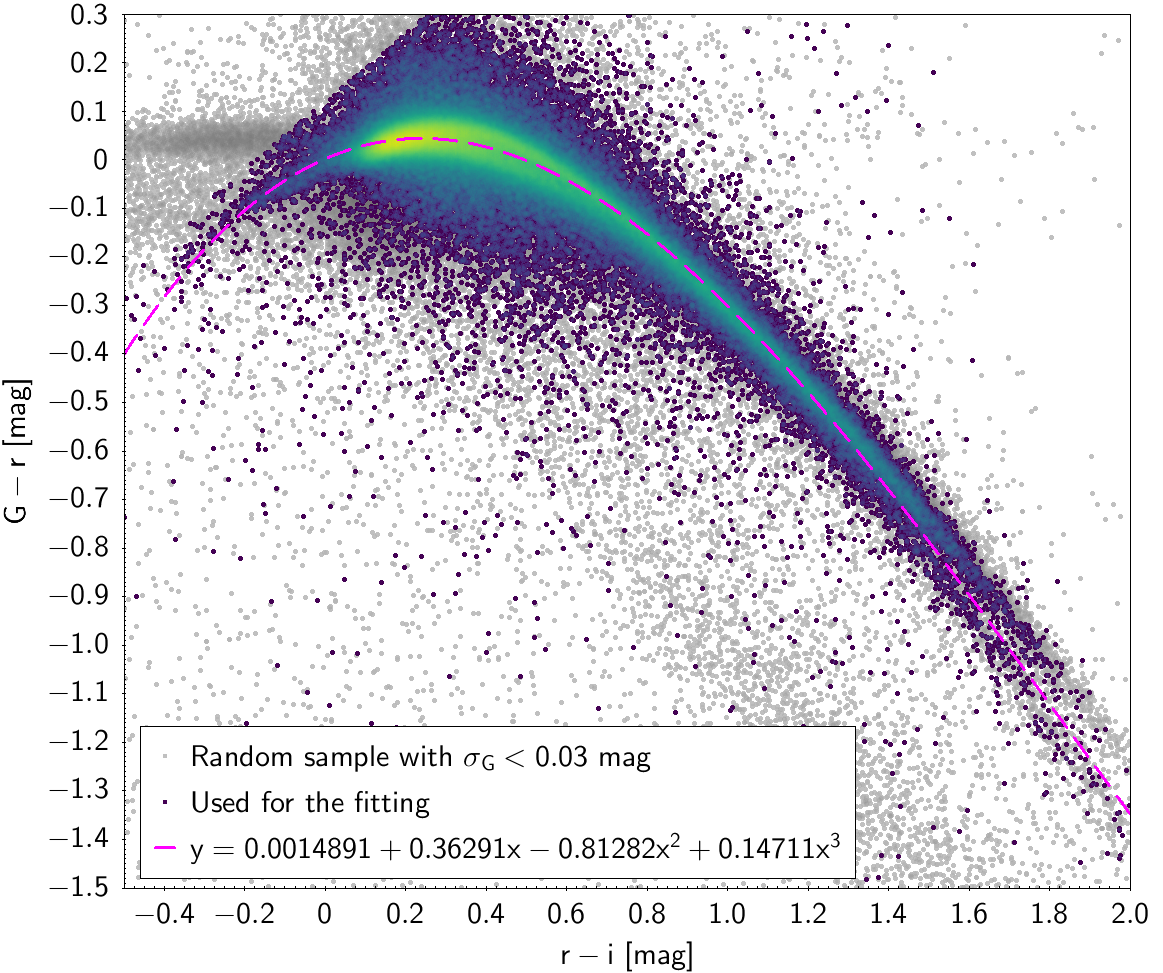}
        \includegraphics[width=0.3\textwidth]{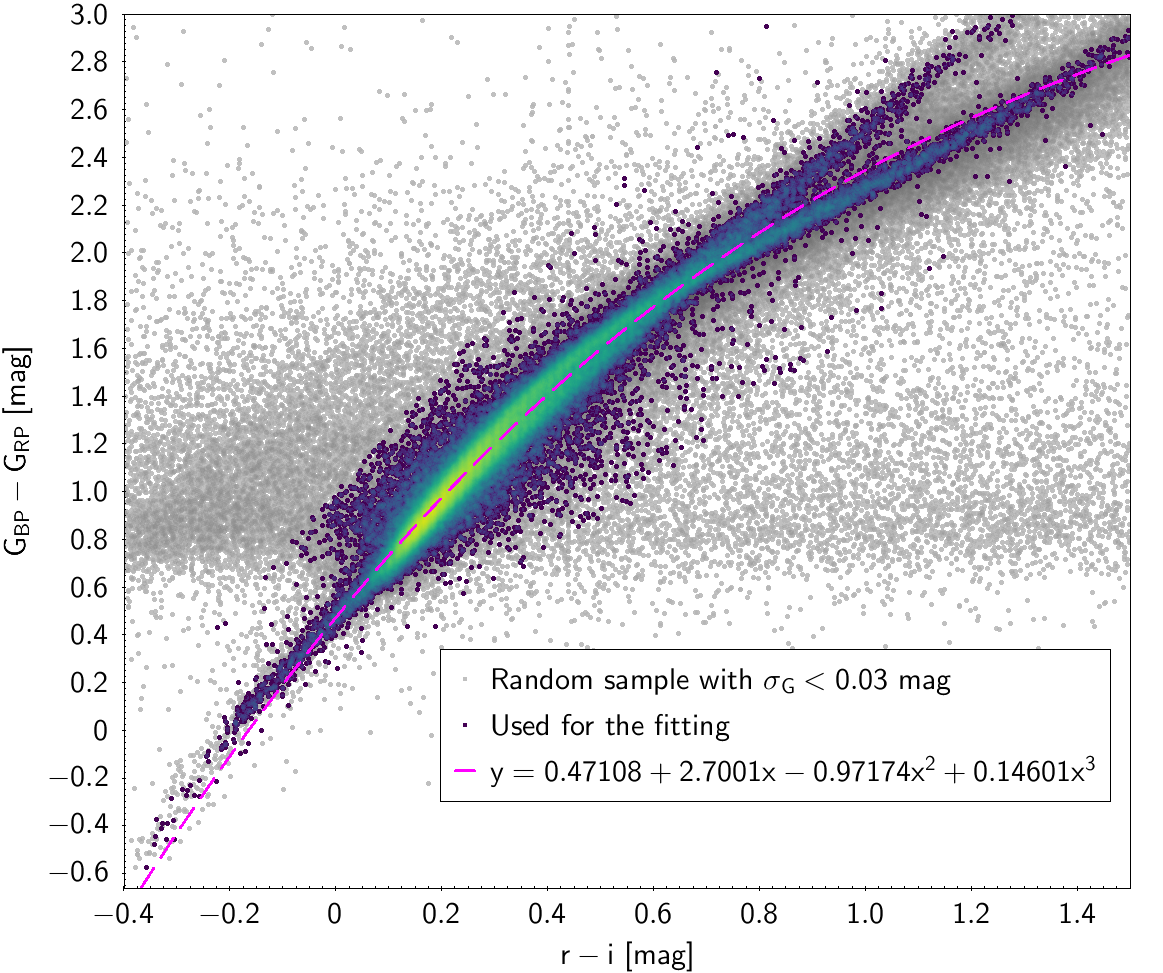}
        \includegraphics[width=0.3\textwidth]{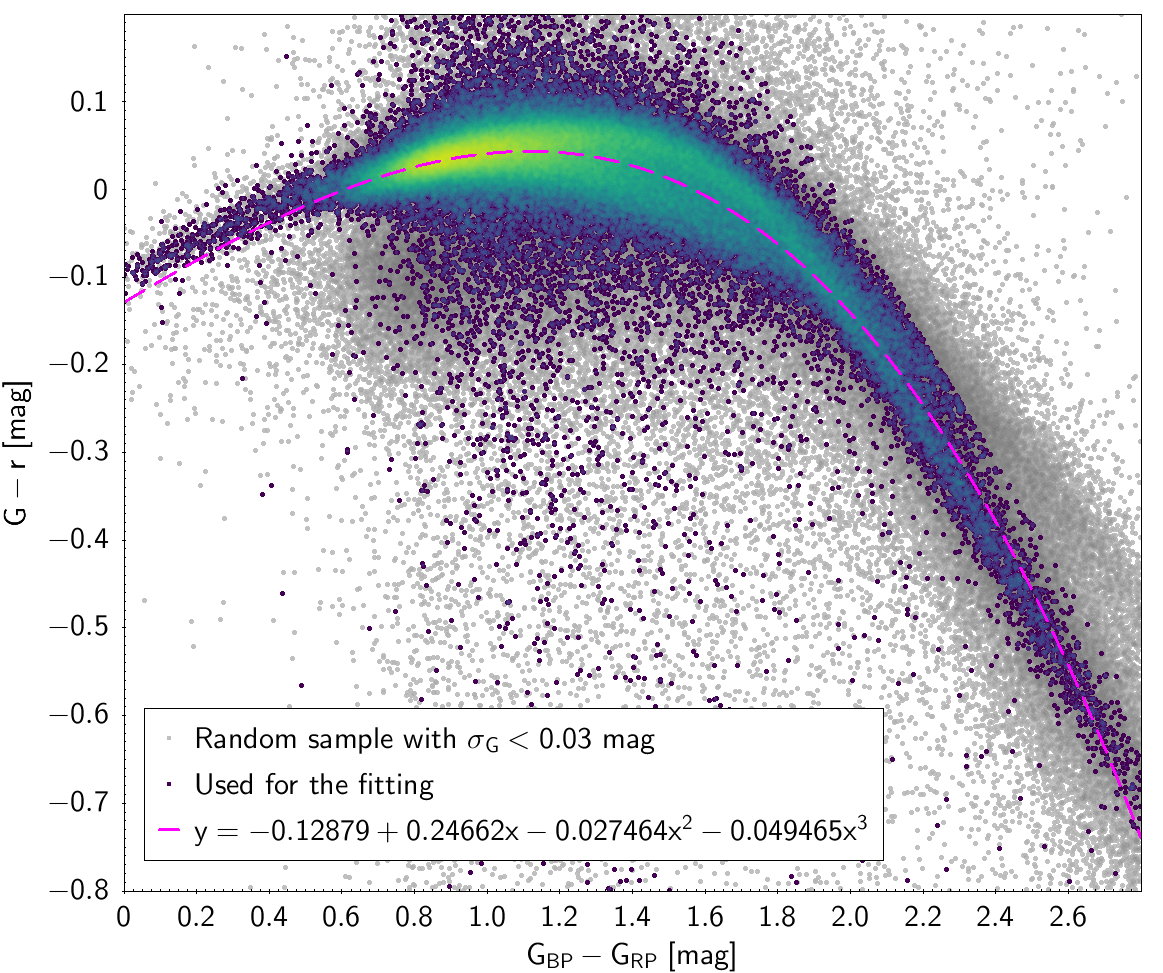}
        \includegraphics[width=0.3\textwidth]{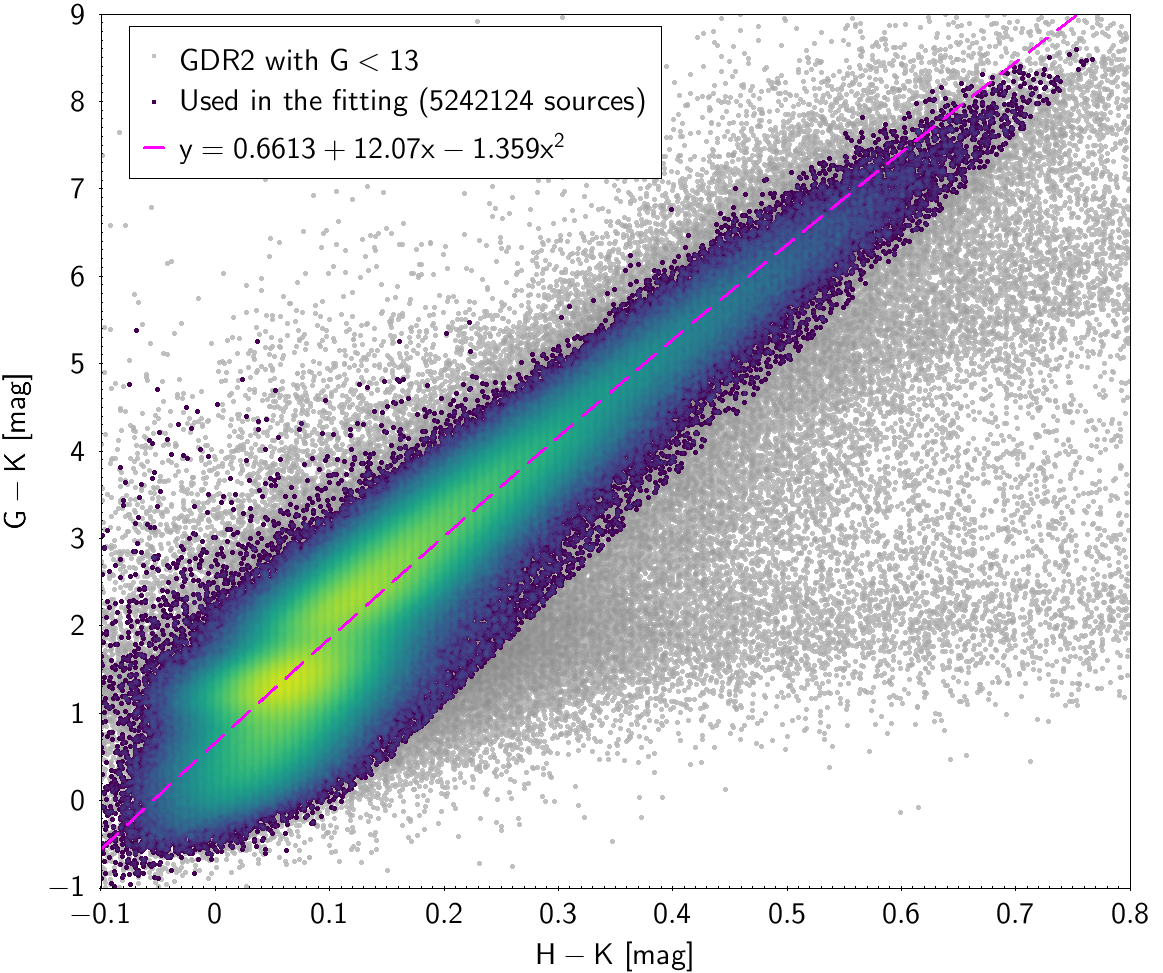}
        \includegraphics[width=0.3\textwidth]{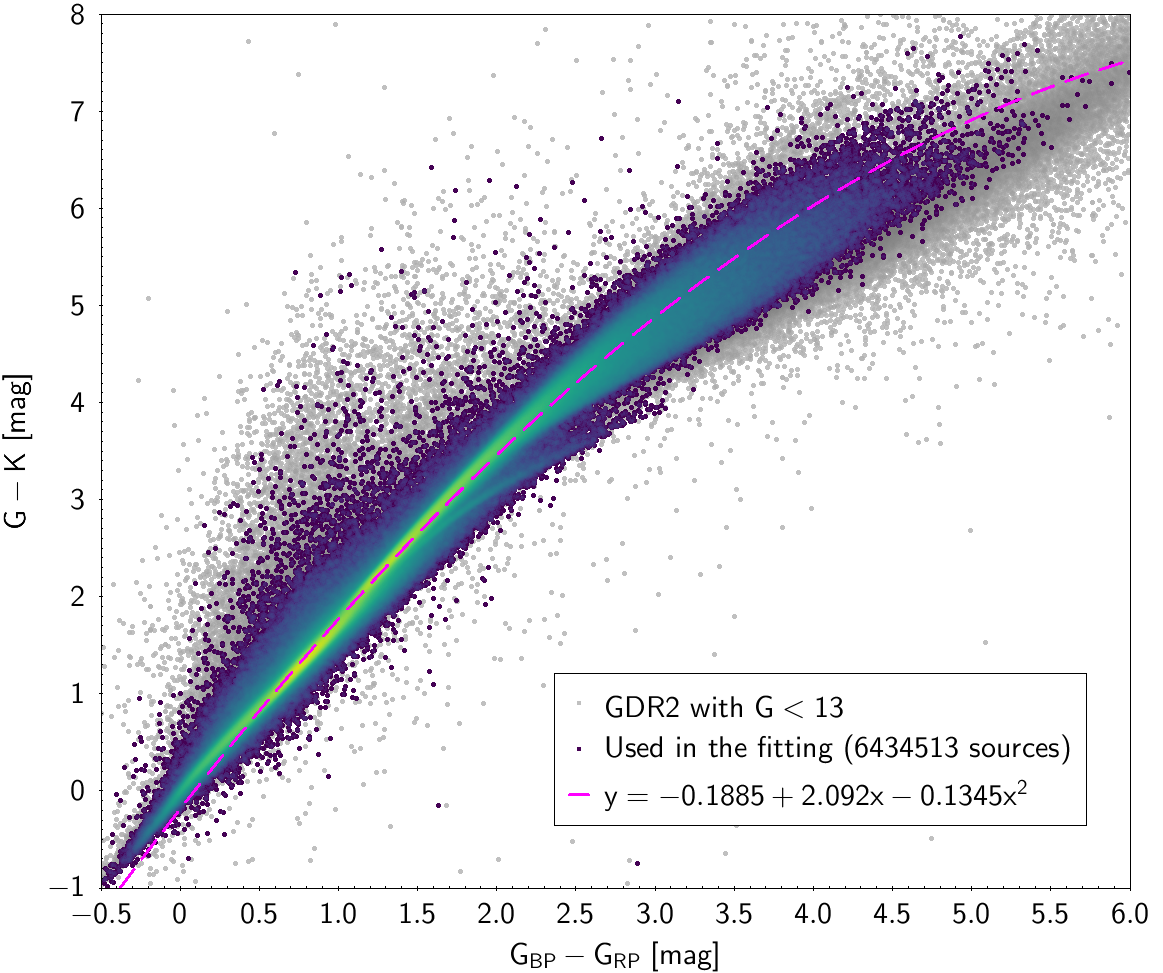}
        \includegraphics[width=0.3\textwidth]{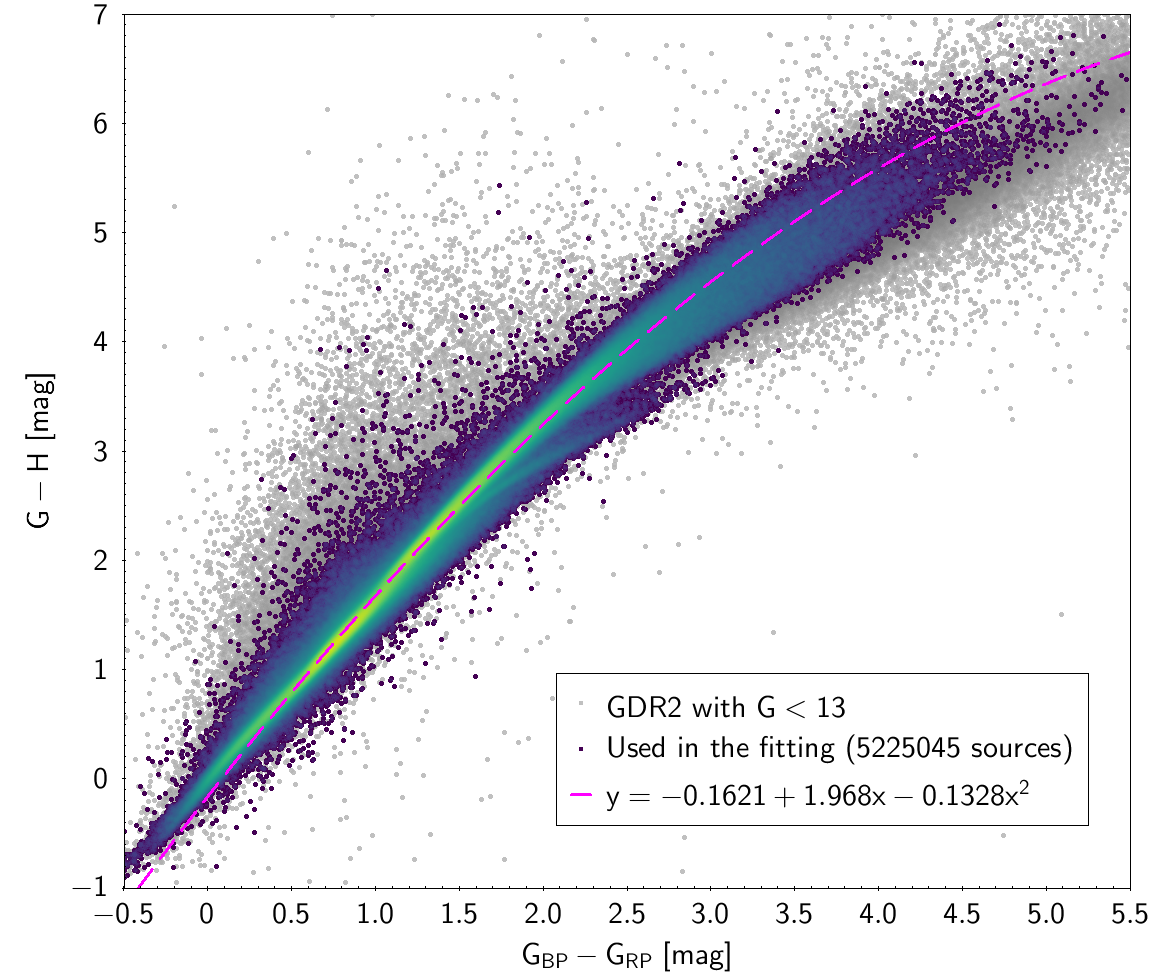}
        \includegraphics[width=0.3\textwidth]{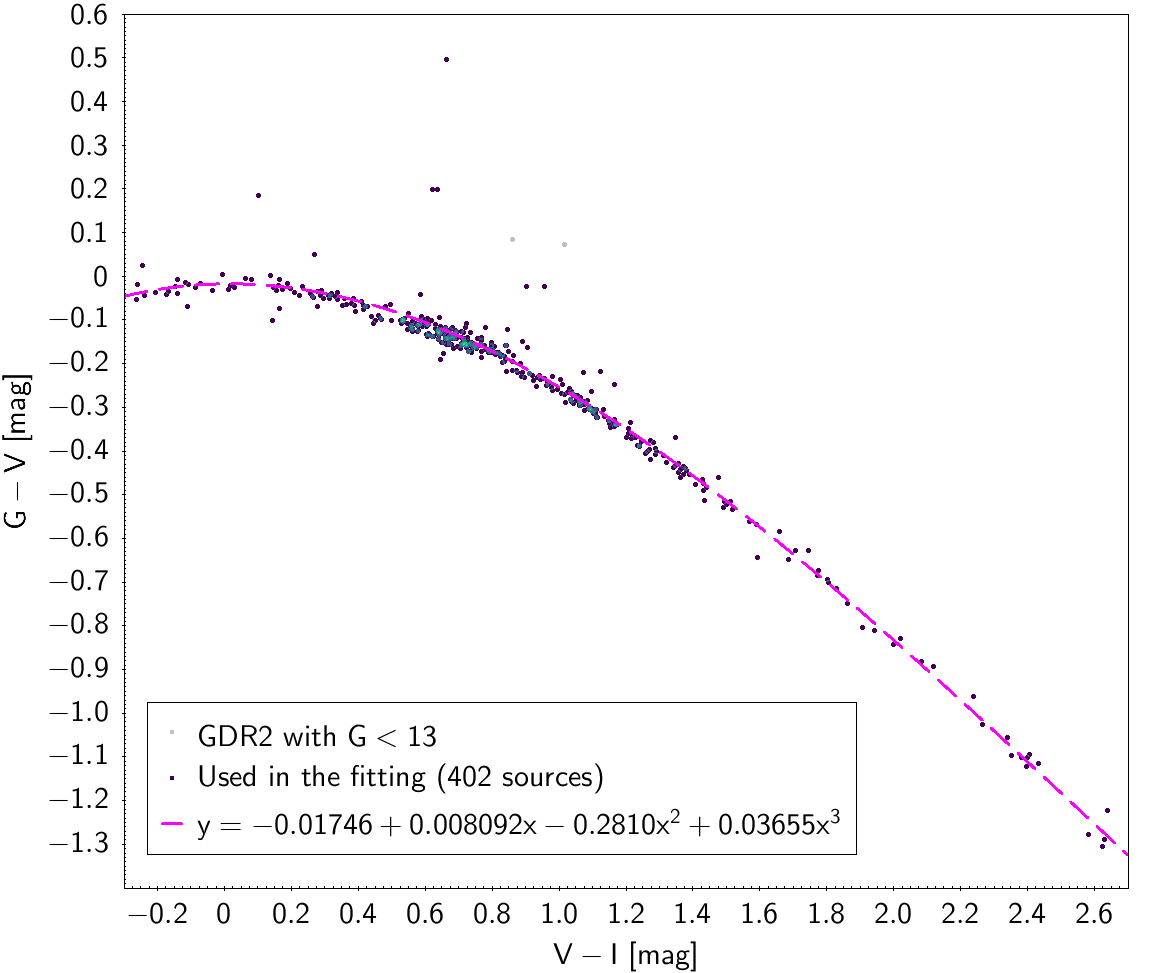}
        \includegraphics[width=0.3\textwidth]{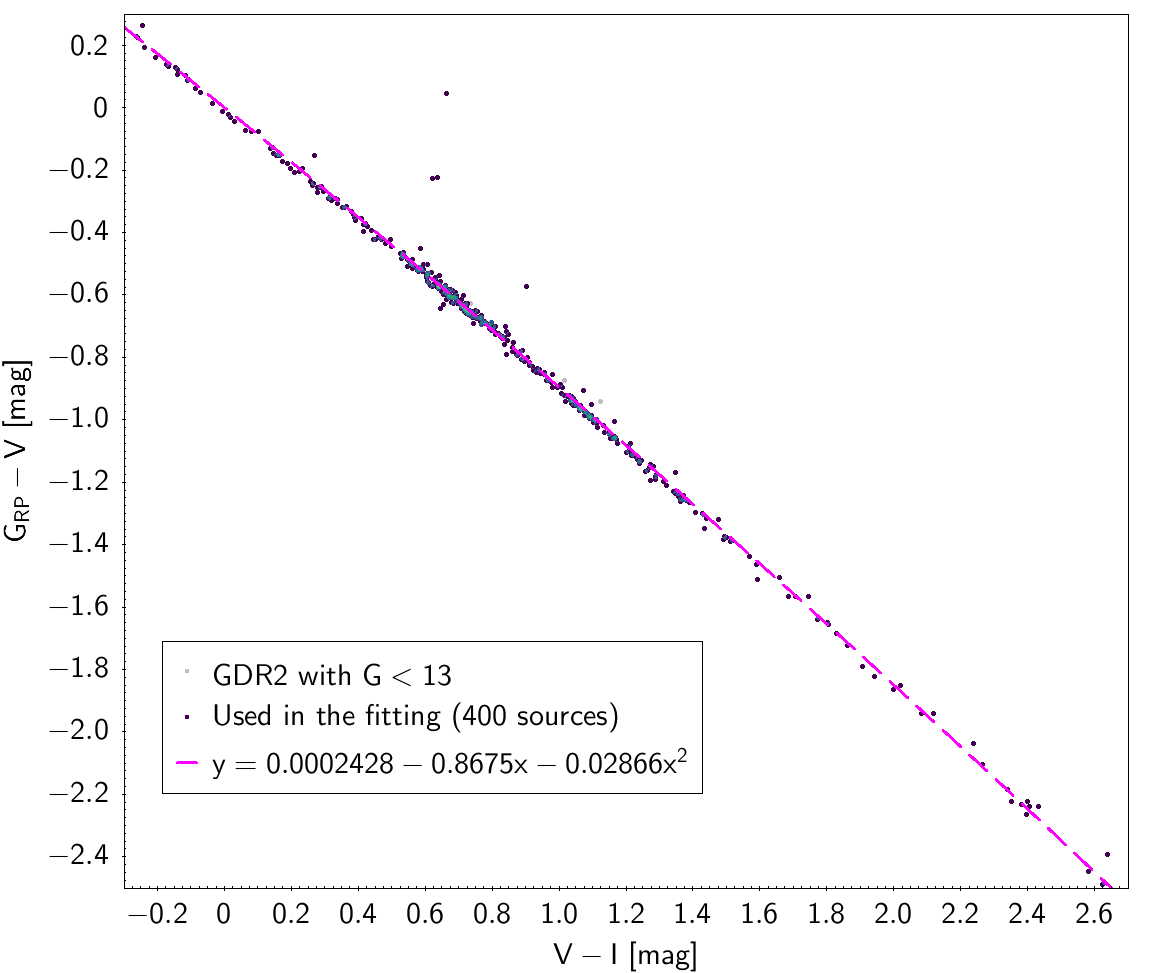}
        \includegraphics[width=0.3\textwidth]{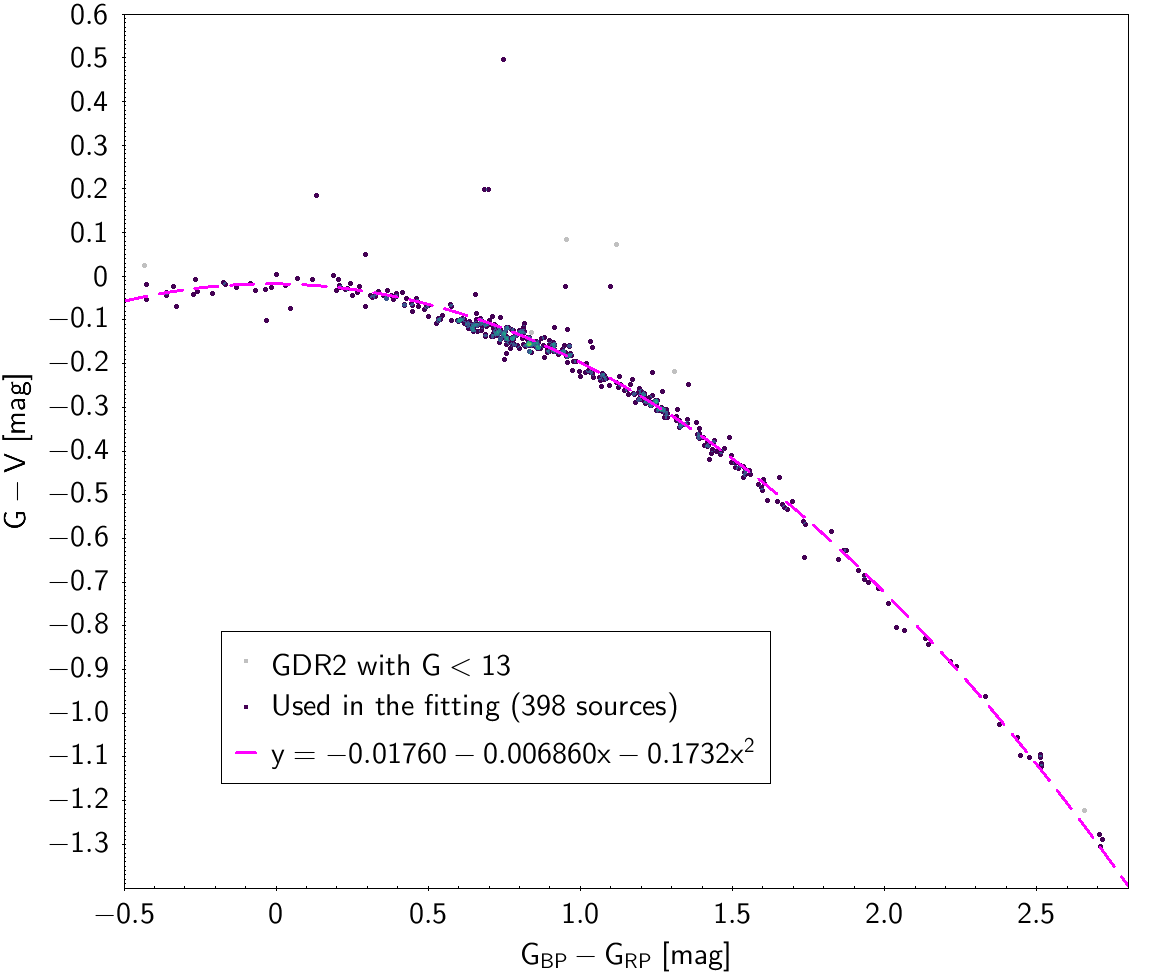}
        \caption[Some diagrams]{Some photometric relationships obtained using {\GDR2} data cross-matched with \Hipparcos\ (top), \TychoTwo\ (second row), SDSS12 (third row), 2MASS (fourth row), and Johnson-Cousins (bottom). 
        \label{fig:some} 
        }
        \end{center}
\end{figure*}

In order to produce the photometric transformations included here, we used only {\GDR2} sources with $G<13$ that could be cross-matched with the external catalogues. As SDSS observations with magnitudes brighter than 14~mag are saturated, we have used a different dataset, extracting from {\GDR2} catalogue some sources with $\sigma_G<0.003$ for all magnitudes and using only those sources with SDSS magnitudes fainter than 15~mag. In order to obtain cleaner transformations, some filtering was made in each colour-colour diagram, which can be found in the {\GDR2} online documentation. 
For the fitting, we used only sources with small magnitude error and small excess flux   ($C$ in Sec.~\ref{Sect:BpRpExcess}) in BP and RP with respect to AF fluxes. We applied the filtering proposed in Eq.~\ref{eq:excessflux} using $a=1.5$ and $b=0.03$ values. 
$ZP_{\rm DR2}$ photometric zeropoints in Table~\ref{Tab:ext_zp} were used.
The polynomial coefficients obtained with the resulting sources are included in Table~\ref{tab:coefs}. The validity of these fittings is, of course, only applicable in the colour intervals used for the fitting (see Table~\ref{tab:applicabilityAll}). Some photometric relationships derived between {\Gaia} and \Hipparcos, \TychoTwo, SDSS12, Johnson-Cousins, and 2MASS are plotted in Fig.~\ref{fig:some}
(for more plots, check the {\GDR2} online documentation).

\begin{table}
\caption[Range of applicability for the relationships]{Range of applicability for the relationships found here between {\Gaia} passbands and other photometric systems.}
\label{tab:applicabilityAll}
\begin{center}
\begin{tabular}{c c c}
\hline
\multicolumn{2}{c}{\textbf{Hipparcos\ relationships}}\\
$G-H_P=f(B-V)$                                   & $-0.2<B-V<1.5$        \\
$G-H_P=f(V-I)$                                   & $-0.3<V-I<3.0$        \\
$G-H_P=f(G_{\rm BP}-G_{\rm RP})$                         & $-0.5<G_{\rm BP}-G_{\rm RP}<4.0$      \\
$G_{\rm BP}-H_P=f(V-I)$                                  & $0.0<V-I<2.5$         \\
$G_{\rm RP}-H_P=f(V-I)$                                  & $-0.3<V-I<2.5$        \\
$G_{\rm BP}-G_{\rm RP}=f(V-I)$                           & $-0.4<V-I<2.5$        \\
\hline
\multicolumn{2}{c}{\textbf{Tycho-2\ relationships}}\\
$G-V_T=f(B_T-V_T)$                               & $-0.1<B_T-V_T<2.0$    \\
$G-V_T=f(G_{\rm BP}-G_{\rm RP})$                 & $-0.3<G_{\rm BP}-G_{\rm RP}<4.0$      \\
$G-B_T=f(G_{\rm BP}-G_{\rm RP})$                 & $0.0<G_{\rm BP}-G_{\rm RP}<2.5$      \\
$G_{\rm BP}-V_T=f(B_T-V_T)$                      & $-0.2<B_T-V_T<2.5$    \\
$G_{\rm RP}-V_T=f(B_T-V_T)$                      & $-0.2<B_T-V_T<2.0$    \\
$G_{\rm BP}-G_{\rm RP}=f(B_T-V_T)$               & $-0.2<B_T-V_T<2.0$    \\
\hline
\multicolumn{2}{c}{\textbf{SDSS relationships}}\\
$G-g=f(g-i)$			 			 & $-1.0<g-i<4.0$	\\ 
$G-g=f(g-r)$			 			 & $-0.6<g-r<1.5$	\\ 
$G-r=f(r-i)$			 			 & $-0.5<r-i<2.0$	\\ 
$G-i=f(r-i)$			 			 & $-0.2<r-i<1.8$	\\ 
$G_{\rm BP}-r=f(r-i)$		 			 & $-0.1<r-i<0.9$	\\ 
$G_{\rm RP}-r=f(r-i)$		 			 & $-0.4<r-i<1.0$	\\ 
$G_{\rm BP}-G_{\rm RP}=f(r-i)$  	 		 & $-0.4<r-i<0.8$	 \\
$G-r=f(G_{\rm BP}-G_{\rm RP})$  	 		 & $0.2<G_{\rm BP}-G_{\rm RP}<2.7$	 \\
$G-i=f(G_{\rm BP}-G_{\rm RP})$  	 		 & $0.0<G_{\rm BP}-G_{\rm RP}<4.5$	 \\
$G-g=f(G_{\rm BP}-G_{\rm RP})$  	 		 & $-0.5<G_{\rm BP}-G_{\rm RP}<2.0$	 \\
\hline
\multicolumn{2}{c}{\textbf{Johnson-Cousins relationships}}\\
$G-V=f(V-I)$                                             & $-0.3<V-I<2.7$       \\
$G_{\rm BP}-V=f(V-I)$                                    & $-0.3<V-I<2.7$       \\
$G_{\rm RP}-V=f(V-I)$                                    & $-0.3<V-I<2.7$       \\
$G_{\rm BP}-G_{\rm RP}=f(V-I)$                           & $-0.3<V-I<2.7$       \\
$G-V=f(V-R)$                                             & $-0.15<V-R<1.4$      \\
$G-V=f(B-V)$                                             & $-0.3<B-V<2.4$       \\
$G-V=f(G_{\rm BP}-G_{\rm RP})$                           & $-0.5<G_{\rm BP}-G_{\rm RP}<2.75$  \\
$G-R=f(G_{\rm BP}-G_{\rm RP})$                           & $-0.5<G_{\rm BP}-G_{\rm RP}<2.75$  \\
$G-I=f(G_{\rm BP}-G_{\rm RP})$                           & $-0.5<G_{\rm BP}-G_{\rm RP}<2.75$  \\
\hline
\multicolumn{2}{c}{\textbf{2MASS relationships}}\\
$G-K_{\rm S}=f(H-K_{\rm S})$                              & $-0.1<H-K_{\rm S}<0.7$ \\
$G_{\rm BP}-K_{\rm S}=f(H-K_{\rm S})$                     & $-0.1<H-K_{\rm S}<0.6$ \\
$G_{\rm RP}-K_{\rm S}=f(H-K_{\rm S})$                     & $-0.1<H-K_{\rm S}<0.7$ \\
$G_{\rm BP}-G_{\rm RP}=f(H-K_{\rm S})$                    & $-0.1<H-K_{\rm S}<0.55$ \\
$G-K_{\rm S}=f(G_{\rm BP}-G_{\rm RP})$                    & $0.25<G_{\rm BP}-G_{\rm RP}<5.5$ \\
$G-H=f(G_{\rm BP}-G_{\rm RP})$                            & $0.25<G_{\rm BP}-G_{\rm RP}<5.0$ \\
$G-J=f(G_{\rm BP}-G_{\rm RP})$                            & $-0.5<G_{\rm BP}-G_{\rm RP}<5.5$ \\
\hline
 \end{tabular}
 \end{center}
 \end{table}
 
\begin{table*}
\caption[Coefficients of the photometric transformations.]{Coefficients of the polynomials 
and the standard deviations ($\sigma$) of the residuals of the fittings for 
sources observed in {\GDR2}.}
\label{tab:coefs}
\footnotesize
\begin{center}
\begin{tabular}{c c c c c c }
\hline
\multicolumn{6}{c}{\textbf{Hipparcos\ relationships}}\\
& & $B-V$ & $(B-V)^{2}$ & $(B-V)^{3}$ &  $\sigma$ \\   
$G-Hp$          & -0.03704 &-0.3915 & 0.01855 &-0.03239 & 0.02502 \\
& & $V-I$ & $(V-I)^{2}$ & $(V-I)^{3}$ &  $\sigma$ \\   
$G-Hp$          & -0.01259 &-0.3336 &-0.1171 & 0.02244 & 0.09610 \\
$G_{\rm BP}-Hp$    & -0.01205 & 0.1105 & & &  0.06624\\
$G_{\rm RP}-Hp$    &  -0.00055 &-1.203 & 0.1078 & & 0.1089     \\
$G_{\rm BP}-G_{\rm RP}$   & -0.01146 & 1.315 &-0.1103 & & 0.09390 \\
& & $G_{\rm BP}-G_{\rm RP}$ & $(G_{\rm BP}-G_{\rm RP})^{2}$ & $(G_{\rm BP}-G_{\rm RP})^{3}$  & $\sigma$ \\   
$G-Hp$          & -0.01968 &-0.2344 &-0.1200 & 0.01490 & 0.06875\\
\hline
\multicolumn{6}{c}{\textbf{Tycho-2\ relationships}}\\
& & $B_T-V_T$ & $(B_T-V_T)^{2}$ & $(B_T-V_T)^{3}$ &  $\sigma$\\   
$G-V_T$          & -0.02051 &-0.2706 & 0.03394 &-0.05937 &  0.06373     \\
$G_{\rm BP}-V_T$   & -0.006739 & 0.2758 &-0.1350 & 0.01098 &  0.04415   \\
$G_{\rm RP}-V_T$   & -0.03757 &-1.188 & 0.4871 &-0.1729 &  0.09637 \\
$G_{\rm BP}-G_{\rm RP}$   &  0.03147 & 1.456 &-0.6043 & 0.1750 &  0.07024 \\
& & $G_{\rm BP}-G_{\rm RP}$ & $(G_{\rm BP}-G_{\rm RP})^{2}$ & $(G_{\rm BP}-G_{\rm RP})^{3}$ &   $\sigma$ \\
$G-V_T$          & -0.01842 &-0.06629 &-0.2346 & 0.02157 &  0.05501    \\
$G-B_T$          & -0.02441 &-0.4899 &-0.9740 & 0.2496 &  0.07293 \\
\hline
\multicolumn{6}{c}{\textbf{SDSS relationships}}\\
& & $g-i$ & $(g-i)^{2}$ & $(g-i)^{3}$ &  $\sigma$\\	
$G-g$           & -0.074189  &-0.51409 &-0.080607 & 0.0016001 & 0.046848 \\ 
& & $g-r$ & $(g-r)^{2}$ & $(g-r)^{3}$ &  $\sigma$\\	
$G-g$           & -0.038025  &-0.76988 &-0.1931  & 0.0060376 & 0.065837 \\
& & $r-i$ & $(r-i)^{2}$ & $(r-i)^{3}$ &  $\sigma$ \\   
$G-r$           &  0.0014891 & 0.36291 &-0.81282  & 0.14711   & 0.043634 \\
$G-i$           & -0.007407  & 1.4337  &-0.95312  & 0.22049   & 0.041165 \\
$G_{\rm BP}-r$       &  0.1463    & 1.7244  &-1.1912   & 0.22004   & 0.059674 \\
$G_{\rm RP}-r$       & -0.29869   &-1.1303  &          &           & 0.039156 \\
$G_{\rm BP}-G_{\rm RP}$   &  0.47108   & 2.7001  &-0.97174  & 0.14601   & 0.067573 \\
& & $G_{\rm BP}-G_{\rm RP}$ & $(G_{\rm BP}-G_{\rm RP})^{2}$ & $(G_{\rm BP}-G_{\rm RP})^{3}$ &  $\sigma$ \\   
$G-r$           & -0.12879   & 0.24662 &-0.027464 &-0.049465  & 0.066739 \\
$G-i$           & -0.29676   & 0.64728 &-0.10141  &	      & 0.098957 \\
$G-g$           &  0.13518   &-0.46245 &-0.25171  & 0.021349  & 0.16497  \\
\hline
\multicolumn{6}{c}{\textbf{Johnson-Cousins relationships}}\\
& & $V-I$ & $(V-I)^{2}$ & $(V-I)^{3}$ &   $\sigma$ \\  
$G-V$           & -0.01746 & 0.008092 &-0.2810 & 0.03655 & 0.04670 \\
$G_{\rm BP}-V$       & -0.05204 & 0.4830 &-0.2001 & 0.02186 & 0.04483 \\
$G_{\rm RP}-V$       & 0.0002428 &-0.8675 &-0.02866 & & 0.04474 \\
$G_{\rm BP}-G_{\rm RP}$   & -0.04212 & 1.286 &-0.09494 & & 0.02366 \\
& & $V-R$ & $(V-R)^{2}$ & $(V-R)^{3}$ &  $\sigma$ \\  
$G-V$           & -0.02269 & 0.01784 &-1.016 & 0.2225 & 0.04895 \\
& & $B-V$ & $(B-V)^{2}$ & $(B-V)^{3}$ &  $\sigma$ \\  
$G-V$           & -0.02907 &-0.02385 &-0.2297 &-0.001768 & 0.06285\\
& & $G_{\rm BP}-G_{\rm RP}$ & $(G_{\rm BP}-G_{\rm RP})^{2}$ &   & $\sigma$ \\  
$G-V$           & -0.01760 &-0.006860 &-0.1732 & & 0.045858 \\
$G-R$           & -0.003226 & 0.3833 &-0.1345 & & 0.04840 \\
$G-I$           &  0.02085 & 0.7419 &-0.09631 & &0.04956\\
\hline
\multicolumn{6}{c}{\textbf{2MASS relationships}}\\
& & $H-K_{\rm S}$ & $(H-K_{\rm S})^{2}$ &  &  $\sigma$ \\
$G-K_{\rm S}$           &  0.6613 & 12.073 &-1.359 & & 0.3692  \\
$BP-K_{\rm S}$          &  0.8174 & 14.66 & 2.711 & & 0.4839\\
$RP-K_{\rm S}$          &  0.3560 & 9.287 &-1.721 & &0.2744 \\
$BP-RP$         &  0.4238 & 6.098 & 1.991 & & 0.2144 \\
& & $G_{\rm BP}-G_{\rm RP}$ & $(G_{\rm BP}-G_{\rm RP})^2$ &  &  $\sigma$ \\  
$G-K_{\rm S}$           & -0.1885 & 2.092 &-0.1345 & & 0.08281 \\
$G-H$           & -0.1621 & 1.968 &-0.1328 & & 0.07103 \\
$G-J$           &  -0.01883 & 1.394 &-0.07893 & & 0.05246\\
\hline
 \end{tabular}
 \end{center}
 \end{table*}

\begin{figure*}[!htbp]
        \begin{center}
        \includegraphics[width=0.32\textwidth]{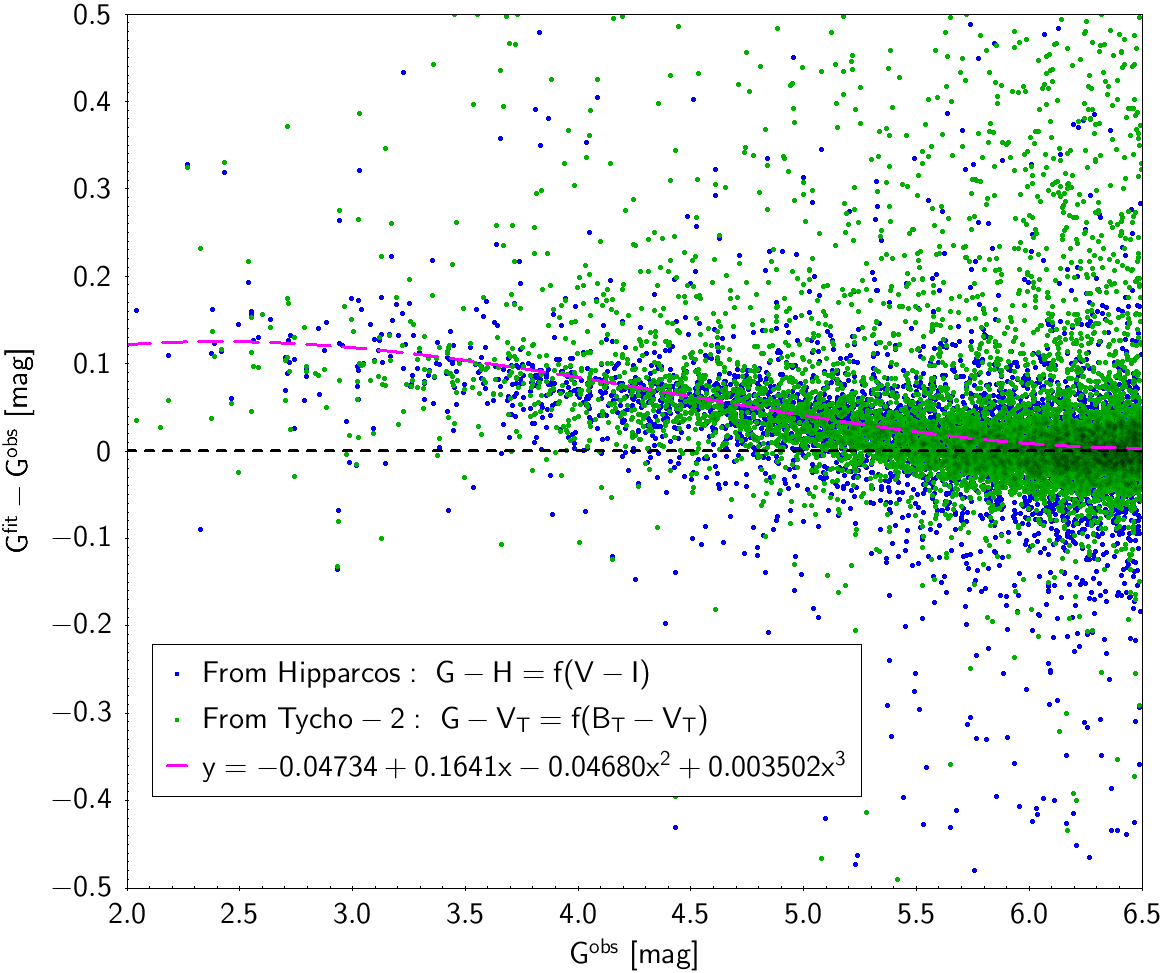}
        \includegraphics[width=0.32\textwidth]{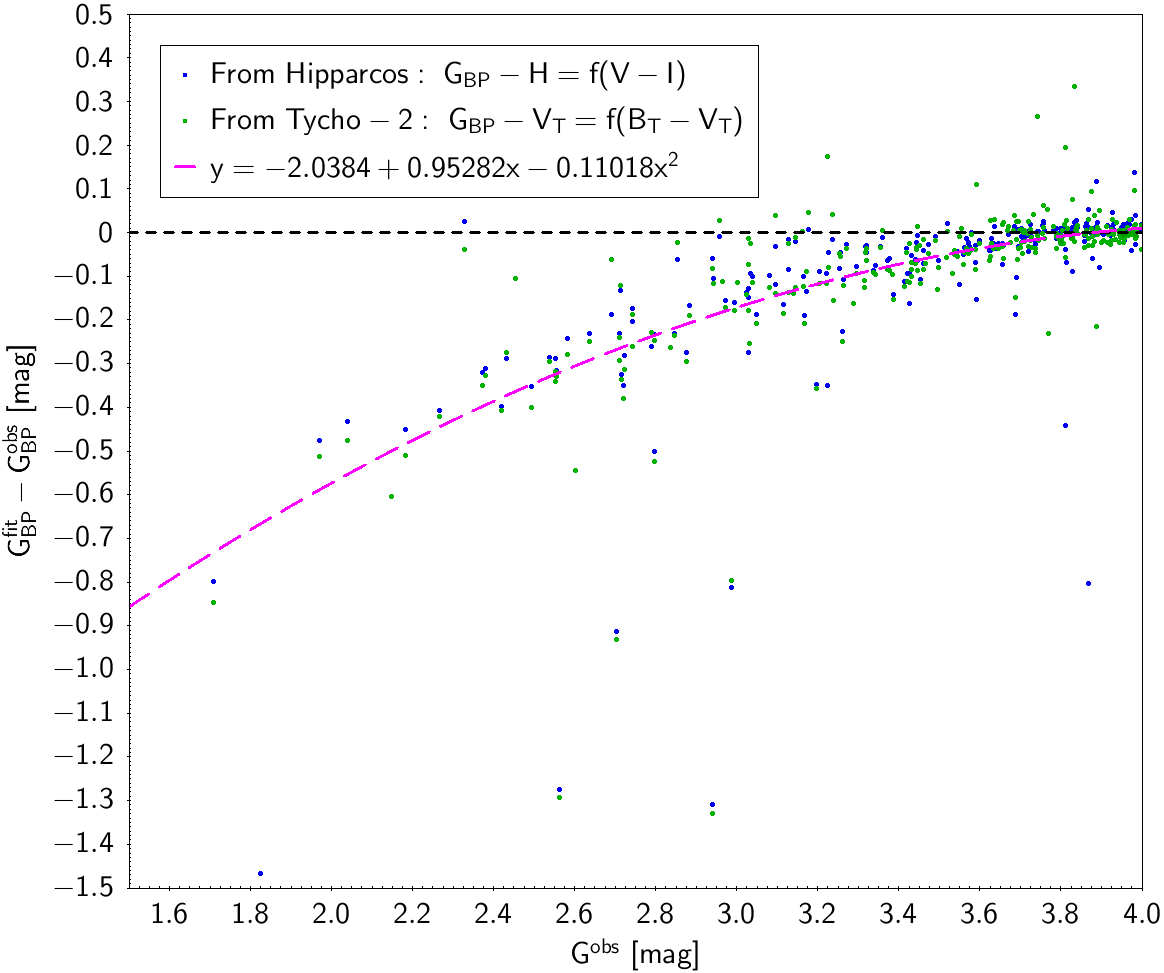}
        \includegraphics[width=0.32\textwidth]{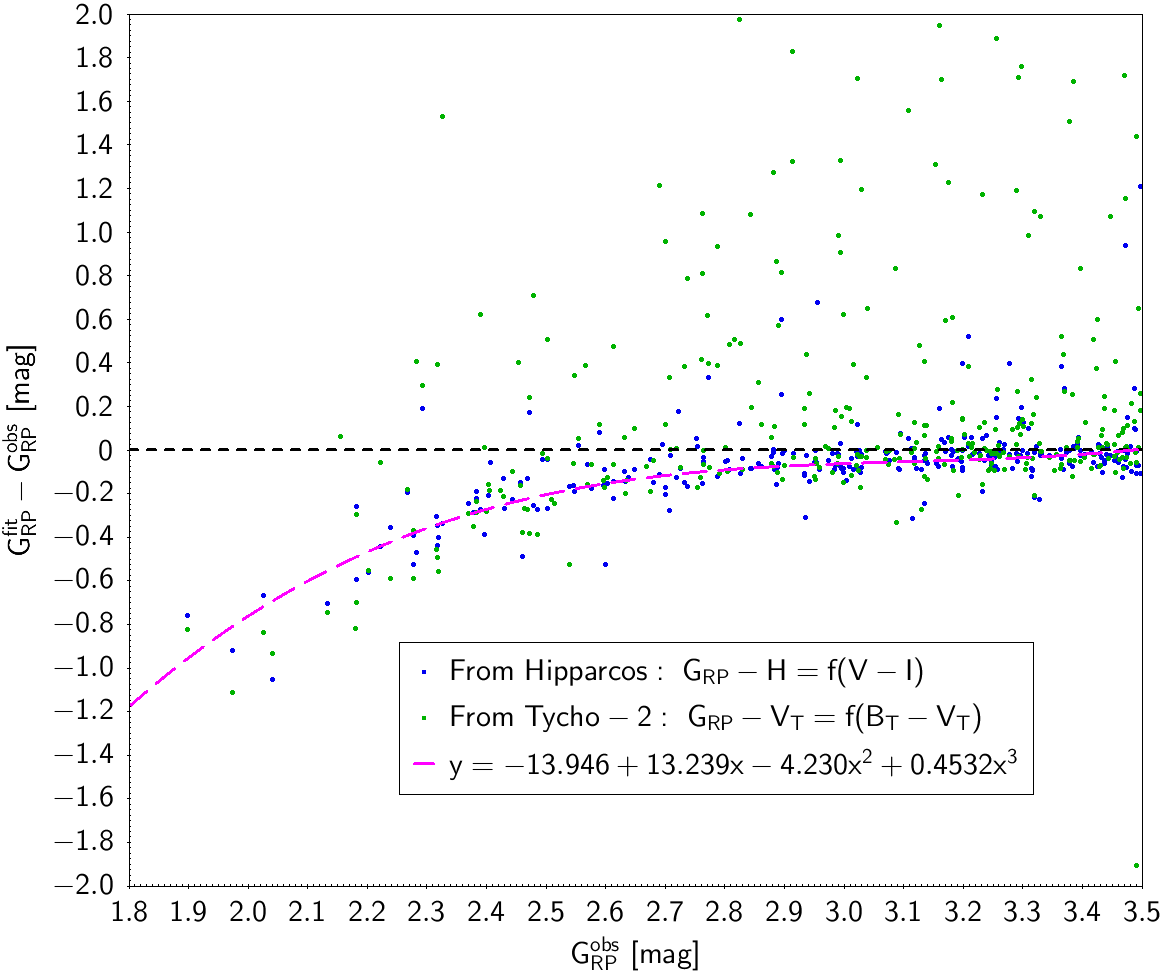}
        \caption{Residuals of the fitted laws for \Hipparcos\ and \TychoTwo\ in Table~\ref{tab:coefs} as a function of the magnitude for $G$ (left), $G_{\rm BP}$ (centre), and $G_{\rm RP}$ (right).
        \label{fig:saturation} 
        }
        \end{center}
\end{figure*}

\section{Saturation}\label{Sect:Saturation}

By analysing the residuals found when fitting $G-H_p$ as a function of $V-I$, or $G-V_T$ as a function of $B_T-V_T$ and plotting them against the magnitude, we can see that the current calibration does not completely account for saturation effects (see~Fig.~\ref{fig:saturation}). 
Fitting a relationship to the trends at the bright end in Fig.~\ref{fig:saturation} can be useful if saturation effects need to be empirically corrected. We combined \TychoTwo\ 
and \Hipparcos\ data to derive these fittings. The corrected magnitudes from the mean magnitudes in {\GDR2}, $G_{\rm XP}^{\rm corr}$ can be obtained with the following equations: 

\begin{eqnarray}
G^{\rm corr}&=&-0.047344+1.16405G-0.046799G^2\nonumber\\
&&+0.0035015G^3\\
G_{\rm BP}^{\rm corr}&=&-2.0384+1.95282 G-0.11018 G^2\\
G_{\rm RP}^{\rm corr}&=&-13.946+14.239 G_{\rm RP}-4.23 G_{\rm RP}^2+0.4532 G_{\rm RP}^3
.\end{eqnarray}

The $G$ relationship should only be used in the range $2.0<G_{\rm obs}<6.5$~mag, the $G_{\rm BP}$ relationship only for $2.0<G_{\rm obs}<4.0$~mag, and the $G_{\rm BP}$ relationship only for $2.0<G_{\rm obs}<3.5$~mag.

\section{Nomenclature}\label{Sect:Acronyms}
Below, we list acronyms and concepts used in this paper.

\small

\begin{tabular}{lp{5cm}}
\hline\hline
2MASS & Two-Micron All Sky Survey\\
AC & ACross scan: direction on the focal plane perpendicular to the scan direction\\ 
AF & Astrometric Field: the 62 astrometric CCDs on the focal plane \\
AL & ALong scan: direction on the focal plane parallel to the scan direction\\
APASS & AAVSO Photometric All-Sky Survey \\
BP & Blue Photometer: the system defined by the blue dispersion prism. Also refers to the associated CCDs.\\ 
Calibration Unit & Set of  parameters and configurations identifying an observation environment, e.g. CCD, passband, Gate, FoV, Window Class.\\
\end{tabular}

\begin{tabular}{lp{5cm}}
CCD & Charge-coupled Device\\
CCD transit & Transit of a source across a single CCD \\
DPAC & Data Processing and Analysis Consortium\\
ESA & European Space Agency \\
\end{tabular}

\begin{tabular}{lp{5cm}}
FoV & \parbox[t]{5cm}{Field of View: one of the two pointing directions of the satellite telescopes. See \citetads{2016A&A...595A...1G} 
for more information regarding the structure of \Gaia.}\\
\end{tabular}

\begin{tabular}{lp{5cm}}
FoV transit & Field-of-view transit, the complete transit of a source across the focal plane\\
Gate & \parbox[t]{5cm}{System designed to reduce the effective exposure time for bright sources that would otherwise suffer heavy saturation. Twelve different
gate configurations can be activated on board in different magnitude ranges \citepads{2016A&A...595A...1G}.}\\
HEALPix & Hierarchical Equal Area isoLatitude Pixelation of a sphere producing a subdivision of a spherical surface in which each pixel covers the same surface area as every other pixel \citepads{2005ApJ...622..759G}.\\ 
IDU & Intermediate Data Update: task operating at the start of each processing cycle on all \Gaia\ data collected from the start of the
mission. This includes a new cross-match \citepads{DR2-DPACP-45}, PSF/LSF calibration and IPD.\\
IPD & \parbox[t]{5cm}{Image Parameter Determination: the task that estimates the fluxes that are then calibrated in PhotPipe as described in \citetads{2016A&A...595A...7C}. 
See \citetads{2016A&A...595A...3F} for more details.}\\
LS & Large-scale: in this paper this refers to one set of photometric calibration describing effects that vary on a large spatial range \\
LSF & Line Spread Function \\
OBMT & \parbox[t]{5cm}{On-board Mission Timeline: the timescale usually used when referring to time in the \Gaia\ context. This scale is
defined in \citetads{2016A&A...595A...1G}.}\\ 
OBMT-Rev & Time in OBMT scale but in units of revolutions rather than nanoseconds (one revolution corresponds to approximately 6 hours)\\ 
Row & A series of CCDs in the AL direction \\
RP & Red Photometer: the system defined by the red dispersion prism. Also refers to the associated CCDs\\
SDSS & Sloan Digital Sky Survey \\
Strip & A series of CCDs in the AC direction\\ 
SS & Small-scale: in this paper this refers to one set of photometric calibration describing effects that vary on a small spatial range\\ 
Window Class & \parbox[t]{5cm}{Configuration defining the AL and AC size of the window centred on each detected source as received on the ground \citepads[see][]{2016A&A...595A...1G}}\\
\hline
\end{tabular}

\end{appendix}

\end{document}